\documentclass[a4paper,11pt]{article}
\pdfoutput=1
\usepackage[no-natbib-sort]{jheppub}

\usepackage[utf8]{inputenc}
\usepackage{etoolbox}
\usepackage{hhline}
\usepackage{array}
\usepackage{amssymb,amsmath,amsthm,amsfonts,graphicx}
\usepackage{latexsym}
\usepackage{bm}
\usepackage[]{hyperref}
\usepackage{cleveref}
\usepackage{nccmath}	

\makeatletter
    \patchcmd{\maketitle}{\@fpheader}{}{}{}
    \makeatother

\def\be{\begin{equation}}
\def\ee{\end{equation}}
\def\ba{\begin{eqnarray}}
\def\ea{\end{eqnarray}}

\def\scalar{{\mathbf S}} 
\def\vector{{\mathbf V}} 
 
\def\A{{\cal A}} 
\def\F{{\cal F}} 
\def\E{{\cal E}} 

\title{Strong cosmic censorship: taking the rough with the smooth}

\author[a]{Oscar~J.~C.~Dias,}
\emailAdd{ojcd1r13@soton.ac.uk}
\affiliation[a]{STAG research centre and Mathematical Sciences, Highfield Campus, University of Southampton, Southampton SO17 1BJ, UK}

\author[b]{Harvey~S.~Reall}
\emailAdd{hsr1000@cam.ac.uk}
\affiliation[b]{Department of Applied Mathematics and Theoretical Physics, University of Cambridge, Wilberforce Road, Cambridge CB3 0WA, UK} 

\author[b]{and Jorge~E.~Santos}
\emailAdd{jss55@cam.ac.uk}

\abstract{
It has been argued that the strong cosmic censorship conjecture is violated by Reissner-Nordstr\"om-de Sitter black holes: for near-extremal black holes, generic scalar field perturbations arising from smooth initial data have finite energy at the Cauchy horizon even though they are not continuously differentiable there. In this paper, we consider the analogous problem for coupled gravitational and electromagnetic perturbations. We find that such perturbations exhibit a much worse violation of strong cosmic censorship: for a sufficiently large near-extremal black hole, perturbations arising from smooth initial data can be extended through the Cauchy horizon in an arbitrarily smooth way. This is in apparent contradiction with an old argument in favour of strong cosmic censorship. We resolve this contradiction by showing that this old argument is valid only for initial data that is not smooth. This is in agreement with the recent proposal that, to recover strong cosmic censorship, one must allow rough initial data. 
}

\begin{document}

\maketitle

\section{Introduction}

The strong cosmic censorship conjecture \cite{penrose} asserts that, in some physically relevant class of initial data for Einstein's equation ({\it e.g.} smooth, complete, asymptotically flat), the maximal Cauchy development is, generically, inextendible. In other words, classical physics is predictable from the initial data. The Reissner-Nordstr\"om and Kerr solutions of the vacuum Einstein equation (with vanishing cosmological constant $\Lambda$) admit Cauchy horizons. Consistency with the conjecture requires that such a Cauchy horizon is non-generic: it is expected that, if the initial data is perturbed, then generically the resulting perturbed spacetime will not admit a Cauchy horizon \cite{Simpson:1973ua,McNamara499,chandra,Poisson:1990eh}. 

Making this conjecture precise is surprisingly subtle.\footnote{See Ref. \cite{Dafermos:2012np} for a more detailed discussion.} Various arguments indicate that, when the initial data is perturbed, the spacetime metric (and other fields) {\it can} be extended continuously across a Cauchy horizon \cite{McNamara121,Ori:1991zz,Dafermos:2003wr,Franzen:2014sqa}. For the Kerr solution, this has been proved recently \cite{Dafermos:2017dbw}. So the ``$C^0$ formulation" of the strong cosmic censorship conjecture (where ``inextendible" means ``inextendible with continuous metric") is false. However, it has also been argued that, generically, curvature invariants diverge at the Cauchy horizon, so the extended spacetime cannot be $C^2$ there \cite{Poisson:1990eh}. Hence the $C^2$ formulation of strong cosmic censorship appears to be true.\footnote{
This formulation of strong cosmic censorship has been proved for spherically symmetric solutions of Einstein-Maxwell theory (with $\Lambda=0$) coupled to a massless scalar field \cite{Luk:2017jxq}.} This is not the end of the story because the total tidal distortion experienced by an observer crossing the Cauchy horizon can remain finite, so the divergence in curvature might not be strong enough to destroy a macroscopic observer \cite{Ori:1991zz}. Therefore demanding a $C^2$ metric appears to be too strong a requirement. 

Ultimately, the question of whether or not an observer can cross the Cauchy horizon depends on the equations of motion for the matter that the observer is made of. And of course the observer will have an effect on the geometry determined by the Einstein equation. This motivates formulating the strong cosmic censorship conjecture as the statement that the maximal Cauchy development should be inextendible {\it as a solution of the equations of motion}. Since the equations of motion are second order, one might think this implies that the fields should be $C^2$. However, one can still make sense of the equations of motion with lower smoothness than this by considering {\it weak} solutions.\footnote{
Given a system of (quasilinear) second order partial differential equations, multiply each equation by a smooth test function of compact support and integrate over spacetime, integrating by parts to eliminate second derivatives of the fields. The fields constitute a weak solution if they satisfy this set of equations for arbitrary test functions. 
} 
Weak solutions have important physical applications {\it e.g.} they describe shocks in a compressible perfect fluid. For the vacuum Einstein equation, a weak solution must have locally square integrable Christoffel symbols in some chart. This leads to Christodoulou's formulation \cite{Christodoulou:2008nj} of the strong cosmic censorship conjecture, that, generically the maximal Cauchy development is intextendible as a spacetime with locally square integrable Christoffel symbols. If this is correct then, generically, one cannot extend beyond the Cauchy horizon consistently with the classical equation of motion. 
 
A popular toy model for studying strong cosmic censorship is a linear massless scalar field. In this case, the analogue of the Christodoulou formulation of strong cosmic censorship is that, for generic smooth initial data, at the Cauchy horizon the scalar field will not belong to the Sobolev space $H^1_{\rm loc}$ of functions that are locally square integrable with a locally square integrable gradient.\footnote{A function is ``locally square integrable" if it is square integrable when multipled by any smooth test function of compact support.}  More informally, the energy of the scalar field will diverge at the Cauchy horizon. Here ``energy" refers to the energy on a spacelike surface intersecting the Cauchy horizon, according to an observer with velocity normal to the surface. This linear version of the Christodoulou formulation of the strong cosmic censorship conjecture has been proved to be true for Reissner-Nordstr\"om \cite{Luk:2015qja} and Kerr \cite{Dafermos:2015bzz} black holes with $\Lambda=0$. 
 
The instability of the Cauchy horizon arises from a blue-shifting of perturbations entering the black hole at late time. It was observed long ago that this effect is weaker for $\Lambda>0$ because there is a competing red-shifting of late time perturbations since such perturbations can disperse by falling across the cosmological horizon.\footnote{
For $\Lambda<0$ one would expect the Cauchy horizon instability to be stronger than for $\Lambda=0$ because perturbations outside a black hole decay very slowly. It has been suggested that the $C^0$ formulation of strong cosmic censorship might be valid for $\Lambda<0$ \cite{Dafermos:2017dbw}.} This led to the claim that the $C^2$ version of strong cosmic censorship is violated for near-extremal Reissner-Nordstr\"om-de Sitter (RNdS)  \cite{Mellor:1989ac} or Kerr-de Sitter (Kerr-dS) \cite{Chambers:1994ap} black holes. However, subsequent work \cite{Brady:1998au} argued that this conclusion is invalid because it neglects backscattering of outgoing radiation just inside the event horizon. It was argued that, in the presence of such outgoing radiation, the Cauchy horizon instability is still strong enough to ensure that the $C^2$ formulation of strong cosmic censorship is respected.\footnote{There is a problem with this claim which we will discuss below.} Nevertheless, it has been conjectured that the {\it Christodoulou} formulation would be violated for near-extremal RNdS and Kerr-dS black holes \cite{Dafermos:2012np}. As we have discussed above, this formulation seems more relevant than the $C^2$ formulation.

Interest in this topic has been revived by recent work of Cardoso {\it et al} \cite{Cardoso:2017soq}. This work considered linear massless scalar field perturbations of a RNdS black hole. It was found that, for a near-extremal black hole, such perturbations have finite energy at the Cauchy horizon and therefore violate the toy model of strong cosmic censorship discussed above. Going beyond the toy model, one can consider the backreaction of the scalar field on the geometry using nonlinear results of Refs. \cite{Hintz:2016gwb,Hintz:2016jak,Costa:2017tjc}. Cardoso {\it et al} argued that, at the nonlinear level, such perturbations would respect the $C^2$ formulation of strong cosmic censorship but, for a near-extremal black hole, the Christodoulou formulation would be violated, in agreement with the conjecture of Ref. \cite{Dafermos:2012np}.

This raises the question of whether the same worrying behaviour is exhibited in the more physical case of Kerr-dS black holes. Surprisingly, the answer appears to be negative: Ref. \cite{Dias:2018ynt} argued that the Christodoulou formulation of strong cosmic censorship is respected by gravitational (or massless scalar field) perturbations of such black holes, even close to extremality. Thus the evidence suggests that, for $\Lambda>0$, the Christodoulou formulation of strong cosmic censorship is respected by the vacuum Einstein equation but not by the Einstein-Maxwell-massless scalar field equations!\footnote{
For massless scalar field perturbations, it has been argued that a near-extremal Kerr-Newman-dS black hole respects strong cosmic censorship provided that it rotates sufficiently rapidly \cite{Hod:2018lmi}. The latter condition cannot be relaxed because the zero rotation limit gives RNdS, for which strong cosmic censorship is violated.}  

Our discussion so far has concerned only perturbations arising from {\it smooth} initial data. Very recently, Dafermos and Shlapentokh-Rothman (DSR) \cite{Dafermos:2018tha} have suggested a way of rescuing strong cosmic censorship with $\Lambda>0$, namely to consider initial perturbations which are not smooth. As discussed above, the equations of motion can be formulated even with low differentiability. For linear massless scalar field perturbations of RNdS, DSR proved that, generically, the solution at the Cauchy horizon is less regular (in the sense of Sobolev spaces) than the initial data. Now there will be some minimum level of regularity which is acceptable, either physically or mathematically, {\it e.g.} for finiteness of energy or (in the nonlinear context) for local well-posedness of the initial value problem. The DSR result suggests a ``rough" ({\it i.e.} non-smooth) formulation of the strong cosmic censorship conjecture: if one has an initial perturbation with the minimum acceptable level of regularity then, generically, the perturbation at the Cauchy horizon will not have this minimum acceptable regularity \cite{Dafermos:2018tha}. 

A lack of smoothness of the initial perturbation was already present, although not noticed, in the earlier work of Ref. \cite{Brady:1998au}. As we will show in section \ref{sec:background}, the argument of Ref. \cite{Brady:1998au} overlooks a subtlety which implies that this argument only works for initial data that is not $C^1$ at the event horizon. Thus the work of Ref. \cite{Brady:1998au} does not establish that the $C^2$ formulation of strong cosmic censorship is respected, because the initial perturbation does not belong to $C^2$. Instead, as we will explain, the argument of Ref. \cite{Brady:1998au} is evidence in favour of the rough version of strong cosmic censorship proposed by DSR.

In this paper, we will hammer a few more nails into the coffin of the smooth versions of strong cosmic censorship for RNdS. We will study linearized electromagnetic and gravitational perturbations of a RNdS black hole. Our results assume that the perturbation arises from smooth initial data. We will show that, near extremality, Christodoulou's formulation of strong cosmic censorship is violated by such perturbations. This is analogous to the massless scalar field results of Ref. \cite{Cardoso:2017soq}. However, in contrast with that case, our results show that, in pure Einstein-Maxwell theory, the $C^2$ version of strong cosmic censorship is also violated near extremality. In fact, {\it generic perturbations arising from smooth initial data can be arbitrarily smooth at the Cauchy horizon}. More precisely, if one desires that every perturbation arising from smooth initial data is $C^r$ at the Cauchy horizon then this can be achieved by taking the black hole to be close enough to extremality and large enough. Hence, in pure Einstein-Maxwell theory with $\Lambda>0$, not only are the Christodoulou and $C^2$ formulations of strong cosmic censorship violated (for smooth initial data), but so is the $C^r$ formulation for any $r \ge 2$!

This paper is organized as follows. In section \ref{sec:background} we review the RNdS solution and discuss the arguments of Refs. \cite{Mellor:1989ac,Brady:1998au}. We will explain the connection between strong cosmic censorship and quasinormal modes of the RNdS solution. In sections \ref{sec:gravbetabound}-\ref{sec:GravResults}  we discuss linearized electromagnetic and gravitational perturbations of RNdS. We will study these perturbations using the Kodama-Ishisbashi (KI) formalism \cite{Kodama:2003kk}. 
In section~\ref{sec:gravbetabound} we determine the condition for a linearized gravitoelectromagnetic perturbation to be extendible across the Cauchy horizon as a weak solution of the equations of motion. In section~\ref{sec:spectralgap} we give the KI master equations and boundary conditions that we later solve analytically and numerically. We also show that vector-type and scalar-type perturbations in RNdS are isospectral, {\it i.e.} they have the same frequency spectrum. In section~\ref{sec:analytics}, we show that RNdS gravitoelectromagnetic quasinormal modes fall into three familes, as in the case of the quasinormal modes of a scalar field discussed in \cite{Cardoso:2017soq}. For all of them, there are regimes in the parameter space where we can derive some analytical approximations. We compare them with the exact numerical data and this proves valuable to identify and classify the quasinormal mode families. Finally, in section~\ref{sec:GravResults} we present our main results for the spectral gap of gravitational and electromagnetic perturbations.
Section \ref{sec:discussion} contains further discussion of the implications of our results.

\section{Background material}

\label{sec:background}

\subsection{The Reissner-Nordstr\"om de Sitter solution}\label{sec:RN} 

Consider Einstein-Maxwell theory with positive cosmological constant $\Lambda$. The action is $S \propto \int d^4 x \sqrt{-g}\left(R -2\Lambda-F^2 \right)$ where $R$ is the Ricci scalar of the metric $g$ and $F=dA$ is the Maxwell field strength associated to the potential 1-form $A$. We define the de Sitter radius $L$ by
\be
\Lambda=\frac{3}{L^2}\,.
\ee
In static coordinates $(t,r,\theta,\phi)$, the Reissner-Nordstr\"om de Sitter (RNdS) solution with mass and charge parameters $M$ and $Q$ is 
\be \label{metricRN}
ds^2=-f\,dt^2+\frac{dr^2}{f}+r^2d\Omega_2^2\,, \qquad F=E_0 \,dt \wedge dr
\ee
with $d\Omega_2^2$ being the line element of a unit radius $S^2$ (parametrized by $\theta$ and $\phi$) and 
\be \label{metricRNaux}
f(r)=1-\frac{r^2}{L^2} -\frac{2M}{r}+\frac{Q^2}{r^2}\,, \qquad E_0(r)=\frac{Q}{r^2} .
\ee
 For an appropriate range of parameters the function $f$ has $3$ positive roots $r_-\le r_+\le r_c$ corresponding to the Cauchy horizon $\mathcal{CH}^+$, event horizon  $\mathcal{H}^+_R$  and cosmological horizon $\mathcal{H}_C^+$ respectively. We will denote the (positive) surface gravities associated to each of these three horizons as $\kappa_-,\kappa_+$ and $\kappa_c$, respectively. For any non-extremal RNdS black hole it can be shown that \cite{Brady:1998au}
\be
\label{kappaineq}
\kappa_- > \kappa_+ \,.
\ee 
The extremal configuration occurs when $\kappa_+$ and $\kappa_-$ vanish. This happens when $Q=Q_{\rm ext}$ where
\be \label{Qext}
Q_{\rm ext}=r_+ \, \frac{\sqrt{2 y_++1}}{\sqrt{3 y_+^2+2 y_++1}}\,,\qquad \hbox{with}\quad y_+=\frac{r_+}{r_c}\,.
\ee
When presenting many of our results and associated plots we will parametrize  the RNdS solution using the dimensionless parameters $Q/Q_{\rm ext}$ and $y_+$.

\begin{figure}[ht]
	\centering
	\includegraphics[width=0.6\textwidth]{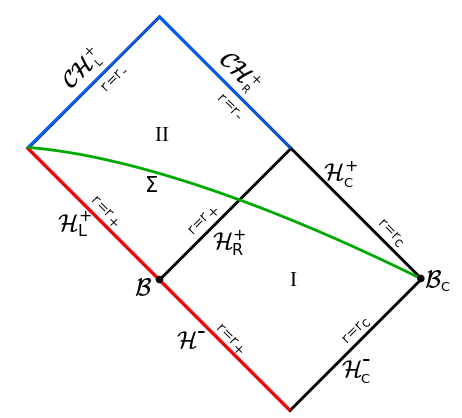}
	\caption{Penrose diagram for Reissner-Nordstr\"om de Sitter. Region I is the black hole exterior and region II the interior. The future event horizon is $\mathcal{H}^+_R$, the future cosmological horizon is $\mathcal{H}_C^+$, and $\mathcal{CH}_{L,R}^+$ are the future Cauchy horizon. $\Sigma$ is a Cauchy surface for regions I and II.}
	\label{fig:penrose}
\end{figure}

The causal structure of a non-extremal RNdS black hole is shown in Fig. \ref{fig:penrose}. Region I is the region with $r_+ < r < r_c$ between the event horizon and cosmological horizon, {\it i.e.} the black hole exterior. Region II is the black hole interior, where $r_-<r<r_+$.   

In region I we define the tortoise coordinate $r_*$ by
\be
 dr_* = \frac{dr}{f}\, 
\ee
and we fix the constant of integration by imposing $r_*=0$ at $r=(r_+ + r_c)/2$. We then define Eddington-Finkelstein coordinates in region I by $u=t-r_*$ and $v=t+r_*$. In ingoing Eddington-Finkelstein coordinates $(v,r,\theta,\phi)$ the metric takes the form
\be
 ds^2 = -fdv^2 + 2 dv dr + r^2 d\Omega_2^2\,.
\ee
This metric can be analytically extended into region II so these coordinates cover regions I and II of Fig. \ref{fig:penrose}. We will also make use of Kruskal coordinates near the event horizon. These are defined in region I by
\be
 U_+ = -e^{-\kappa_+ u}\,,  \qquad V_+ = e^{\kappa_+ v}\,, 
\ee
and these coordinates also allow the metric to be analytically extended into region II (where $U_+>0$, $V_+>0$) as well as two further regions not shown in Fig. \ref{fig:penrose}. The future event horizon ${\cal H}_R^+$ is the surface $U_+ = 0$. In region II, we have $V_+ = e^{\kappa_+ v}$ and we define the coordinate $u$ in this region by
\be
U_+ = +e^{- \kappa_+ u}.
\ee 
Note that $u \rightarrow + \infty$ as we approach ${\cal H}_R^+$ in either region I or region II. In region II we define $t$ and $r_*$ by $u=t-r_*$ and $v=t+r_*$. The coordinate $r_*$ ranges from $-\infty$ at the event horizons ${\cal H}^+_L$ and ${\cal H}^+_R$ to $+\infty$ at the Cauchy horizons ${\cal CH}_L^+$ and ${\cal CH}_R^+$ (see Fig. \ref{fig:penrose}). 

In region II, the ingoing Eddington-Finkelstein coordinates are smooth at the ``left" component ${\cal CH}_L^+$ of the Cauchy horizon. We will be interested in the ``right" component of the Cauchy horizon ${\cal CH}_R^+$. To introduce coordinates regular there, we use outgoing Eddington-Finkelstein coordinates $(u,r,\theta,\phi)$. The metric is
\be
ds^2 = -f du^2 - 2 du dr + r^2 d\Omega_2^2\,.
\ee
The Cauchy horizon ${\cal CH}_{R}^+$ is the surface $r=r_-$ in these coordinates. 

In region II, we define Kruskal coordinates near the Cauchy horizon as 
\be
 U_- = - e^{ \kappa_- u}\,,  \qquad V_- = - e^{- \kappa_- v}\,,
\ee 
The Cauchy horizon ${\cal CH}_R^+$ is the surface $V_-=0$ in these coordinates. 

Finally, in region I, we define Kruskal coordinates at the cosmological horizon by
\be
\label{Vcdef}
 U_c = e^{\kappa_c u} \,,  \qquad V_c = - e^{-\kappa_c v}  \,.
\ee
The future cosmological horizon ${\cal H}_C^+$ is the surface $V_c=0$. 

\subsection{The work of Moss {\it et al.}}

Strong cosmic censorship for RNdS black holes was first studied by Moss and collaborators in a series of papers. In this section we will review the arguments of Moss {\it et al} presented in Refs. \cite{Mellor:1989ac,0264-9381-9-4-001,Brady:1998au}. The analysis of \cite{Mellor:1989ac} concluded that strong cosmic censorship is violated by some RNdS black holes. However, this conclusion was modified in Refs. \cite{0264-9381-9-4-001,Brady:1998au}, resulting in the revised conclusion of Ref. \cite{Brady:1998au} that in fact (the $C^2$ version of) strong cosmic censorship is never violated by RNdS black holes. We will explain why this latter conclusion is valid only if one allows {\it non-smooth} initial data. 

We will consider perturbations by a scalar field $\Phi$ although the results of Moss {\it et al}\, apply also to the case of coupled electromagnetic and gravitational perturbations, which we will study later. One can prescribe initial data for the scalar field on the surface $\Sigma$ of Fig. \ref{fig:penrose} since this is a Cauchy surface for regions I and II. Equivalently, one can prescribe (characteristic) initial data for the scalar field on the null surface ${\cal H}_L^+ \cup {\cal H}^- \cup {\cal H}_c^-$. We will follow the latter approach. Given generic initial data for the scalar field, we want to know how the field behaves at the Cauchy horizon ${\cal CH}_R^+$.

This problem was first investigated by Mellor and Moss (MM) \cite{Mellor:1989ac}. Their results (rederived below) indicate that the scalar field will fail to be $C^1$ at the Cauchy horizon if there exists a sufficiently slowly decaying quasinormal mode. More precisely, let $\alpha$ be the {\it spectral gap}, {\it i.e.} the distance from the real axis in frequency space to the lowest (slowest decaying) quasinormal frequency. Define
\be
\label{betadef}
 \beta = \frac{\alpha}{\kappa_-}\,.
\ee
MM showed that if $\beta<1$ then the scalar field fails to be $C^1$ at ${\cal CH}_R^+$. When gravitational backreaction is included, the blow-up of the derivatives of $\Phi$ at ${\cal CH}_R^+$ is expected to cause a blow up of curvature. Thus if $\beta<1$ for all black holes then the $C^2$ version of strong cosmic censorship is expected to hold. However, by studying quasinormal modes, MM argued that RNdS black holes with $|Q| \approx M$ have $\beta>1$, so the scalar field is $C^1$ at ${\cal CH}_R^+$, 
which is evidence for a violation of the $C^2$ version of strong cosmic censorship.

MM modified this claim in Ref. \cite{0264-9381-9-4-001}. They were motivated by earlier work on a toy model (null dust) \cite{0264-9381-9-1-011} which suggested that the analysis of Ref. \cite{Mellor:1989ac} missed an important effect arising from late-time ingoing radiation propagating along the cosmological horizon ${\cal H}_c^+$. MM argued that, in the presence of such radiation, the scalar field will fail to be $C^1$ at ${\cal CH}_R^+$ if $\beta'<1$ where $\beta' = {\rm min}(\alpha,\kappa_c)/\kappa_-$. Subsequently, Brady, Moss and Myers (BMM) \cite{Brady:1998au} argued that one must also include the effect of scattering of outgoing radiation propagating near ${\cal H}_R^+$ and in this case the scalar field will fail to be $C^1$ at ${\cal CH}_R^+$ if $\beta''<1$ where $\beta'' = {\rm min}(\alpha,\kappa_+,\kappa_c)/\kappa_-$. In view of (\ref{kappaineq}), this gives $\beta''<1$ for any non-extremal RNdS black hole and so BMM concluded that the $C^2$ version of strong cosmic censorship is always respected.

We will show that the arguments of Refs \cite{0264-9381-9-4-001,Brady:1998au} are valid only for initial data which is {\it not smooth}, in fact not even $C^1$, at, respectively, the future cosmological horizon ${\cal H}_C^+$ or future event horizon ${\cal H}_R^+$. Hence this work cannot be regarded as evidence in favour of the $C^2$ version of strong cosmic censorship because the initial data is not in $C^2$. However, we will show that these arguments can be reinterpreted as evidence in favour of the rough version of strong cosmic censorship proposed in Ref.  \cite{Dafermos:2018tha}. If one insists on smooth initial data then the original conclusion of MM is still valid: it is simply the quasinormal modes which determine whether or not strong cosmic censorship (in either the $C^2$ or Christodoulou formulation) is violated. 

We will consider solutions which can be written as superpositions of mode solutions. A mode solution has the separable form
\be\label{ansatzScalarField}
 \Phi = \frac{1}{r}e^{-i \omega t} R(r) Y_{\ell m}(\theta,\phi)\,,
\ee 
where $Y_{\ell m}$ is a spherical harmonic. Substituting this into the wave equation or Klein Gordon equation (if the field is massive) one finds that the function $R$ satisfies an equation of the form 
\be
\label{Req}
 -\frac{d^2 R}{dr_*^2} + V_\ell(r) R = \omega^2 R\,,
\ee
where the potential $V_\ell(r)$ is independent of $\omega$ and vanishes exponentially fast as a function of $r_*$ as $r_* \rightarrow \pm \infty$ in either region I or II. 

We will start by considering solutions in region II. For real $\omega$, by reformulating \eqref{Req} as an integral equation, one can define two linearly independent solutions with the following behaviour as $r_* \rightarrow -\infty$ in region II
 \cite{chandra,Mellor:1989ac,Kehle:2018upl,Dafermos:2018tha}\footnote{We use the notation of Ref. \cite{Dafermos:2018tha} although our mode functions differ from theirs by a factor of $r$. For us, $R_{{\rm out},+} \sim e^{i \omega r_*}$ means $R_{{\rm out},+} = e^{i \omega r_*} \hat{R}_{{\rm out},+}$ where $\hat{R}_{{\rm out},+}$ is a real analytic function of $r$ for $r_-<r<r_c$ with $\hat{R}_{{\rm out},+}(r_+)=1$.}:
\be
\label{Rpluscond}
 R_{{\rm out},+} \sim e^{i \omega r_*}\,, \qquad R_{{\rm in},+} \sim e^{-i \omega r_*}\,.
\ee
$R_{{\rm out},+}$ gives a scalar field solution $\Phi$ smooth on ${\cal H}^+_L$ and $R_{{\rm in},+}$ gives a solution smooth on ${\cal H}^+_R$ (see Fig. \ref{fig:penrose}). Similarly as $r_* \rightarrow \infty$ we can define two linearly independent solutions by
\be
 R_{{\rm out},-} \sim e^{i \omega r_*}\,, \qquad R_{{\rm in},-} \sim e^{-i \omega r_*}\,,
\ee
and these give scalar field solutions that are smooth at ${\cal CH}_R^+$ and ${\cal CH}_L^+$, respectively. We can now write
\begin{subequations}
\begin{align} 
 & R_{{\rm out},+} = {\cal A}(\omega) R_{{\rm out},-} + {\cal B}(\omega) R_{{\rm in},-} 
\,,  \label{ABdef} \\
&R_{{\rm in},+} = \tilde{\cal A}(\omega) R_{{\rm in},-} + \tilde{\cal B}(\omega) R_{{\rm out},-}  \,,  \label{cAdef}
\end{align}
\end{subequations}
where ${\cal A}$ and ${\cal B}$ are the transmission and reflection coefficients for fixed frequency scattering of waves propagating out from ${\cal H}_L^+$ and $\tilde{\cal A}$, $\tilde{\cal B}$ are the transmission and reflection coefficients for scattering of waves propagating in from ${\cal H}_R^+$.

In region II, initial data can be specified on the characteristic hypersurface ${\cal H}_L^+ \cup {\cal H}_R^+$. We assume that the data on ${\cal H}_L^+$ is a wavepacket with Fourier transform $Z(\omega)$:
\be
 \Phi |_{{\cal H}_L^+} =  \int d\omega e^{-i \omega u} Z(\omega) Y_{\ell m}(\theta,\phi)\,,
\ee
and the data on ${\cal H}_R^+$ is a wavepacket with Fourier transform $\tilde{Z}(\omega)$:
\be
\label{Ztdef}
 \Phi |_{{\cal H}_R^+} =  \int d\omega e^{-i \omega v} \tilde{Z}(\omega) Y_{\ell m}(\theta,\phi)\,.
\ee
It follows that the solution in region II is
\ba
 \Phi &=& \int d\omega e^{-i \omega t} \left[ Z(\omega)R_{{\rm out},+}(\omega,r) + \tilde{Z}(\omega) R_{{\rm in},+}(\omega,r) \right] Y_{\ell m}(\theta,\phi) \nonumber \\
 &=&  \Phi_{\rm out} + \Phi_{\rm in}
\ea
where
\begin{subequations}
\begin{align} 
 &  \Phi_{\rm out} \equiv \int d\omega e^{-i \omega t} \left[ Z(\omega) {\cal A}(\omega) + \tilde{Z}(\omega) \tilde{\cal B}(\omega) \right] R_{{\rm out},-} (\omega,r) Y_{\ell m}(\theta,\phi)\,,\\
& \Phi_{\rm in} \equiv  \int d\omega e^{-i \omega t} \left[ Z(\omega) {\cal B}(\omega) + \tilde{Z}(\omega) \tilde{\cal A}(\omega) \right] R_{{\rm in},-} (\omega,r) Y_{\ell m}(\theta,\phi)\,.
\end{align}
\end{subequations}
These are, respectively, the parts of $\Phi$ that are outgoing and ingoing near the Cauchy horizon. The outgoing part is smooth at ${\cal CH}_R^+$ and the ingoing part is smooth at ${\cal CH}_L^+$. We are interested in how smooth the ingoing part is at ${\cal CH}_R^+$ where $r_* \rightarrow \infty$ and we have
\be
 \Phi_{\rm in} \approx \int d\omega \, e^{-i \omega v} \left[ Z(\omega) {\cal B}(\omega) + \tilde{Z}(\omega) \tilde{\cal A}(\omega) \right]  Y_{\ell m}(\theta,\phi)
\ee
and hence, taking a derivative w.r.t. the Kruskal coordinate $V_-$ that is smooth at ${\cal CH}_R^+$,
\be
\label{Fint}
 \partial_{V_-} \Phi_{\rm in} \approx e^{\kappa_- v} \int d\omega e^{-i \omega v} {\cal F}(\omega) Y_{\ell m}(\theta,\phi)
\ee
where
\be
 {\cal F}(\omega) = - i \omega \left[ Z(\omega) {\cal B}(\omega) + \tilde{Z}(\omega) \tilde{\cal A}(\omega) \right]. 
\ee
We now want to examine whether $\partial_{V_-} \Phi$ diverges at ${\cal CH}_R^+$, where $v \rightarrow \infty$. To do this we need to determine whether or not the integral decays faster than $e^{-\kappa_- v}$ as $v \rightarrow \infty$. To determine the decay of the integral, we can deform the contour of integration into a line of constant ${\rm Im}(\omega)$ in the lower half complex $\omega$ plane. How far we can deform the contour depends on the analyticity properties of the quantity ${\cal F}(\omega)$, which we will now investigate, following \cite{chandra,Mellor:1989ac,Kehle:2018upl}. 

First, to calculate ${\cal B}$ and $\tilde{\cal A}$ we proceed as follows. For functions $f(r_*)$ and $g(r_*)$ the Wronskian is (a prime denotes a derivative w.r.t. $r_*$)
\be
 W[f,g] = f' g - f g'\,,
\ee
and this is constant (in $r$) if $f,g$ are solutions of (\ref{Req}). We now have
\be
 \tilde{\cal A}(\omega) = \frac{W[R_{{\rm in},+},R_{{\rm out},-}]}{W[R_{{\rm in},-},R_{{\rm out},-}]}=-\frac{W[R_{{\rm in},+},R_{{\rm out},-}]}{2i \omega}\,,
\ee
where the latter expression follows from evaluating the Wronskian in the denominator at $r_* \rightarrow \infty$. Similarly,
\be
 {\cal B}(\omega) = \frac{W[R_{{\rm out},+},R_{{\rm out},-}]}{W[R_{{\rm in},-},R_{{\rm out},-}]}=-\frac{W[R_{{\rm out},+},R_{{\rm out},-}]}{2i\omega}\,.
\ee
The analyticity properties of the solutions of the radial equation have been determined in Refs. \cite{chandra,Kehle:2018upl,Dafermos:2018tha}. The result is that $R_{{\rm in},+}(\omega,r)$ can be analytically continued to the complex $\omega$ plane, except for simple poles at negative integer multiples of $ i \kappa_+$. Similarly, $R_{{\rm out},+}$ has simple poles at positive integer multiples of $i \kappa_+$ and $R_{{\rm out},-}$ has simple poles at negative integer multiples of $i \kappa_-$. Using (\ref{kappaineq}), it follows that, in the lower half-plane, the first pole of $\tilde{\cal A}$ is at $-i \kappa_+$ and the first pole of ${\cal B}$ is at $-i \kappa_-$. 

Consider first the case in which the wavepackets on ${\cal H}_L^+$ and ${\cal H}_R^+$ are {\it compactly supported} in $u$ and $v$ respectively. Then $Z(\omega)$ and $\tilde{Z}(\omega)$ are entire functions. Using (\ref{kappaineq}) it then follows that we can deform our contour of integration to a line of constant ${\rm Im}(\omega)$ in the lower half-plane, until we hit a pole in $\tilde{\cal A}(\omega)$ at $\omega = - i \kappa_+$. Now, {\it if} the wavepacket on ${\cal H}_R^+$ is generic then  $\tilde{Z}(-i \kappa_+) \ne 0$ and so this pole will also be a pole of ${\cal F}(\omega)$ with residue proportional to $\tilde{Z}(-i \kappa_+)$. Hence
\be
  \partial_{V_-} \Phi_{\rm in} \propto e^{(\kappa_- - \kappa_+) v} \tilde{Z}(-i \kappa_+) \propto \tilde{Z}(-i \kappa_+) (-V_-)^{\kappa_+/\kappa_- - 1}\,.
\ee
Using (\ref{kappaineq}), the above quantity diverges at ${\cal CH}_R^+$ where $v \rightarrow \infty$. Hence, for generic compactly supported smooth initial data prescribed on ${\cal H}_L^+ \cup {\cal H}_R^+$ the solution will not be $C^1$ at ${\cal CH}_R^+$, in apparent support of strong cosmic censorship.

It turns out that this argument is too quick because, in the problem of interest, we are {\it not} free to prescribe the initial data on ${\cal H}_R^+$. Instead, this data is determined by the solution outside the black hole, {\it i.e.} in region I. We will now review the argument of Ref. \cite{Mellor:1989ac} that shows that in fact $\tilde{Z}(-i \kappa_+)$ {\it vanishes}, invalidating the above argument. This analysis will reveal instead that the question of strong cosmic censorship depends on quasinormal modes of the black hole. 

First we need to define the mode functions in region I. We define two linearly independent solutions $R_{{\rm in},+}$ and $R_{{\rm out},+}$ of equation (\ref{Req}) using exactly the same conditions (\ref{Rpluscond}) as before except that now these conditions are being applied in region I instead of region II. We define a second pair of linearly independent solutions $R_{{\rm in},c}$ and $R_{{\rm out},c}$ in region I in terms of their behaviour at the cosmological horizon $r_* \rightarrow \infty$
 \be
 R_{{\rm out},c} \sim e^{i \omega r_*}\,, \qquad R_{{\rm in},c} \sim e^{-i \omega r_*}\,.
\ee
We can expand $R_{{\rm in},+}$ in terms of these solutions as
\be
\label{TRdef}
  R_{{\rm in},+} = \frac{1}{{\cal T}(\omega)} R_{{\rm in},c} + \frac{{\cal R}(\omega)}{{\cal T}(\omega)} R_{{\rm out},c}\,.
\ee
Here, ${\cal T}$ and ${\cal R}$ are the transmission and reflection coefficients for scattering of waves incident from ${\cal H}_c^-$. Similarly, we can write
\be
 R_{{\rm out},c} =\frac{1}{\tilde{\cal T}(\omega)} R_{{\rm out},+}+ \frac{\tilde{\cal R}(\omega)}{\tilde{\cal T}(\omega)} R_{{\rm in},+} \,,
\ee
where $\tilde{\cal T}$ and $\tilde{\cal R}$ are the transmission and reflection coefficients for waves progating out of ${\cal H}^-$. 

In region I, initial data can be specified on the characteristic hypersurface ${\cal H}^- \cup {\cal H}_c^-$. We assume that the data on ${\cal H}^-$ is a wavepacket with Fourier transform $X(\omega)$:
\be
 \Phi |_{{\cal H}^-} =  \int d\omega e^{-i \omega u} X(\omega) Y_{\ell m}(\theta,\phi)
\ee
and the data on ${\cal H}_c^-$ is a wavepacket with Fourier transform $\tilde{X}(\omega)$:
\be
 \Phi |_{{\cal H}_c^-} =  \int d\omega e^{-i \omega v} \tilde{X}(\omega) Y_{\ell m}(\theta,\phi)\,.
\ee
It follows that the solution in region I is
\ba
 \Phi &=& \int d\omega e^{-i \omega t} \left[ X(\omega)\tilde{\cal T}(\omega) R_{{\rm out},c}(\omega,r) + \tilde{X}(\omega) {\cal T}(\omega) R_{{\rm in},+}(\omega,r) \right] Y_{\ell m}(\theta,\phi) \\
 &=&   \int d\omega e^{-i \omega t} \left\{X(\omega) R_{{\rm out},+}(\omega,r) + \left[ \tilde{X}(\omega) {\cal T}(\omega) + X(\omega) \tilde{\cal R}(\omega) \right] R_{{\rm in},+}(\omega,r)\right\} Y_{\ell m}(\theta,\phi)\,. \nonumber
\ea
We can now evaluate this on the event horizon ${\cal H}_R^+$, where $r_* \rightarrow - \infty$. The first term vanishes there provided our initial outgoing wavepacket on ${\cal H}^-$ vanishes on the black hole bifurcation sphere, as it must for the Fourier transform to be well-defined. This leaves
\be
 \Phi|_{{\cal H}_R^+} = \int d \omega e^{-i \omega v}  \left[ \tilde{X}(\omega) {\cal T}(\omega) + X(\omega) \tilde{\cal R}(\omega) \right] Y_{\ell m}(\theta,\phi)
\ee
so from (\ref{Ztdef}) we can read off
\be
 \tilde{Z}(\omega) =  \tilde{X}(\omega) {\cal T}(\omega) + X(\omega) \tilde{\cal R}(\omega). 
\ee
From (\ref{TRdef}) we obtain
\be
{\cal T}(\omega) = \frac{W[R_{{\rm in},c},R_{{\rm out},c}]}{W[R_{{\rm in},+},R_{{\rm out},c}]}= -\frac{2i \omega}{W[R_{{\rm in},+},R_{{\rm out},c}]}
\ee 
where in the final step we evaluated the numerator at $r_* \rightarrow \infty$. Similarly,
\be
 \tilde{\cal R}(\omega) = - \frac{W[R_{{\rm out},+},R_{{\rm out},c}]}{W[R_{{\rm in},+},R_{{\rm out},c}]}\,.
 \ee
Consider initial data which is compactly supported on ${\cal H}^-$ and ${\cal H}_c^-$ (w.r.t. $u$, $v$ respectively), so $X(\omega)$ and $\tilde{X}(\omega)$ are entire functions. Recall that the analytic continuation of $R_{{\rm in},+}$ has simple poles at negative integer multiples of $i \kappa_+$. It follows that the analytic continuations of ${\cal T}$ and $\tilde{\cal R}$ have zeroes at these locations. Hence, for this initial data, $\tilde{Z}(\omega)(-i \kappa_+) =0$, as first explained by Mellor and Moss \cite{Mellor:1989ac}.

Recall that the behaviour of $\Phi$ near ${\cal CH}_R^+$ is determined by the analyticity properties of ${\cal F}(\omega)$. From the above we have
\ba
\label{Fres}
 {\cal F}(\omega) &=& \frac{1}{2}  W[R_{{\rm out},+},R_{{\rm out},-}] Z(\omega) + \frac{1}{2} W[R_{{\rm in},+},R_{{\rm out},-}] \left(\tilde{X}(\omega) {\cal T}(\omega) + X(\omega) \tilde{\cal R}(\omega) \right)  \\
 &=& \frac{1}{2}  W[R_{{\rm out},+},R_{{\rm out},-}] Z(\omega) - \frac{W[R_{{\rm in},+},R_{{\rm out},-}]}{2W[R_{{\rm in},+},R_{{\rm out},c}]} \left( 2i \omega \tilde{X}(\omega) + W[R_{{\rm out},+},R_{{\rm out},c}] X(\omega) \right). \nonumber 
 \ea
We start by considering the case in which the initial data on ${\cal H}_L^+$ and ${\cal H}^-$ are compactly supported functions of $u$, and the initial data on ${\cal H}_c^-$ is a compactly supported function of $v$, so $Z(\omega)$, $X(\omega)$ and $\tilde{X}(\omega)$ are entire functions. In the above expression, the mode functions with poles in the lower half plane are $R_{{\rm in},+}$ (at negative integer multiples of $i \kappa_+$), $R_{{\rm out},-}$ (at negative integer multiples of $i \kappa_-$) and $R_{{\rm out},c}$ (at negative integer multiples of $i \kappa_c$). However, in ${\cal F}(\omega)$ the poles associated to $R_{{\rm in},+}$ and $R_{{\rm out},c}$ will cancel out in the ratios of Wronskians. Therefore singularities of ${\cal F}(\omega)$ in the lower half plane can only arise from the poles in $R_{{\rm out},-}$ and where 
\be
 W[R_{{\rm in},+},R_{{\rm out},c}] = 0\,.
\ee
This is the condition for $R_{{\rm in},+}$ and $R_{{\rm out},c}$ to be linearly dependent, the defining condition of a {\it quasinormal mode}. The corresponding values of $\omega$ are called quasinormal frequencies. We see that, for compactly supported initial data, ${\cal F}(\omega)$ is analytic in the lower half-plane except for poles at quasinormal frequencies and at negative integer multiples of $i \kappa_-$. 

As discussed above, the {\it spectral gap} $\alpha$ is defined as the infimum (smallest value) of $- {\rm Im}(\omega)$ over all quasinormal modes. Deform the contour of integration in (\ref{Fint}) the line ${\rm Im}(\omega)=-\alpha+\epsilon$ for arbitrarily small $\epsilon>0$. In other words, we push the contour of integration down until just before it hits the ``lowest" ({\it i.e.} slowest decaying) quasinormal mode(s). In doing this we may pick up contributions from poles at multiples of $-i\kappa_-$ if these lie closer to the real axis than the lowest quasinormal mode. However, the contribution from such poles to the integral of (\ref{Fint}) will have $v$-dependence $e^{- n \kappa_-}$ (for positive integer $n$), and the contribution to (\ref{Fint}) will behave as $e^{(1-n)\kappa_- v} = (-V_-)^{n-1}$, which is {\it smooth} at ${\cal CH}_R^+$. The non-smooth part of (\ref{Fint}) arises from the integral along the new contour of integration. This integral decays as $e^{-\alpha v}$ for large $v$. Hence the non-smooth part of (\ref{Fint}) is proportional to
\be
 e^{(\kappa_- - \alpha) v} = (-V_-)^{\beta-1}
\ee
where $\beta$ is defined in (\ref{betadef}). If $\beta<1$ then the scalar field is not $C^1$ at the Cauchy horizon ${\cal CH}_R^+$ (where $V_-=0$). If $\beta<1/2$ then it does not even have locally square integrable derivatives, {\it i.e.} it does not have locally finite energy. On the other had, if $\beta \ge r$ for some positive integer $r$ then the above result is consistent with the scalar field being $C^r$ at the Cauchy horizon. So, for compactly supported initial data, the question of strong cosmic censorship reduces to identifying the most slowly decaying quasinormal modes of the black hole \cite{Mellor:1989ac}.

We now investigate what happens when we relax the condition that the initial data on ${\cal H}_c^-$ has compact support. For now we continue to assume compact support on ${\cal H}^-$ and ${\cal H}_L^-$. Solutions arising from such initial data were first considered by Mellor and Moss \cite{0264-9381-9-4-001}. They argued that late time ingoing radiation propagating along ${\cal H}_c^+$ will lead to an additional pole in ${\cal F}(\omega)$ at $\omega = - i \kappa_c$. Their argument goes as follows. Assume that the wavepacket on ${\cal H}_c^-$ is smooth at the cosmological bifurcation sphere ${\cal B}_c$ ($U_c=V_c=0$). The wavepacket must vanish there (otherwise it cannot be built as a superposition of modes as assumed above). Demanding that it does so smoothly leads to the condition $\Phi \propto V_c$ as $V_c \rightarrow 0$ on ${\cal H}_c^-$ ({\it i.e.} on $U_c=0$), which implies $\Phi \propto e^{-\kappa_c v}$ for large $v$ on ${\cal H}_c^-$. This implies that the Fourier transform $\tilde{X}(\omega)$ is analytic in the strip $-i \kappa_c < {\rm Im}(\omega) \le 0$ but $\tilde{X}(\omega)$ generically has a simple pole at $\omega =-i\kappa_c$. This is the basis of the claim in Ref. \cite{0264-9381-9-4-001} that ${\cal F}(\omega)$ has a pole at $\omega = -i\kappa_c$. However, this claim is incorrect because, in (\ref{Fres}), this pole in $\tilde{X}(\omega)$ is cancelled by a corresponding pole in $W[R_{{\rm in},+},R_{{\rm out},c}]$ arising from the pole in $R_{{\rm out},c}$ at $\omega = -i \kappa_c$. In other words, {\it this pole is cancelled by a corresponding zero in the transmission coefficient} ${\cal T}(\omega)$.\footnote{\label{smoothpole}
More generally, writing the initial data on ${\cal H}_c^-$ as $\Phi = f(V_c) = f(e^{-\kappa_c v})$, for smooth $f$, taking the Fourier transform and repeatedly integrating by parts one can see that $\tilde{X}(\omega)$ can have poles at negative integer multiples of $i \kappa_c$. These are all cancelled by corresponding zeros in ${\cal T}(\omega)$.}
 
We see that considering this data with non-compact support on ${\cal H}_c^-$ does not change our conclusions above: it is still the quasinormal modes which determine whether or not strong cosmic censorship is violated. However, in making this statement we have assumed that our initial data is {\it smooth} at the cosmological bifurcation sphere. If we allow non-smooth data, as advocated in Ref. \cite{Dafermos:2018tha}, then late-time ingoing radiation {\it does} lead to a new effect. Consider $\Phi \propto e^{-\gamma \kappa_c v}$ for large $v$ on ${\cal H}_c^-$, {\it i.e.} $\Phi \propto V_c^\gamma$ on ${\cal H}_c^-$, with $0< \gamma <1$. Clearly such data is not differentiable at $V_c=0$, but it has locally finite energy if $\gamma>1/2$ since this is the condition for the gradient of $\Phi$ to be locally square integrable. For such data, $\tilde{X}(\omega)$ has a pole at $\omega = - i \gamma \kappa_c$ and, in the expression for ${\cal F}(\omega)$, this is {\it not} cancelled by a zero of ${\cal T}(\omega)$. Hence at ${\cal CH}_R^+$ we have $\partial_{V_-} \Phi_T \propto e^{(\kappa_- - \gamma \kappa_c)v}=(-V_-)^{\delta -1}$, where $\delta = \gamma \kappa_c /\kappa_-$. Locally finite energy at the Cauchy horizon requires $\delta>1/2$. If $\kappa_c < 2 \kappa_-$ then we can choose $\gamma>1/2$ such that $\delta<1/2$.\footnote{More mathematically, the initial data is such that the solution initially belongs to $H^1_{\rm loc}$ but the solution at ${\cal CH}_R^+$ does not belong to $H^1_{\rm loc}$.}   In other words, for a RNdS black hole with $\kappa_c< 2 \kappa_-$, ingoing wavepackets with locally finite energy on ${\cal H}_c^-$ give solutions whose energy is not locally finite at ${\cal CH}_R^+$. However, we emphasize that such wavepackets are not smooth at the cosmological bifurcation sphere.
 
Next we consider relaxing the condition that the wavepacket on ${\cal H}_L^+$ has compact support. This was important in the argument of Ref. \cite{Brady:1998au} asserting that (the $C^2$ version of) strong cosmic censorship is respected for any RNdS black hole. Once again, we will first consider the case of smooth initial data. On ${\cal H}_L^+ \cup {\cal H}^-$ ({\it i.e.} the line $V_+=0$) we will assume that the data vanishes at the bifurcation sphere ${\cal B}$, {\it i.e.} at $U_+=0$, (which is required for the Fourier transforms $Z(\omega)$ and $X(\omega)$ to exist as functions) but has non-vanishing derivative there, so for small $U_+$ we have, for some constant $k$
\be
 \Phi \big|_{V_+=0} \approx k \,U_+ \,Y_{\ell m}(\theta,\phi)\,.
\ee
In region II this gives $\Phi|_{{\cal H}_L^+} \approx k e^{-\kappa_+ u} Y_{\ell m}$ as $u \rightarrow \infty$. Similarly in region I we have $\Phi|_{{\cal H}_L^+} \approx -k e^{-\kappa_+ u} Y_{\ell m}$  as $u \rightarrow \infty$. It follows that $Z(\omega)$ and $X(\omega)$ both have poles at $\omega = - i \kappa_+$, with equal and opposite residues. Hence it appears that ${\cal F}(\omega)$ will have a pole at $\omega = - i \kappa_+$ \cite{Brady:1998au}. But we will now show that the poles in $Z$ and $X$ cancel out in the expression for ${\cal F}(\omega)$. First, note that if there is a pole at $\omega=-i\kappa_+$ in ${\cal F}$ then the residue of this pole is proportional to 
\be
\label{resdef}
 \lim_{\omega \rightarrow -i \kappa_+} \left( W[R_{{\rm out},+},R_{{\rm out},-}] + \frac{W[R_{{\rm in},+},R_{{\rm out},-}]W[R_{{\rm out},+},R_{{\rm out},c}] }{W[R_{{\rm in},+},R_{{\rm out},c}]} \right).
\ee 
Recall that $R_{{\rm in},+}$ has a simple pole at $\omega = -i \kappa_+$, {\it i.e.}
\be
 R_{{\rm in},+}(\omega,r) = \frac{h(r)}{\omega + i \kappa_+} + g(\omega,r)
\ee
where $g(\omega,r)$ is analytic at $\omega =-i\kappa_+$. The solution $R_{{\rm in},+}$ is obtained by converting (\ref{Req}) to an integral equation, and solving by iteration \cite{chandra,Kehle:2018upl}. Indeed this is how one sees that it has a simple pole at $\omega = -i \kappa_+$. One can also see from this procedure that the residue $h(r)$ can be expressed as a series in $e^{\kappa_+ r_*}$, and is proportional to $e^{\kappa_+ r_*}$ as $r_* \rightarrow - \infty$. Now, $h(r)$ must satisfy (\ref{Req}) with $\omega =-i\kappa_+$. But the solution of (\ref{Req}) with behaviour $e^{\kappa r_*} = e^{i \omega r_*}$ as $r_* \rightarrow -\infty$ is $R_{{\rm out},+}(-i \kappa_+,r)$. Hence we have\footnote{
At the special values $\omega = -i n\kappa_+$ ($n=1,2,3,\ldots$), $R_{{\rm out},+}$ gives mode solutions that can be smoothly extended through ${\cal H}_R^+$, proportional to $U_+^n$ near ${\cal H}_R^+$, and the second linearly independent solution of (\ref{Req}) gives non-smooth mode solutions involving $\log U_+$.}
\be
 h(r) = c \,R_{{\rm out},+}(-i \kappa_+,r)
\ee 
for some constant of proportionality $c$. It turns out that $c$ has opposite signs in regions I and II because of the way we defined $R_{{\rm out},+}$. To see this, note that $e^{-i \omega t} R_{{\rm in},+}$ is smooth at ${\cal H}_R^+$ hence $e^{-\kappa_+ t} h(r)$ should be smooth at ${\cal H}_R^+$. But in region I near ${\cal H}_R^+$ we have $e^{- \kappa_+ t} R_{{\rm out},+}(-i\kappa_+,r) \sim e^{-\kappa_+ u} = -U_+$ whereas in region II we have $e^{- \kappa_+ t} R_{{\rm out},+}(-i\kappa_+,r) \sim e^{-\kappa_+ u} = +U_+$. Hence smoothness implies that the constant $c$ has equal magnitude but opposite sign in regions I and II. It follows that, since the numerator is evaluated in region II and the denominator in region I, we have 
\be
 \lim_{\omega \rightarrow -i \kappa_+} \frac{W[R_{{\rm in},+},R_{{\rm out},-}]}{W[R_{{\rm in},+},R_{{\rm out},c}]} =- \left( \frac{W[R_{{\rm out},+},R_{{\rm out},-}]}{W[R_{{\rm out},+},R_{{\rm out},c}]} \right)_{\omega = -i \kappa_+}
\ee
and so the residue (\ref{resdef}) {\it vanishes}. Hence ${\cal F}(\omega)$ does not have a pole at $\omega = -i \kappa_+$. Similarly, it does not have a pole at any negative integer multiple of $i \kappa_+$.\footnote{
One can argue as in footnote \ref{smoothpole} that $X(\omega)$ and $Z(\omega)$ can have poles $\omega = - i n\kappa_+$ for $n=1,2,3,\ldots$. Their residues are related by a factor of $(-1)^n$. This is cancelled by a corresponding factor of $(-1)^n$ relating the constant $c$ in regions I and II. The residue in ${\cal F}(\omega)$ then vanishes exactly as for the $n=1$ case.} Hence, once again, we find that for smooth initial data, relaxing the condition of compact support does not lead to anything new, in contrast to the claim of Ref. \cite{Brady:1998au}. 

The reason that the argument of Ref. \cite{Brady:1998au} fails is that the poles in $Z$ and $X$ at $\omega = - i\kappa_+$ cancelled out in ${\cal F}(\omega)$. This cancellation arose because we assumed that the first derivative of $\Phi$ was continuous at ${\cal B}$, {\it i.e.} that the inital data is $C^1$ there. In order to avoid such a cancellation we have to consider initial data that is not $C^1$, {\it i.e.} we have to consider rough initial data, as proposed in Ref. \cite{Dafermos:2018tha}. 
For example, consider initial data which vanishes on ${\cal H}^-$ and ${\cal H}_c^-$, {\it i.e.} $X(\omega)=\tilde{X}(\omega)=0$. It follows that the resulting solution will vanish throughout region I. On ${\cal H}_L^+$ we take initial data $\Phi |_{{\cal H}_L^+}\propto U_+^\gamma$ as $U_+ \rightarrow 0+$, where $0<\gamma \le 1$ and hence
\be
\label{initgamma}
 \partial_{U_+} \Phi |_{V_+=0}\propto U_+^{\gamma-1} \; {\rm for} \; U_+>0\,, \qquad  \partial_{U_+} \Phi |_{V_+=0}=0 \; {\rm for}\;  U_+<0\,.
\ee
Clearly our initial data is continuous, but not $C^1$, at $U_+=0$. The resulting solution will fail to be $C^1$ at $U_+=0$, {\it i.e.} along the event horizon ${\cal H}_R^+$. In terms of $u$, our data behaves as $e^{-\gamma \kappa_+ u}$ as $u \rightarrow \infty$ on ${\cal H}_L^+$ so $Z(\omega)$ has a pole at $\omega = - i \gamma \kappa_+$ and hence, even for $\gamma=1$, ${\cal F}(\omega)$ has a pole at the same location. It then follows that at ${\cal CH}_R^+$ we have
\be
\label{CHbeta}
 \partial_{V_-} \Phi \sim e^{(\kappa_- - \gamma \kappa_+) v} = (-V_-)^{\delta - 1} 
\ee
where
\be
\label{deltadef}
 \delta = \gamma \frac{\kappa_+}{\kappa_-} \,,
\ee
and hence from (\ref{kappaineq}) we have
\be
 \delta < \gamma\,.
\ee
Comparing (\ref{initgamma}) and (\ref{CHbeta}), we see that the solution at ${\cal CH}_R^+$ is less smooth than the initial data. In particular, the condition for the initial data to have locally square integrable first derivatives ({\it i.e.} finite energy) is $\gamma>1/2$ whereas the condition for the solution at ${\cal CH}_R^+$ to have locally square integrable first derivatives is $\delta>1/2$. For any non-extremal RNdS black hole, we can choose our initial data so that $\gamma>1/2$ but $\delta<1/2$. Hence one can find an initial wavepacket with finite energy that has infinite energy at the Cauchy horizon. So if we allow such rough initial data then the Christodoulou version of strong cosmic censorship is respected, as argued in Ref. \cite{Dafermos:2018tha}.

Once one is prepared to contemplate non-smooth initial data, there is no reason to work with wavepackets to show that this version of strong cosmic censorship is respected. One can work just as well with an outgoing mode solution in region II with {\it complex} frequency $\omega=\omega_1 - i \gamma \kappa_+$ (as was done in Ref. \cite{Dafermos:2018tha} for {\it ingoing} mode solutions in region I). In region II, $R_{{\rm out,}+}$ can be analytically continued to complex $\omega$, as long as $\omega$ is not a positive integer multiple of $i \kappa_+$. These mode solutions behave as $e^{-i \omega u}$ near ${\cal H}_R^+$. Now
\be
 e^{-i \omega u} = U_+^{i \omega_1/\kappa_++ \gamma}
\ee 
hence such modes vanish on ${\cal H}_R^+$ ({\it i.e.} $U_+=0$) if $\gamma>0$. We extend the mode into region I simply by taking it to vanish in region I, {\it i.e.} we take vanishing initial data on ${\cal H}^-$ and ${\cal H}_c^-$. At the Cauchy horizon the reflected part of the mode behaves as
\be
{\cal B}(\omega) e^{-i \omega v} = {\cal B}(\omega_1-i \gamma \kappa_+) (-V_-)^{i \omega_1/\kappa_- + \delta}
\ee 
with $\delta$ given by (\ref{deltadef}). As before, (\ref{kappaineq}) implies that we can choose $\gamma>1/2$ such that $\delta<1/2$. The initial data then has locally finite energy but the energy diverges at the Cauchy horizon.\footnote{Of course one also has to check that ${\cal B}(\omega_1 - i \gamma \kappa_+) \ne 0$ but one can probably prove as in Ref. \cite{Dafermos:2018tha} that ${\cal B}(\omega)$ has only isolated zeros and hence one can ensure ${\cal B}(\omega_1 - i \gamma \kappa_+) \ne 0$ by adjusting $\omega_1$ if necessary.}

In summary, we have seen that the argument of Ref. \cite{Brady:1998au} does not support the strong cosmic censorship conjecture for smooth initial data. However, a modification of this argument can be viewed as supporting the strong cosmic censorship conjecture for rough initial data formulated in Ref. \cite{Dafermos:2018tha}: initial data with locally finite energy generically gives a solution whose energy is not locally finite at the Cauchy horizon. 

\subsection{Recent work on strong cosmic censorship with $\Lambda>0$}

For smooth initial data, we have explained why the conclusion of Ref. \cite{Mellor:1989ac} remains valid and so whether or not strong cosmic censorship is respected can decided by looking at quasinormal modes. However, one deficiency of the above analysis is the assumption that the initial data vanishes at the bifurcation spheres ${\cal B}$ and ${\cal B}_c$. This assumption is required so that the Fourier transforms $Z(\omega)$, $X(\omega)$ and $\tilde{X}(\omega)$ are functions, rather than distributions. This assumption has been eliminated by more recent work in the mathematics literature  \cite{Hintz:2015jkj}, which proves that, for {\it any} smooth initial data, if $\beta>1$ then the scalar field is $C^1$ at the Cauchy horizon and if $\beta>1/2$ then the scalar field has finite local energy at ${\cal CH}_R^+$. 

The recent numerical study of Ref. \cite{Cardoso:2017soq} showed that massless scalar field perturbations of RNdS black holes always have $\beta<1$ so generic scalar field perturbations are not $C^1$ at the Cauchy horizon, which supports the $C^2$ formulation of strong cosmic censorship for the Einstein-Maxwell-massless scalar field theory. However, it was also found that near-extremal RNdS black holes have $\beta>1/2$ and so, for smooth initial data, the Christodoulou version of strong cosmic censorship is violated in this theory. 

Surprisingly, this conclusion does not hold for Kerr-dS black holes. Indeed, Ref. \cite{Dias:2018ynt} showed that Kerr-dS black holes always have $\beta<1/2$ and so, for smooth initial data, the Christodoulou version of strong cosmic censorship is respected for such black holes in Einstein gravity coupled to a massless scalar field. In fact, it was shown that the same result holds for linearized gravitational perturbations so it was argued that the Christodoulou version of strong cosmic censorship, with smooth initial data, is satisfied by the vacuum Einstein equations. 

Finally, we should mention the work of Ref. \cite{Costa:2017tjc}. This studies spherically symmetric perturbations of RNdS in the {\it nonlinear} Einstein-Maxwell-scalar field system. For this system, it is proved that the smoothness at ${\cal CH}_R^+$ is determined by how fast perturbations decay at late time along the event horizon ${\cal H}_R^+$. 
Since linear theory should be reliable for determining the latter, this work provides justification for believing that nonlinear effects will not invalidate the conclusions of a linear analysis of the behaviour near ${\cal CH}_R^+$.

\section{Bound for weak solutions of linearized Einstein-Maxwell equations} 
\label{sec:gravbetabound} 

As discussed previously, the {\it spectral gap} $\alpha$ is defined as the infimum (smallest value) of $- {\rm Im}(\omega)$ over all quasinormal frequencies $\omega$. Defining $\beta = \alpha/\kappa_-$ as in \eqref{betadef}, we showed above that if $\beta<1/2$ then generic scalar field perturbations arising from smooth initial data do not have locally square integrable derivatives ({\it i.e.} locally finite energy) at the Cauchy horizon. What about gravitoelectromagnetic modes? What condition yields a linearized gravitoelectromagnetic perturbation that constitutes a weak solution of the equations of motion at the Cauchy horizon? Is the critical value still $\beta=1/2$? In this section we will show that the answer to the latter question is positive. The analysis is rather technical so the reader may wish to skip to the summary in subsection \ref{sec:weak_conclude}.

Coupled linear gravitational and electromagnetic perturbations of RNdS can be studied using the Kodama-Ishisbashi (KI) formalism \cite{Kodama:2003kk}.
This formalism divides linearized gravitoelectromagnetic perturbations into perturbations arising from vector spherical harmonics and those arising from scalar spherical harmonics (there are no tensor spherical harmonics in 4d). We will consider the vector sector first (subsection~\ref{sec:KIvectorBound}) and then the scalar sector (subsection~\ref{sec:KIscalarBound}). The main conclusions are summarized in subsection \ref{sec:weak_conclude}.

\subsection{Vector-type gravitoelectromagnetic perturbations of RNdS} 
\label{sec:KIvectorBound} 

Vector perturbations of the background \eqref{metricRN} are described by  \cite{Kodama:2003kk}
\begin{subequations}\label{pert:vector}
\begin{align} 
&  \delta g_{ab}=0 \,, \quad 
  \delta g_{ai}=r f_a \vector_i ,\quad 
  \delta g_{ij} = -\frac{2}{k_V}r^2 H_T  D_{(i}\vector_{j)} \,;  \label{pert:vectorG}\\
 & \delta F_{ab}=0,\quad  
  \delta F_{ai}=D_a\A \, \vector_i,\quad  
  \delta F_{ij}=\A \left(D_i \vector_j- D_j \vector_i \right). \label{pert:vectorM} 
  \end{align} 
\end{subequations} 
where $f_a$, $H_T$ and $\A$ are functions of $\{x^a\}=\{t,r\}$. Additionally, $D_j$ is the covariant derivative with respect to 
the unit $S^2$ metric $\gamma_{ij}$ and $\vector_i$ is a vector spherical harmonic, {\it i.e.}  a regular solution of
\begin{equation}\label{harmonic:vector}
\left (\triangle_{2} + k_V^2 \right) \vector_i=0 \,, \qquad 
   D_i \vector^i=0 \,. 
\end{equation} 
Here, $\triangle_{2} \equiv \gamma^{ij}  D_i  D_j$ and regularity requires that the  eigenvalues $k_V^2$ are quantized as  \be
k_V^2=\ell_{\rm v}(\ell_{\rm v} + 1) -1\,, \qquad \ell_{\rm v} = 1,2,3, \ldots
\ee
The case $\ell_{\rm v}=1$   ($k_V^2=1$) is special case since in this case $\vector_i$ is a Killing vector on the $S^{2}$ and thus $ D_{(i}\vector_{j)}=0$. Consequently,  from \eqref{pert:vector} it follows that the metric components $\delta g_{ij}$ on $S^2$ are not perturbed.

For  $\ell_{\rm v}>1$, all the information about the perturbations can be encoded in two gauge invariant variables $ \Omega $ and $\A$. The latter was introduced in \eqref{pert:vector}  while the former is defined in terms of   $f_a, H_T$ via  
\begin{eqnarray} \label{FfunctionPhi}
   \frac{1}{r}\,\epsilon_{ab} D^b \Omega=f_a + \frac{r}{k_V} D_a H_T\,,
\end{eqnarray}    
where $\epsilon_{ab}$ denotes the anti-symmetric tensor 
on the 2-dimensional orbit spacetime. 
These two gauge invariant variables obey a coupled system of two master equations \cite{Kodama:2003kk}\footnote{In our conventions the parameter $\kappa$ of  \cite{Kodama:2003kk} is equal to $\sqrt{2}$.
}
\begin{subequations}\label{vector:MasterEqs}
\begin{align} 
 & r^2D_a\left( \frac{1}{r^2}D^a\Omega \right)
  -\frac{k_V^2-1}{r^2}\,\Omega 
  =\frac{4 Q}{r^2}\,\A \,,  \label{vector:MasterEqsO} \\
&D_aD^a\A-\frac{1}{r^2}
 \left(k_V^2+1+ \frac{4Q^2}{r^{2}}\right)\A 
=(k_V^2-1)\frac{Q}{r^{4}}\,\Omega \,.   \label{vector:MasterEqsA}
\end{align}
\end{subequations}
Once we have solved \eqref{vector:MasterEqs}, we can reconstruct the original metric perturbations \eqref{pert:vectorG} using the map  \cite{Kodama:2003kk}
\be
f_t(t,r)=\frac{f}{r}\partial_r \Omega-\frac{r}{k_V}\partial_t H_T\,, \qquad f_r(t,r)=-\frac{1}{r f}\partial_t \Omega-\frac{r}{k_V}\partial_r H_T\,.
\ee
This determines $\delta g_{\mu\nu}$ up to a gauge transformation (infinitesimal diffeomorphism) corresponding to $H_T$. A convenient choice of gauge is $H_T=0$. Note that the Maxwell perturbation is gauge invariant in the vector sector \cite{Kodama:2003kk}.

We will be interested in quasinormal modes, for which we have
\be
\label{vector:ansatz}
 \Omega(t,r) = e^{-i\omega t}   \Omega_{\omega \ell}(r)\,, \qquad \A(t,r)= e^{-i\omega t} \A_{\omega \ell}(r)
\ee
where $\ell \equiv \ell_{\rm v}$ and the frequency $\omega$ is determined in terms of $\ell_{\rm v}$ and a radial ``overtone" number $n=0,1,2,\ldots$. The quantized spectrum of frequencies is determined requiring that the perturbations are ingoing at the future event horizon $\mathcal{H}^+_R$ and outgoing at the future cosmological horizon $\mathcal{H}_c^+$ (see Fig.~\ref{fig:penrose}). In general, quasinormal frequencies are complex, $\omega = \omega_R + i \omega _I$, with $\omega_I <0$ so that quasinormal modes decay exponentially with time outside the black hole. 

For the regularity analysis at $\mathcal{H}^+_R$ it is convenient to work in ingoing coordinates $(v,r,\theta,\phi)$ since they are regular both in regions I and II of Fig. \ref{fig:penrose}. Then, a quasinormal mode is an analytic function of these coordinates in region I and can be analytically continued into region II. In these ingoing coordinates, a quasinormal mode has time dependence $e^{-i \omega v}$, and thus it diverges as $v \rightarrow - \infty$, {\it i.e.} along the red line on Fig. \ref{fig:penrose}. We will determine the frequency spectrum of vector quasinormal modes in Section \ref{sec:spectralgap}.

As reviewed above, the behaviour at the Cauchy horizon $\mathcal{CH}_{R}^+$ of a generic perturbation arising from smooth initial data is determined by the lowest quasinormal mode \cite{Mellor:1989ac}. Therefore we need to determine the smoothness at $\mathcal{CH}_{R}^+$ of the metric and Maxwell perturbations of our quasinormal modes. To do this, it is convenient to use outgoing coordinates in the black hole interior. Converting \eqref{vector:ansatz} to these outgoing coordinates in region II yields
\be
\label{vector:ansatzOut}
 \Omega(u,r) = e^{-i\omega u}   \tilde\Omega_{\omega \ell}(r)\,, \qquad \A(u,r) = e^{-i\omega u} \tilde\A_{\omega \ell}(r)\,,
\ee
for some functions $\tilde\Omega_{\omega \ell}$ and $\tilde\A_{\omega \ell}$.
A Frobenius analysis of \eqref{vector:MasterEqs} about the right Cauchy horizon $\mathcal{CH}_{R}^+$, dictates that there is a pair $\{ \Omega^{(1)},\Omega^{(2)} \}$ of linearly independent solutions for  $\Omega$ and another pair  $\{ \A^{(1)},\A^{(2)} \}$ for $\A$. These two pairs of linearly independent solutions behave as
\begin{subequations}\label{vectorCauchy}
\begin{align}
& \Omega^{(1)} = e^{-i \omega u}  \widehat{\Omega}^{(1)}_{\omega \ell}(r)\,, \qquad  \Omega^{(2)} = e^{-i \omega u}  (r-r_-)^{i \omega/\kappa_-} \, \widehat{\Omega}^{(2)}_{\omega \ell }(r)\,; \label{vectorCauchy:a}\\
& \A^{(1)} = e^{-i \omega u}  \widehat{\A}^{(1)}_{\omega \ell}(r)\,, \qquad  \A^{(2)} = e^{-i \omega u}  (r-r_-)^{i \omega/\kappa_-}  \, \widehat{\A}^{(2)}_{\omega \ell }(r)\,; \label{vectorCauchy:b}
\end{align}
\end{subequations}
where $\widehat{\Omega}^{(1,2)}$ and $\widehat{\A}^{(1,2)}$ denote non-vanishing smooth functions at $r=r_-$. The solutions labelled $(1)$ are outgoing at ${\cal CH}_R^+$. These are smooth at ${\cal CH}_R^+$. The solutions labelled $(2)$ are ingoing at ${\cal CH}_R^+$. These are not smooth at ${\cal CH}_R^+$. Our quasinormal mode will be a superposition of the ingoing and outgoing solutions at ${\cal CH}_R^+$.

Given the behaviours \eqref{vectorCauchy} for the master variables, what is the corresponding behaviour of the metric and Maxwell perturbations at the Cauchy horizon? Again, we work in outgoing coordinates $\{\tilde x^\mu\}=\{u,r,\theta,\phi\}$ and write the metric perturbation in these coordinates as $\delta \tilde{g}_{\mu\nu}$. The KI formalism maintains covariance w.r.t. diffeomorphisms on $S^2$ and on the transverse 2d orbit space. Hence $\delta \tilde{g}_{\mu\nu}$ takes the same form as in (\ref{pert:vectorG}) with $f_a$ replaced by the quantity $\tilde{f}_a$ obtained from $f_a$ by the 2d coordinate transformation $(t,r) \rightarrow (u,r)$, and $\tilde{H}_T = H_T$. Choosing the gauge $\tilde{H}_T =0$,  we find that the two linearly independent solutions for $\tilde{f}_a$ have the following behaviour near Cauchy horizon
\begin{equation}\label{vectorCauchy2}
 \tilde{f}_a^{(1)} = e^{-i \omega u}\sum_{j \ge 0} \hat{f}_a^{(1;j)}(r-r_-)^j , \qquad  \tilde{f}_a^{(2)} = e^{-i \omega u} (r-r_-)^{\hat{\alpha}_{a}+i \omega/\kappa_-}   \sum_{j \ge 0} \hat{f}_a^{(2;j)}(r-r_-)^j\,,
\end{equation}
where $\hat{\alpha}_{a}=\{0,-1\}$ for $a=\{u,r\}$, respectively, for constant coefficients $\hat{f}_a^{(1;j)}$ and $\hat{f}_a^{(2;j)}$. The behaviour at the Cauchy horizon of the Maxwell perturbation $\delta\tilde{F}_{\mu\nu}$ follow straightforwardly from \eqref{pert:vectorM}  and  \eqref{vectorCauchy:b}. 

Note that the outgoing solutions $\tilde{f}_a^{(1)}$ in \eqref{vectorCauchy2} are smooth, but the ingoing solutions  $\tilde{f}_a^{(2)}$ are not. This holds in the gauge $\tilde{H}_T =0$. We will now determine how much smoother we can make the solution using a gauge transformation.

In the vector sector, an infinitesimal gauge vector $\xi$ has a harmonic decomposition
\begin{equation}\label{gauge:vector}
 \xi_{a}=0 \,, \qquad 
 \xi_i =e^{-i \omega u} r L(r) \vector_{i} \,.
\end{equation} 
Under such gauge transformation the metric perturbation transforms according to
\begin{equation}\label{gaugeTransf}
\delta \tilde{g}_{\mu\nu} \to \delta \widetilde{g}_{\mu\nu} = \delta \tilde{g}_{\mu\nu} - 2\nabla_{(\mu}\xi _{\nu)}\,,
\end{equation}
and the Maxwell perturbation $\delta \tilde{F}$ is invariant: see \eqref{pert:vectorM} and recall that $\A$ is, by construction, a gauge invariant variable.

We now assume
\be
L(r)=\sum_{k \ge 0} L^{(k)}(r-r_-)^{i\omega/\kappa_- +k}
\ee
and we want to choose the coefficients $L^{(k)}$ to make \eqref{vectorCauchy2} as smooth as possible at $r=r_-$. We find that $L^{(0)}$ can be chosen to set $\hat{f}_r^{(2;0)}=0$ in \eqref{vectorCauchy2}. 
We can then choose  $L^{(1)}$ to set $\hat{f}_r^{(2;1)}=0$. But this choice then dictates that $\widetilde{f}_u^{(2)}$ and $\widetilde{H}_T^{(2)}$ behave as  $(r-r_-)^{i\omega/\kappa_-}$ because the gauge parameters $L^{(k)}$ with $k\geq 2$ do not appear at this order. Altogether, we can find a gauge where
the two linearly independent gravitoelectromagnetic solutions at the Cauchy horizon have the leading behaviour 
\begin{subequations}\label{vectorCauchy3}
\begin{align}
& \widetilde{f}_a^{(1)} = e^{-i \omega u}  \widehat{f}_a^{(1)}(r)\,, \qquad  \widetilde{f}_a^{(2)} = e^{-i \omega u}  (r-r_-)^{ \alpha_a+ i \omega/\kappa_-  }  \widehat{f}_a^{(2)}(r)\,;\\
& \widetilde{H}_T^{(1)} = e^{-i \omega u}  \widehat{H}_T^{(1)}(r)\,, \qquad \widetilde{H}_T^{(2)} = e^{-i \omega u}  (r-r_-)^{i \omega/\kappa_-}  \widehat{H}_T^{(2)}(r)\,; \\
&  \delta\widetilde{F}_{ai}^{(1)} = e^{-i \omega u} \delta\widehat{F}_{ai}^{(1)}(r) , \qquad  \delta\widetilde{F}_{ai}^{(2)} = e^{-i \omega u}  (r-r_-)^{-\alpha_a+i \omega/\kappa_-} \delta\widehat{F}_{ai}^{(2)}(r)\,; \\
&  \delta\widetilde{F}_{ij}^{(1)} = e^{-i \omega u} \delta\widehat{F}_{ij}^{(1)}(r) , \qquad  \delta\widetilde{F}_{ij}^{(2)} = e^{-i \omega u}  (r-r_-)^{i \omega/\kappa_-} \delta\widehat{F}_{ij}^{(2)}(r)
\end{align}
\end{subequations}
where $\alpha_a=\{0,1\}$  for $a=\{u,r\}$, respectively and $ \widehat{f}_a, \widehat{H}_T$ and $\delta\widehat{F}_{ai},\delta\widehat{F}_{ij}$ are functions that are smooth at $r=r_-$ (recall that $\delta\widetilde{F}_{ab}$ components are not excited in the vector sector; see \eqref{pert:vector}). 
 
 At ${\cal CH}_R^+$, our gravitoelectromagnetic quasinormal mode is some linear combination of the smooth outgoing solution (1) and the non-smooth ingoing solution (2). There is no reason for the coefficients in this linear combination to vanish. Therefore, the regularity of the quasinormal mode is determined by the ingoing solutions. 

For the vacuum Einstein equation, the regularity of the metric required for a weak solution is that the Christoffel symbols should be square integrable in some chart \cite{Christodoulou:2008nj}. By linearizing this condition, or by considering second order perturbation theory \cite{Dias:2018ynt}, the corresponding condition for a {\it linearized} metric perturbation to constitute a weak solution is that, in some gauge, the perturbation, and its first derivatives, should be locally square integrable, {\it i.e.} the perturbation should belong to the Sobolev space $H^1_{\rm loc}$. In Einstein-Maxwell theory, the corresponding statement is that, in some gauge, the metric perturbation should belong to $H^1_{\rm loc}$ and the Maxwell field strength perturbation should be locally square integrable ({\it i.e.} belong to $L^2_{\rm loc}$). 

From  \eqref{vectorCauchy3} we see that we can reach a gauge for which the least smooth components of the metric perturbation behave as $\delta \widetilde{g}^{(2)} \sim (r-r_-)^p$ with $p = i \omega/\kappa_-$. Hence $\partial_r \delta \widetilde{g}^{(2)} \sim (r-r_-)^{p-1}$, which is square integrable if, and only if, $2(\gamma -1)>-1$ where $\gamma = {\rm Re}(p)$. Similarly, \eqref{vectorCauchy3} shows that the least smooth components of the Maxwell field strength perturbation behave as $\delta \widetilde{F}^{(2)}\sim (r-r_-)^{p-1}$ (again with $p = i \omega/\kappa_-$). Once again this is  locally square integrable if, and only if, $2(\gamma -1)>-1$ (again with $\gamma = {\rm Re}(p)$). Hence, the condition for a vector-type gravitoelectromagnetic quasinormal mode to constitute a weak solution at the Cauchy horizon is $\gamma>1/2$, {\it i.e.}
\be\label{betaBound}
  -\frac{{\rm Im}(\omega)}{\kappa_-}>\frac{1}{2}\,.
\ee 
The above analysis shows that this condition is sufficient for the mode to constitute a weak solution at the Cauchy horizon. We believe it is also a necessary condition, and this can probably be proved along similiar lines to the argument in Ref. \cite{Dias:2018ynt}, exploiting gauge invariance of the KI variables. However, since we are mainly interested in {\it violation} of strong cosmic censorship, we will not perform such an analysis here. 

The above analysis was for the case $\ell_{\rm v}>1$. For the special case $\ell_{\rm v}=1$, the field $H_T$ is not defined since $ D_{(i}\vector_{j)}=0$. It follows that the two quantities defined by the RHS of \eqref{FfunctionPhi}  are no longer gauge invariant (and thus, neither is $\Omega$). There is a single gauge invariant quantity (denoted by $F^{1}=\epsilon^{ab} rD_a\left(f_b/r \right)$ in (4.8) of \cite{Kodama:2003kk}) and the map that reconstructs $\delta g_{\mu\nu}$ and $\delta F_{\mu\nu}$ from the gauge invariant quantity is (necessarily) different from the one described above for the $\ell>1$ case: in the end of the day  $\A$ is the only dynamical field although it still obeys the wave equation \eqref{vector:MasterEqsA} (with $k_V^2=1$) \cite{Kodama:2003kk}. We have done this analysis and gravitoelectromagnetic field reconstruction\footnote{The reader can find the full details in the discussions (4.8)-(4.15) and (4.31)-(4.33) of \cite{Kodama:2003kk}.} and we find that the condition  for a $\ell_{\rm v}=1$ vector-type gravitoelectromagnetic quasinormal mode to constitute a weak solution at the Cauchy horizon is still given by \eqref{betaBound}.

\subsection{Scalar-type gravitoelectromagnetic perturbations of RNdS} 
\label{sec:KIscalarBound} 

Scalar perturbations of the background \eqref{metricRN} take the form \cite{Kodama:2003kk}
\begin{subequations}\label{pert:scalar}
\begin{align} 
& \delta g_{ab}= f_{ab} \scalar, \quad  
 \delta g_{ai}= rf_a \scalar_i  , \quad 
 \delta g_{ij}= 2r^2 \left( H_L\gamma_{ij}\scalar + H_T \scalar_{ij} \right), \\
&  \delta F_{ab}=\left[\E-D_c(E_0X^c)\right] \epsilon_{ab}\scalar,  \quad
\delta F_{ai}=\epsilon_{ab} \left( r \E^b + k_S E_0 X^b\right)\scalar_i\,,\quad \delta F_{ij}=0,
\end{align} 
\end{subequations}
with $f_{ab},f_a,H_T,H_L, \E$ and  $\E^b$ being functions of $\{x^a\}=\{t,r\}$ and $\epsilon_{ab}$ is the anti-symmetric unit tensor. Moreover, $E_0=Q/r^2$ was introduced in \eqref{metricRNaux} and we have defined 
\begin{equation}
X_a = \frac{r}{k_S}\left( f_a + \frac{r}{k_S} D_a H_T \right).
\end{equation}
 The scalar spherical harmonics $\scalar$, and the associated scalar-type  vector harmonic $\scalar_i$ and traceless scalar-type tensor harmonic $\scalar_{ij}$ are defined by (note that $\scalar_{i}^{\phantom{i}i}=0$)
\begin{equation}
  (D_i D^i + k_S^2 ) \scalar =0 \,, \quad
  \scalar_i = -\frac{1}{k_S}  D_i \scalar \,, \quad 
  \scalar_{ij} = \frac{1}{k_S^2}  D_i D_j\scalar  
                 + \frac{1}{2}\gamma_{ij}\scalar \,.        
\end{equation}
The eigenvalues are quantized as 
\be
k_S^2 = \ell_{\rm s}(\ell_{\rm s}+1) \qquad \ell_{\rm s} = 1,2,3, \ldots. 
\ee
Harmonics with $\ell_{\rm s} = 0$ are non dynamical -- they correspond to variations of the black hole parameters $M,Q$. Harmonics with $\ell_{\rm s} = 1$ are special because $ \scalar_{ij}$ vanishes for these harmonics. For now we assume $\ell_{\rm s} > 1$ and comment on the case $\ell_{\rm s} =1$ at the end of this section. 

Gauge invariant variables for the scalar perturbations are $\E, \E_a$ $-$ already introduced in \eqref{pert:scalar} $-$ and, for $\ell_{\rm s} > 1$, $\F$ and $\F_{ab}$ defined as \cite{Kodama:2003kk}
\begin{equation} \label{scalarFsGaugeinv}
   \F = H_L + \frac{1}{2} H_T + \frac{1}{r} \,X_a D^a r \,, \quad
   \F_{ab} = f_{ab} + D_a X_b + D_b X_a \,, 
\end{equation}
The Bianchi identity requires that $\F_{a}^{\phantom{a}b}$ is traceless,
\begin{equation}\label{bianchi}
\F_{a}^{\phantom{a}a} = 0\,. 
\end{equation}

The equations of motion imply that the gauge invariant quantities $\E$ and  $\E_a$ can be expressed in terms of a single KI master variable $\A$ as 
\be
\E=-\frac{1}{k_S}D_c(r\E^c)\,,\qquad  \E_a =\frac{k_S}{r}\,D_a \A\,. 
\ee
On the other hand, introducing 
\be\label{XYX} 
  X = \F^t_t -2\F  \,, \qquad 
  Y = \F^r_r -2\F \,, \qquad 
  Z =  \F^r_t \,,    
\ee
a second gauge invariant master variable $\Phi$ can be defined as  \cite{Kodama:2003kk}
\be
\label{PhiS} 
\Phi= \frac{  2 Z/(i\omega) - r(X+Y)}{H}\,, \qquad \hbox{with} \quad H=k_S^2 -2+ 6M/r-4Q^2/r^2 \,,
\ee 
where here and henceforward, we assume that all perturbed quantities $Q(t,r)$ have the Fourier decomposition $Q(t,r)=e^{-i\omega t}Q(r)$ with $\omega$ being the associated frequency. 

The KI master variables $\Phi$ and $\A$ obey the following coupled system of equations \cite{Kodama:2003kk}
\begin{subequations}\label{scalar:MasterEqs}
\begin{align} 
& f(f\Phi')' +(\omega^2-V_S)\Phi=S_\Phi(\Phi,\A)\,, \\
& D_aD^a\A  -\frac{1}{r^2}\left(k_S^2 
        +\frac{8Q^2f/r^2}{H}\right)\A =\frac{Q}{r^3}\left( \frac{4H^2-2P_Z}{8H}\Phi +fr\partial_r\Phi \right),
\end{align} 
\end{subequations}
where $f$ is defined in \eqref{metricRNaux}. The potential $V_S$ and source term $S_\Phi(\Phi,\A)$ are lengthy expressions given in equations (5.42)-(5.44) of \cite{Kodama:2003kk}. The auxiliary quantity $P_Z$ is given in (C.8) of \cite{Kodama:2003kk}.

Given a solution of the above equations we will need to reconstruct the metric and Maxwell field perturbations in terms of the master variables $\Phi$ and $\A$. For that, we first write the variables $X$, $Y$ and $Z$ in terms of $\Phi$ and $\A$ and their  derivatives as \cite{Kodama:2003kk}
\begin{eqnarray}\label{scalar:XYZ} 
&& X=\frac{1}{r}\left[ \left( \frac{\omega^2r^2}{f}
     -\frac{P_{X0}}{16H^2}\right)\Phi
     +\frac{P_{X1}}{4H}r\partial_r\Phi \right] 
     + 2 E_0\left( \frac{P_{XA}}{2H^2}\A
         -\frac{4rf}{H}\,\partial_r\A \right),\nonumber \\
&& Y=\frac{1}{r}\left[ \left( -\frac{\omega^2r^2}{f}
     -\frac{P_{Y0}}{16H^2}\right)\Phi
     +\frac{P_{Y1}}{4H}r\partial_r\Phi \right] 
     +2 E_0\left(\frac{P_{YA}}{2H^2}\A
         +\frac{4rf}{H}\,\partial_r\A \right),\\
&& Z=i\omega \left( \frac{P_Z}{4H}\Phi
        -fr\partial_r\Phi -8 E_0  \frac{r f}{H} \A \right), \nonumber
\end{eqnarray}
where the coefficients $P_{X0},P_{X1},P_{XA},P_{Y0},P_{Y1},P_{YA}$ and $P_Z$ are functions of $r$ that can be found in equations~(C.4)-(C.10) and (C.11)-(C.16)  of \cite{Kodama:2003kk}. It follows from the equations of motion, including the Bianchi identity \eqref{bianchi}, that $f_{ab}$ and $H_L$ can be written as a function of $X,Y,Z$ ({\it i.e.} of $\Phi, \A$, their radial first derivative and $\omega$) and of $f_a, H_T$ and their radial derivatives. To simplify our task (and without prejudice since we will consider gauge transformations later) we can fix the gauge as
 \begin{equation}\label{scalar:gaugeFix}
  f_a=0,\, \quad H_T=0 \,.
\end{equation}
Then, the metric functions $f_{ab}$ and $H_L$ depend only on $X, Y, Z$. That is to say,  via \eqref{scalar:XYZ} and the Bianchi identity \eqref{bianchi} they can be written solely in terms of the master variables $\Phi, \A$ and their radial derivative as
\begin{equation}\label{scalar:ffromPhi}
 \hspace{-1cm} f_{tt}=\frac{f}{2}(X-Y)\,,\quad f_{tr}=-i\omega \,\frac{Z}{f}\,, \quad
f_{rr}=\frac{X-Y}{2f}\,,\quad  H_L=\frac{X+Y}{4}\,.
\end{equation}

We will find the frequency spectrum of scalar quasinormal modes in Section \ref{sec:spectralgap}. But first we must discuss the behaviour of the scalar-type perturbations at the Cauchy horizon $\mathcal{CH}_{R}^+$. The discussion of regularity at this null hypersurface proceeds very similarly to the vector sector case. Namely, the master variables $\Omega$ and $\A$ for the scalar quasinormal modes also admit the Fourier decomposition \eqref{vector:ansatz} and, when analytically continued into region II and converted to outgoing coordinates, these master variables also behave as  \eqref{vector:ansatzOut}. Moreover, a Frobenius analysis of  \eqref{scalar:MasterEqs} around $\mathcal{CH}_{R}^+$, dictates that there is a pair $\{ \Omega^{(1)},\Omega^{(2)} \}$ of linearly independent solutions for  $\Omega$ and another pair  $\{ \A^{(1)},\A^{(2)} \}$ for $\A$. These two pairs of linearly independent solutions still behave as described in \eqref{vectorCauchy} (with the identification $\ell\equiv\ell_{\rm s}$). 

Given these behaviours for the KI scalar master variables $\Omega$ and $\A$,  we can now find the behaviour of the metric and Maxwell perturbations for the outgoing and ingoing modes near ${\cal CH}_R^+$. Just as we did for vector-type perturbations, in region II we transform to outgoing coordinates $\{\tilde x^\mu\}=\{u,r,\theta,\phi\}$ in which the metric perturbation $\delta \tilde{g}_{ab}$ takes the same form as in \eqref{pert:scalar}, with $f_a$ replaced by the quantity $\tilde{f}_a$ obtained from $f_a$ via the coordinate transformation from $(t,r)$ to $(u,r)$, $f_{ab}$ is similarly replaced by $\tilde{f}_{ab}$, but $\tilde{H}_L = H_L$ and $\tilde{H}_T = H_T$ are unchanged. Similarly, the Maxwell perturbation $\delta \tilde{F}_{\mu\nu}$ is written in terms of $\tilde{\E}_a$ and $\tilde{\E}= \E$.

Choosing the gauge \eqref{scalar:gaugeFix} (which translates into $\tilde{f}_a =0, \tilde{H}_T =0$),  we find that the outgoing (smooth) and ingoing (non-smooth) solutions for $\tilde{f}_{ab}$, $\tilde{H}_L$, $\tilde{\E}$ and $\tilde{\E}_a$ have, respectively, the following expansions about the Cauchy horizon:
\begin{subequations}\label{scalarCauchy2}
\begin{align}
&\hspace{-0.3cm} \tilde{f}_{ab}^{(1)} = e^{-i \omega u} \sum_{j \ge 0}\hat{f}_{ab}^{(1;j)}(r-r_-)^{j+\hat{\gamma}_{ab}} , \quad  \tilde{f}_{ab}^{(2)} = e^{-i \omega u}   \sum_{j\ge 0}\hat{f}_{ab}^{(2;j)}(r-r_-)^{j+\hat{\alpha}_{ab}+i \omega/\kappa_-} \,; \label{vectorCauchy2a}\\
&\hspace{-0.3cm} \tilde{H}_{L}^{(1)} = e^{-i \omega u} \sum_{j \ge 0}\hat{H}_{L}^{(1;j)}(r-r_-)^j , \qquad   \tilde{H}_{L}^{(2)} = e^{-i \omega u}  \sum_{j\ge 0}\hat{H}_{L}^{(2;j)}(r-r_-)^{j+i \omega/\kappa_-} \,; \\
&\hspace{-0.3cm}  \tilde{\E}^{(1)} = e^{-i \omega u} \sum_{j \ge 0} \hat{\E}^{(1;j)}(r-r_-)^j , \qquad   \tilde{\E}^{(2)} = e^{-i \omega u}  \sum_{j \ge 0} \hat{\E}^{(2;j)}(r-r_-)^{j+i \omega/\kappa_-} \,; \\
&\hspace{-0.3cm}  \tilde{\E}_a^{(1)} = e^{-i \omega u} \sum_{j \ge 0} \hat{\E}_a^{(1;j)}(r-r_-)^{j}, \qquad   \tilde{\E}^{(2)}_a = e^{-i \omega u}  \sum_{j \ge 0} \hat{\E}_a^{(2;j)}(r-r_-)^{j+\hat{\epsilon}_a+i \omega/\kappa_-} \,;
 \label{scalarCauchy2d}
\end{align}
\end{subequations}
where $\hat{\gamma}_{ab}=\{0,0,-1\}$ for $ab=\{uu,ur,rr\}$,  $\hat{\alpha}_{ab}=\{0,-1,-2\}$ for $ab=\{uu,ur,rr\}$ and $\hat{\epsilon}_a=\{-1,0\}$ for $a=\{u,r \}$, respectively, and 
 $ \hat{f}_{ab}^{(1;j)}$, etc are constants that depend on $\omega$ and  $\ell_{\rm s}$. 
 
 The behaviour \eqref{scalarCauchy2} is valid in the particular gauge $\tilde{f}_a =0, \tilde{H}_T =0$. Even the outgoing solution (1) is not regular at the Cauchy horizon in this gauge (due to the component $\tilde{f}_{rr}^{(1)}$). We will now show that we can make the outgoing solution smooth, and the ingoing solutions smoother at the Cauchy horizon with a gauge transformation. In the scalar sector, an infinitesimal gauge vector $\xi$ has the harmonic decomposition
\begin{equation}\label{gauge:scalar}
 \xi_{a}=e^{-i \omega u} P_a(r) \scalar \,, \qquad 
 \xi_i =e^{-i \omega u} r L(r) \scalar_{i}  \,.
\end{equation} 
Under such gauge transformation the metric and Maxwell perturbations transform according to
\begin{equation}\label{scalar:gaugeTransf}
\delta \tilde{g}_{\mu\nu} \to \delta \widetilde{g}_{\mu\nu} = \delta \tilde{g}_{\mu\nu} - 2\nabla_{(\mu}\xi _{\nu)}\,, \qquad  \delta\tilde{F}_{\mu\nu} \to \delta \widetilde{F}_{\mu\nu} = \delta \tilde{F}_{\mu\nu}-\xi^\alpha \nabla_\alpha F_{\mu\nu}+ 2F_{\alpha [\mu}\nabla_{\nu ]}\xi^\alpha\,.
\end{equation}
We assume the following expansions for the functions appearing in the gauge transformation
\begin{subequations}
\begin{align}
&P_a(r)=\sum_{k \ge 0} \left[N_a^{(k)}+\bar{N}_a^{(k)}\log(r-r_-)\right](r-r_-)^{k}+\sum_{k \ge 0} P_a^{(k)}(r-r_-)^{i\omega/\kappa_- +k-1} \\
&L(r)=\sum_{k \ge 0} \left[M_a^{(k)}+\bar{M}_a^{(k)}\log(r-r_-)\right](r-r_-)^{k}+\sum_{k \ge 0} L^{(k)}(r-r_-)^{i\omega/\kappa_- +k}
\end{align}
\end{subequations}
and we now try to choose the constants $\{N_a^{(k)},\bar{N}_a^{(k)},P_a^{(k)},M^{(k)},\bar{M}_a^{(k)},L^{(k)}\}$ to eliminate as much as we can the leading terms in \eqref{scalarCauchy2} that are responsible for the lack of smoothness at the Cauchy horizon. Consider first the ingoing solution (labelled by superscript $^{(2)}$). 
We find that a choice of  $P_a^{(0)}$ (with $P_u^{(0)}=0$), $P_a^{(1)}$ and $L^{(0)}$ allows us to set 
\be
\hat{f}_{rr}^{(2;0)}=\hat{f}_{rr}^{(2;1)} =\hat{f}_{ur}^{(2;0)}=0,
\ee
and also to eliminate the leading term, proportional to $(r-r_-)^{i\omega/\kappa_--1}$, in $\widetilde{f}_r^{(2)}$ (note that $\widetilde{f}_a^{(2)}$ becomes non-zero as a result of the gauge transformation; the term  $(r-r_-)^{i\omega/\kappa_--1}$ in $\widetilde{f}_r^{(2)}$ has a contribution due to $P_r^{(0)}$ and another due to $L^{(0)}$). We can now choose $P_a^{(2)}$ and $L^{(1)}$ to set
\be
\hat{f}_{rr}^{(2;2)}=\hat{f}_{ur}^{(2;1)} = 0
\ee
and to eliminate the term proportional to $(r-r_-)^{i\omega/\kappa_-}$ in $\widetilde{f}_r^{(2)}$. But this choice then dictates that the leading term of $\widetilde{f}_{uu}^{(2)},\widetilde{f}_u^{(2)}$ and $\widetilde{H}_{L,T}^{(2)}$ is $(r-r_-)^{i\omega/\kappa_-}$ because these terms in these quantities do not depend on the higher order  gauge parameters and we have no more gauge freedom to avoid such powers. 

Consider now the outgoing solution (labelled by superscript $^{(1)}$) in \eqref{scalarCauchy2}. With a choice of gauge parameters $\{N_a^{(k)},\bar{N}_a^{(k)},M^{(k)},\bar{M}_a^{(k)}\}$ we must be able to eliminate the non-smooth terms $\hat{f}_{rr}^{(1;j)}$ and $ \tilde{\E}_u^{(1)}$ in \eqref{scalarCauchy2} that are unphysical and just due to our `bad' choice of gauge $\tilde{f}_a =0, \tilde{H}_T =0$.  
A choice of  $\bar{N}_a^{(0)}$, $\bar{N}_a^{(1)}$ and $\bar{M}^{(0)}$, $\bar{M}^{(1)}$ (with $\bar{N}_u^{(0)}=\bar{N}_r^{(1)}=\bar{M}^{(0)}=0$) allows to set 
\be
\hat{f}_{rr}^{(1;0)}=0
\ee
and also to eliminate all the terms $(r-r_-)^0 \log(r-r_-)$ that typically appear in the fields $\widetilde{f}_{ab}^{(1)},\widetilde{f}_a^{(1)},\widetilde{H}_{L,T}^{(1)}$  and $\delta\widetilde{F}_{ab}^{(1)}, \delta\widetilde{F}_{ai}^{(1)} $ as a result of the gauge transformation. We have now the freedom to choose $N_a^{(0)}$, $N_a^{(1)}$ and $M^{(0)}=0$ to set
\be
\hat{f}_{rr}^{(1;1)}=\hat{f}_{ur}^{(1;0)} =\hat{H}_{L}^{(1;0)}= 0
\ee
and to eliminate the term proportional to $(r-r_-)^0$ in $\widetilde{f}_r^{(1)}$ and $\widetilde{H}_T^{(1)}$ (these fields become non-zero as a result of the gauge transformation). But with this choice it follows  that the leading term of $\widetilde{f}_{uu}^{(1)}$ and $\widetilde{f}_u^{(1)}$ is $(r-r_-)^0$ since these terms do not depend on the higher order gauge parameters, {\it i.e.} we have no further gauge freedom to eliminate such terms. After these gauge transformations, the electromagnetic fields $\delta\widetilde{F}_{ab}^{(1)}, \delta\widetilde{F}_{ai}^{(1)} $ also behave as $(r-r_-)^0$.

Altogether, our analysis shows that we can find a gauge where the two linearly independent gravitoelectromagnetic solutions at the Cauchy horizon have the leading behaviour 
\begin{subequations}\label{scalarCauchy3}
\begin{align}
& \widetilde{f}_{a}^{(1)} = e^{-i \omega u} (r-r_-)^{\alpha_a} \widehat{f}_{a}^{(1)}(r) , \qquad  \widetilde{f}_{a}^{(2)} = e^{-i \omega u}  (r-r_-)^{\alpha_a+i \omega/\kappa_-} \widehat{f}_{a}^{(2)}(r)\,; \label{vectorCauchy3a}\\
& \widetilde{f}_{ab}^{(1)} = e^{-i \omega u} (r-r_-)^{\alpha_{ab}} \widehat{f}_{ab}^{(1)}(r) , \qquad  \widetilde{f}_{ab}^{(2)} = e^{-i \omega u}  (r-r_-)^{\alpha_{ab}+i \omega/\kappa_-} \widehat{f}_{ab}^{(2)}(r)\,; \label{vectorCauchy3b}\\
&  \widetilde{H}_{L,T}^{(1)} = e^{-i \omega u}(r-r_-) \widehat{H}_{L,T}^{(1)}(r) , \qquad  \widetilde{H}_{L,T}^{(2)} = e^{-i \omega u}  (r-r_-)^{i \omega/\kappa_-} \widehat{H}_{L,T}^{(2)}(r)\,; \label{vectorCauchy3c}\\
&  \delta\widetilde{F}_{ab}^{(1)} = e^{-i \omega u} \delta\widehat{F}_{ab}^{(1)}(r) , \qquad  \delta\widetilde{F}_{ab}^{(2)} = e^{-i \omega u}  (r-r_-)^{-1+i \omega/\kappa_-} \delta\widehat{F}_{ab}^{(2)}(r)\,; \label{vectorCauchy3d}\\
&  \delta\widetilde{F}_{ai}^{(1)} = e^{-i \omega u}\delta\widehat{F}_{ai}^{(1)}(r) , \qquad  \delta\widetilde{F}_{ai}^{(2)} = e^{-i \omega u}  (r-r_-)^{-\alpha_a+i \omega/\kappa_-} \delta\widehat{F}_{ai}^{(2)}(r)\,; \label{vectorCauchy3e}
\end{align}
\end{subequations}
where $\alpha_a=\{0,1\}$ for $a=\{u,r \}$ (respectively) and $\alpha_{ab}=\{0,1,1\}$ for $ab=\{uu,ur,rr \}$ (respectively), and 
 $\widehat{f}_a, \widehat{f}_{ab},\widehat{H}_{L,T}$ and $\delta\widehat{F}_{ab},\delta\widehat{F}_{ai}$ are smooth functions that depend on $\omega$ and  $\ell_{\rm s}$ (recall that $\delta \widetilde{F}_{ij}$ is not excited in the scalar sector; see \eqref{pert:scalar}). Note that the outgoing solution is manifestly smooth at the Cauchy horizon.
 
As explained above, for a weak solution we need the metric perturbation and its first derivative to be locally square integrable, and the Maxwell field strength perturbation to be locally square integrable. Using the above results, we can repeat the argument we used for vector-type perturbations to see that the condition for a scalar-type quasinormal mode to be extendible as a weak solution across the Cauchy horizon is exactly the same condition \eqref{betaBound} that we obtained for vector-type perturbations. 

Finally, in this section we have so far assumed $\ell_{\rm s} > 1$. Harmonics with $\ell_{\rm s} = 1$ are special because $ \scalar_{ij}$ vanishes for these harmonics; as a consequence, the field $H_T$ is not defined. It follows that, for $\ell_{\rm s}=1$, the fields ${\cal F}$ and ${\cal F}_{ab}$ defined in \eqref{scalarFsGaugeinv} are no longer gauge invariant \cite{Kodama:2003kk}. Additionally, the Bianchi identity no longer implies \eqref{bianchi} and it turns out that only the electromagnetic field is dynamical \cite{Kodama:2003kk}. For our purposes, a pragmatic way to deal with this $\ell_{\rm s} = 1$ case, as suggested in \cite{Kodama:2003kk}, is to impose \eqref{bianchi} as a gauge condition and then fix a residual gauge freedoom at our convenience.\footnote{For further details see the discussions below (5.8) and (5.28) and, specially,  Appendix D of \cite{Kodama:2003kk}.} We can then reconstruct the gravitoelectromagnetic fields $\delta g_{\mu\nu}$ and $\delta F_{\mu\nu}$ in this particular gauge following steps similar to those described above for the $\ell_{\rm s} > 1$. Finally, we add again gauge transformations to make our solutions smoother. In the end of the day, we find that the  condition  for a $\ell_{\rm s}=1$ scalar-type gravitoelectromagnetic quasinormal mode to constitute a weak solution at the Cauchy horizon is still given by \eqref{betaBound}.

\subsection{Conclusions} \label{sec:weak_conclude}

We have shown that the condition for a linearized gravitoelectromagnetic mode solution to be extendible as a weak solution across the Cauchy horizon is \eqref{betaBound}. We define $\beta$ in terms of the spectral gap $\alpha$ as in \eqref{betadef}. If $\beta <1/2$ then there exists a quasinormal mode which violates \eqref{betaBound}. One can add an arbitrary multiple of this quasinormal mode to any other linear perturbation. Hence if $\beta<1/2$ then a generic linear perturbation cannot be extended as a weak solution across the Cauchy horizon. So if $\beta<1/2$ then the Christodoulou formulation of strong cosmic censorship is respected.

Conversely, if $\beta>1/2$ then all quasinormal modes respect \eqref{betaBound}. Since the behaviour at the Cauchy horizon is determined by the slowest decaying quasinormal mode, in this case, any linearized gravitoelectromagnetic perturbation arising from smooth initial data can be extended across ${\cal CH}_R^+$ as a weak solution of the equation of motion, so the Christodoulou version of strong cosmic censorship is violated for smooth initial data. 

Finally, we can consider extendibility in $C^r$. By this we mean that there exists a gauge so that, at ${\cal CH}_R^+$, the metric is $C^r$ and the Maxwell field strength is $C^{r-1}$ (so the Maxwell potential is $C^r$ in some gauge). It is easy to see from the above analysis that a quasinormal mode is extendible in $C^r$ across ${\cal CH}_R^+$ if $-{\rm Im}(\omega)/\kappa_-  \ge r$. Thus, in Einstein-Maxwell theory, the $C^r$ version of strong cosmic censorship is respected if $\beta < r$ and violated if $\beta > r$.

\section{Computing the gravitoelectromagnetic quasinormal modes} 
\label{sec:spectralgap} 

In this section, we first discuss (subsection~\ref{sec:KI}) the Kodama-Ishisbashi (KI) master equations  \cite{Kodama:2003kk} and boundary conditions of the quasinormal mode problem that we later solve analytically and numerically. We will also prove that vector-type and scalar-type modes of RNdS have the same frequency spectrum, {\it i.e.} they are isospectral (subsection~\ref{sec:iso}). 

\subsection{Master equations and boundary conditions} 
\label{sec:KI} 

\subsubsection{Vector-type modes} 
\label{sec:KIvector} 
The vector equations \eqref{vector:MasterEqs} describe a pair of coupled ODEs for the gauge invariant variables $\Omega$ and $\A$. They can be rewritten as a pair of two decoupled ODEs for a pair of master variables $\Phi_\pm$. These are linear combinations of the original gauge invariant variables, namely
\begin{equation}\label{vector:masterV}
\Phi_\pm =a_\pm r^{-1}\Omega + b_\pm \A
\end{equation}
where $a_\pm$ and $b_\pm$ are functions of $M,Q,\ell$ given in equations (4.35)-(4.36) of \cite{Kodama:2003kk}. 
Under \eqref{vector:masterV}, \eqref{vector:MasterEqs} tranform into the KI vector master equations 
\begin{equation}\label{vector:masterEOM}
   f\left(f\,\Phi_\pm' \right)'+\left(\omega^2-V_{\rm{v}\pm}\right)\Phi_\pm
  =0\,,
\end{equation}
where the potentials  are given by
\begin{equation}\label{vector:masterEOMaux}
V_{\rm{v}\pm}=\frac{f}{r^2}\left[k_V^2 +1
  +\frac{4Q^2}{r^2} 
   +\frac{1}{r}\left(-3M \pm \sqrt{9M^2+4(k_V^2-1)Q^2} \right)\right]\,.
 \end{equation}   
 When $Q= 0$, $\Phi_-$ and $\Phi_+$ are simply
proportional to $\Omega$ and $\A$, respectively.  Thus, in the neutral limit, $\Phi_-$ and $\Phi_+$ represent, respectively, the gravitational and electromagnetic modes of the  Schwarzschild black hole.  Note that $\Phi_+$ modes have $\ell_V = 1,2,3 \ldots$ whereas $\Phi_-$ modes have $\ell_V = 2,3,4,\ldots$.

Vector quasinormal modes are solutions of \eqref{vector:masterEOM} that obey ingoing   boundary conditions at the black hole horizon and outgoing boundary conditions at the cosmological horizon. More concretely, at the black hole horizon  $r=r_+$ a Frobenius analysis yields the expansion 
\be \label{vector:BCp}
\Phi(r)=(r-r_+)^{\pm\,\frac{i \omega}{2\,\kappa_+}}\sum_{n=0}^{+\infty}(r-r_+)^n\,\Phi^{(n,+)}
\ee
where $\Phi$ is either $\Phi_+$ or $\Phi_-$. Regularity at the event horizon, which follows from demanding a smooth expansion in ingoing coordinates $(v,r,\theta,\phi)$ around $\mathcal{H}^+_R$, requires that we discard the solution with the positive sign. Similarly, a Frobenius expansion at the cosmological horizon $r=r_c$ yields the two possible solutions
\be
\Phi(r)=(r_c-r)^{\pm\,\frac{i \omega}{2\,\kappa_c}}\sum_{n=0}^{+\infty}(r_c-r)^n\,\Phi^{(n,c)}\,,
\ee
and imposing outgoing boundary conditions at the cosmological horizon $\mathcal{H}^c_R$ requires that we discard the irregular solution with plus sign. We are thus lead to introduce the field redefinition:
\be
\Phi_\pm(r)=(r-r_+)^{-\,\frac{i \omega}{2\,\kappa_+}}(r_c-r)^{-\,\frac{i \omega}{2\,\kappa_c}}\tilde{\Phi}_\pm(r)
\label{eq:Phire}
\ee
where $\tilde{\Phi}_\pm(r)$ is a smooth function at $r=r_+$ and at $r=r_c$. This effectively imposes the desired boundary conditions since our numerical method can only search for smooth functions $\tilde{\Phi}_\pm(r)$.

Inserting \eqref{eq:Phire} into \eqref{vector:masterEOM} we get a pair of decoupled ODEs for $\tilde{\Phi}_\pm$. Each of these ODEs is quadratic in the frequency $\omega$. That is to say, for each $\ell$ we have to solve a quadratic eigenvalue problem to find the eigenvalue $\omega$ and the associated eigenfunction $\tilde{\Phi}_-$ (or $\omega$ and $\tilde{\Phi}_+$). The boundary conditions for  $\tilde{\Phi}_\pm(r)$ follow directly from doing a Taylor  expansion of the master equation about the black hole and cosmological horizons. These reveals that at both horizons we have a Robin boundary condition, {\it i.e.} of the type 
\be  \label{vector:BCs}
\mathcal{Q}^{+,1}(\omega)\tilde{\Phi}_\pm^\prime(r_+)=\mathcal{Q}^{+,0}(\omega)\tilde{\Phi}_\pm(r_+)\,,\quad\mathcal{Q}^{c,1}(\omega)\tilde{\Phi}_\pm^\prime(r_c)=\mathcal{Q}^{c,0}(\omega)\tilde{\Phi}_\pm(r_c)\,.
\ee
where $\mathcal{Q}^{+,1},\mathcal{Q}^{+,0},\mathcal{Q}^{c,1}$ and $\mathcal{Q}^{c,0}$ are known functions which are at most second order polynomials in $\omega$. 

It is also convenient to use a radial coordinate whose range is independent of the black hole parameters. We define 
\be \label{def:y}
 y=\frac{r-r_+}{r_c-r_+}\,,
\ee
such that $y\in [0,1]$ with $y=0$ ($y=1$) corresponding to the event (cosmological) horizon.\footnote{
Note that in later sections we will often work with a quantity $y_+ \equiv r_+/r_c$. We emphasize that this is {\it not} related to the coordinate $y$.}

The resulting equation for $\tilde{\Phi}_-$ (or $\tilde{\Phi}_+$) can now be solved using a pseudospectral grid discretization (with the methods reviewed in \cite{Dias:2015nua}) as a standard quadratic eigenvalue problem or employing a Newton-Raphson algorithm.
In the former method one writes the equation as a quadratic eigenvalue problem for the frequency $\omega$, which is then solved using \emph{Mathematica}'s built-in routine \emph{Eigensystem}. More details of this method and the discretization scheme can be found {\it e.g.} in \cite{Dias:2010eu}. The second method is based on an application of the Newton-Raphson root-finding algorithm, and is detailed in \cite{Cardoso:2013pza,Dias:2015nua}. The advantage of the first method is that it gives all modes simultaneously. The second method computes a single mode at a time, and only when a seed is known that is sufficiently close to the true answer. However, this method is much quicker as both the size of the grid and numerical precision increases, and can be used to push the numerics to extreme regions of the parameter space. 

\subsubsection{Scalar-type modes} 
\label{sec:KIscalar} 
The  pair of coupled ODEs  \eqref{scalar:MasterEqs} for the scalar gauge invariant variables $\Phi$ and $\A$ can be rewritten as a pair of two decoupled ODEs for a pair of scalar master variables $\Phi_\pm$. The latter are given by the linear combinations
\begin{equation}\label{scalar:masterV}
\Phi_\pm =a_\pm \,\Phi + b_\pm \,\A
\end{equation}
where $a_\pm$ and $b_\pm$ are functions of $M,Q,\ell$ given in equations (5.57)-(5.58) of \cite{Kodama:2003kk}.
Inserting  \eqref{scalar:masterV} into \eqref{scalar:MasterEqs}  yields the KI scalar master equations
\begin{equation}\label{scalar:masterEOM}
 f\left(f\,\Phi_\pm' \right)'+\left(\omega^2-V_{\rm{s}\pm}\right)\Phi_\pm
  =0\,,
\end{equation}
where the potentials  $V_{\rm{s}\pm}$ are given by equations (5.60)-(5.63) of \cite{Kodama:2003kk}.
When $Q= 0$, $\Phi_-$ is proportional to $\Phi$ and $\Phi_+$ is proportional to $\A$.  Hence, in the neutral limit, $\Phi_-$ and $\Phi_+$ represent, respectively, the gravitational and electromagnetic scalar modes of the  Schwarzschild black hole. Note that $\Phi_+$ modes have $\ell_S = 1,2,3 \ldots$ whereas $\Phi_-$ modes have $\ell_S = 2,3,4,\ldots$. 

Scalar quasinormal modes are solutions of \eqref{scalar:masterEOM} that obey ingoing   boundary conditions at the black hole horizon and outgoing boundary conditions at the cosmological horizon. The analysis of these boundary conditions is very much similar to the one done for the KI vector sector. In fact equations  \eqref{vector:BCp} to \eqref{vector:BCs} and the subsequent discussion apply without change to the scalar sector of perturbations. 

\subsection{Isospectrality \label{sec:isospectral}} 
\label{sec:iso} 
As discussed in previous sections, gravitoelectromagnetic perturbations of RNdS black holes come in two classes: vector-type and scalar-type. Although they obey two seemingly distinct equations of motion, it turns out they have the same quasinormal mode spectra. For this reason, the spectrum of quasinormal modes of RNdS black holes is said to be isospectral. This is a classical result in the context of asymptotically flat RN black holes, which was first uncovered by Chandrasekhar in \cite{Chandrasekhar:579245}. It turns out the same result applies in the context of RNdS black holes, but with more involved algebra.

Just as in \cite{Chandrasekhar:579245}, we start by noting that the scalar potential $V_{\rm{s}\pm}(r)$ $-$ introduced in \eqref{scalar:masterEOM} $-$ can be written in the following compact manner
\be
V_{\rm{s}\pm}(r)=\beta_{\pm} f(r)\frac{\mathrm{d}\check{F}_{\pm}(r)}{\mathrm{d}r}+\beta_{\pm}^2 \check{F}_{\pm}(r)^2+\tilde{\kappa}\,\check{F}_{\pm}(r)\,,
\ee
where
\begin{subequations}\label{auxiso}
\begin{align}
&\beta_{\pm}=3\,M\mp\sqrt{9M^2+4\,Q^2\,(\ell-1)(\ell+2)}\,,
\\
&\tilde{\kappa}=(\ell-1)(\ell+2)\left[(\ell-1)(\ell+2)+2\right]\,,
\\
&\check{F}_{\pm}(r)=\frac{f(r)}{r\left[(\ell-1)(\ell+2)\,r+\beta_{\pm}\right]}\,,
\end{align}
\end{subequations}
and $f(r)$ is given in \eqref{metricRNaux}. Rather remarkably, the vector potential \eqref{vector:masterEOMaux} takes a similar form
\be
V_{\rm{v}\pm}(r)=-\beta_{\pm} f(r)\frac{\mathrm{d}\check{F}_{\pm}(r)}{\mathrm{d}r}+\beta_{\pm}^2 \check{F}_{\pm}(r)^2+\tilde{\kappa}\,\check{F}_{\pm}(r)\,,
\ee
with the same quantities defined in \eqref{auxiso}.

Because of this simple relation between the scalar and vector potentials, one can relate solutions of the vector equation to solutions of the scalar equation (and vice versa), via the map
\begin{subequations}\label{map:iso}
\begin{align}
&\Phi_{\rm{s}\pm}(r)=\frac{1}{\tilde{\kappa}+2\,i\,\omega\,\beta_{\pm}}\left[\left(\tilde{\kappa}+2\,\beta_{\pm}^2F_{\pm}(r)\right)\Phi_{\rm{v}\pm}(r)+2\,\beta_{\pm}\,f(r)\,\frac{\mathrm{d}\Phi_{\rm{v}\pm}(r)}{\mathrm{d}r}\right]\,,
\\
&\Phi_{\rm{v}\pm}(r)=\frac{1}{\tilde{\kappa}-2\,i\,\omega\,\beta_{\pm}}\left[\left(\tilde{\kappa}+2\,\beta_{\pm}^2F_{\pm}(r)\right)\Phi_{\rm{s}\pm}(r)-2\,\beta_{\pm}\,f(r)\,\frac{\mathrm{d}\Phi_{\rm{s}\pm}(r)}{\mathrm{d}r}\right]\,,
\end{align}
\end{subequations}
where, momentarily, we added the subscripts $\rm{s}$ and $\rm{v}$ to distinguish between scalar and vector perturbations.

Maps between solutions might not take physical solutions into physical solutions since one has to check that the maps preserve the relevant boundary conditions. This is the case ({\it i.e.} the map \eqref{map:iso} preserves the boundary conditions) for asymptotically flat RN black holes and RNdS black holes, but it is \emph{not} the case for RN black holes with anti de-Sitter boundary conditions \cite{Dias:2013sdc}. For this reason isospectrality occurs in the former two cases, but not in the latter. Note that the differential map  \eqref{map:iso}  alone is not \emph{enough} to guarantee that the critical $\beta$ bound  \eqref{betaBound} found for vector-type modes also holds for scalar-type perturbations, since the two types of metric perturbations are orthogonal to each other. For this reason, in Section \ref{sec:gravbetabound} we had to do the analysis that finds the bound \eqref{betaBound} for the vector and scalar-type of perturbations independently. We concluded that it turns out that \eqref{betaBound} holds for both sectors. 

\section{Classifying the families of quasinormal modes and analytical results} 
\label{sec:analytics} 

Cardoso {\it et al} found that massless scalar field quasinormal modes of RNdS can be classified into three families  \cite{Cardoso:2017soq}. We find that the same is true for gravitoelectromagnetic quasinormal modes. The three families are 1) ``photon sphere" modes, 2) ``de Sitter" modes and 3)  ``near-extremal" modes. The ``photon sphere" modes are identified in the geometric optics limit,  $\ell\gg 1$, and  are related to the properties of the unstable circular photon orbits in the equatorial plane of the black hole background (subsection \ref{sec:geo}).

The de Sitter modes reduce, when $M$ and $Q$ vanish, to the gravitational and electromagnetic quasinormal modes of de Sitter spacetime (subsection \ref{sec:dSqnm}). 
Finally, the ``near-extremal" modes have their wavefunction peaked near the horizon and an approximate expression for these modes (strictly valid in the extremal limit) can be obtained analysing the perturbations in the near-horizon geometry of a near-extremal RNdS black hole (subsection \ref{sec:NHqnm}). 

In the previous Section \ref{sec:isospectral} we found that the spectra of vector-type and scalar-type of quasinormal modes is isospectral. It follows that for each family of modes we just have to consider two sectors (not four) of perturbations corresponding to perturbations for each of the gauge invariant variables $\Phi_-$ and $\Phi_+$. As a test of our numerical code, we did several checks ({\it i.e.} for different black holes) that the frequency eingenvalues of the vector-type equation of motion are indeed the same as those that solve the scalar-type equation of motion.

In this section we will obtain approximate analytical expressions for the three families of modes (that are valid at least in a certain region of the RNdS parameter space). Then we compare these analytical results with the exact data that results from our numerical search of the frequency spectra in the full RNdS parameter space $0\le y_+ \le 1$ and $0<Q/Q_{\rm ext}\le 1$.

\subsection{\label{sec:geo}Photon sphere family of modes and its geometric optics limit}

In this subsection we will find an analytical expression for the photon sphere quasinormal modes in the geometric optics limit, {\it i.e.} in the WKB limit $\ell\to \infty$. We find that this analytical expression gives an imaginary part of the frequency that matches very well the numerical results even for $\ell=1$ (the real part is not such a good approximation for low $\ell$). Our geometric optics results are independent of the spin of the perturbing field and so they should agree with the geometric optics results for massless scalar field photon sphere modes in Ref. \cite{Cardoso:2017soq}.

Consider a null geodesic $x^\mu(\tau)$ of a RNdS black hole. By spherical symmetry there is no loss of generality in assuming that the geodesic is confined to the equatorial plane $\theta= \pi/2$. There are conserved quantities associated to the Killing fields $K = \partial/\partial t$ and $\chi = \partial/\partial \phi$:
\be
e\equiv - K_\mu \dot{x}^\mu\qquad \text{and}\qquad j\equiv \chi_\mu \dot{x}^\mu\,,
\label{eq:conserved}
\ee
where the dot represents derivative with respect to the affine parameter $\tau$. This gives
\begin{equation}\label{tphidot}
\dot{t}=\frac{e}{f}\,,\qquad \dot{\phi}=\frac{j}{r^2}\,.
\end{equation}
The radial motion is governed by
\be
\label{eq:geodesic}
\dot{r}^2+V(r;b)=0\,,
\ee
where
\be
V(r;b)=\frac{j^2}{b^2} \left[\frac{b^2}{r^2}\left(1-\frac{r^2}{L^2}-\frac{2 M}{r}+\frac{Q^2}{r^2}\right)-1\right].
\label{eq:pot}
\ee
and we have defined the geodesic impact parameter as
\be \label{def:b}
b\equiv\frac{j}{e}\,.
\ee
Now, we want to find the photon sphere, where null particles are trapped on unstable circular orbits. This occurs for values $r=r_s$ and $b=b_s$ such that
\be
V(r_s,b_s)=0\quad\text{and}\quad \left.\partial_r V(r,b)\right|_{r=r_s,b=b_s}=0.
\ee
This gives
\be \label{photonsphere}
r_s=\frac{1}{2} \left(\sqrt{9 M^2-8 Q^2}+3 M\right)\, \quad \hbox{and}\quad b_s(r_s)=\frac{L r_s^2}{\sqrt{L^2 \left(r_s (r_s-2 M)+Q^2\right)-r_s^4}}\,,
\ee
where we can check that $r_+\leq r_s \leq r_c$.

The orbital angular velocity (Kepler frequency) of our null circular photon orbit can now be computed using \eqref{tphidot}, \eqref{def:b} and \eqref{photonsphere} yielding
\be
\Omega_c \equiv \frac{\dot{\phi}}{\dot{t}}=\frac{1}{b_s}\,.
\ee
We now have to  compute the largest Lyapunov exponent $\lambda_L$, measured in units of $t$, associated with perturbations of an unstable circular photon orbit $r(\tau)=r_s$. This is done considering  perturbations  $r(\tau)=r_s+\delta r(\tau)$ of the radial geodesic equation \eqref{eq:geodesic}. Small deviations obey the linearized equation
\be
\delta r'(t)-\frac{\sqrt{r_s^2-2 Q^2}}{b_s r_s} \,\delta r(t)=0\,
\ee
which has solution 
\be
 \delta r(t)=C \,e^{\lambda_L t} \qquad \hbox{with} \quad \lambda_L=\frac{\sqrt{r_s^2-2 Q^2}}{b_s r_s}
\ee
being the desired (largest) Lyapunov exponent. Note that $C$ is an integration constant and the unstable photon orbit parameters $r_s$ and $b_s$ are given in terms of the RNdS parameters $\{L,M,Q\}$ by  \eqref{photonsphere}.

Finally, one can reconstruct the spectrum of the photon sphere family of quasinormal modes with $\ell\gg 1$ using \cite{Goebel:1972,Ferrari:1984zz,Ferrari:1984ozr,Mashhoon:1985cya,Bombelli:1991eg,Cornish:2003ig,Cardoso:2008bp,Dolan:2010wr,Yang:2012he} 
\be \label{wWKB}
\omega_{\mathrm{WKB}} \approx \ell\,\Omega_c-i\left(n+\frac{1}{2}\right)\lambda_L\,,  
\ee
where $n=0,1,2,\ldots$ is the radial overtone. Note that this geometric optics/WKB approximation is universal in the sense that it is blind the particular sector of perturbations we look at. That is, it is expected to be a good approximation to both photon sphere modes $\Phi_\pm$ (or for a massless scalar field \cite{Cardoso:2017soq}). 

Note that, at this order, ${\rm Im}(\omega_{\mathrm{WKB}})$ is independent of $\ell$ (assuming $\ell\gg 1$) while ${\rm Re}(\omega_{\mathrm{WKB}})$ does depend on $\ell$. 
One might wonder whether next-to-leading order corrections to this result might change significantly \eqref{wWKB}, especially near extremality. However, the corrections to ${\rm Im}(\omega)$ are of order $1/\ell$ so, for any fixed background, the corrections to ${\rm Im}(\omega)$ can be made arbitrarily small by taking $\ell$ sufficiently large.\footnote{
In fact for vanishing $\Lambda$ the corrections to ${\rm Im}(\omega_{\mathrm{WKB}})$ are ${\cal O}(1/\ell^2)$ \cite{Dolan:2010wr} and we expect that the same is true with $\Lambda >0$.} So the WKB results for ${\rm Im}(\omega)$ should be reliable for sufficiently large $\ell$. 

\begin{figure}[ht]
	\centering
	\includegraphics[width=0.54\textwidth]{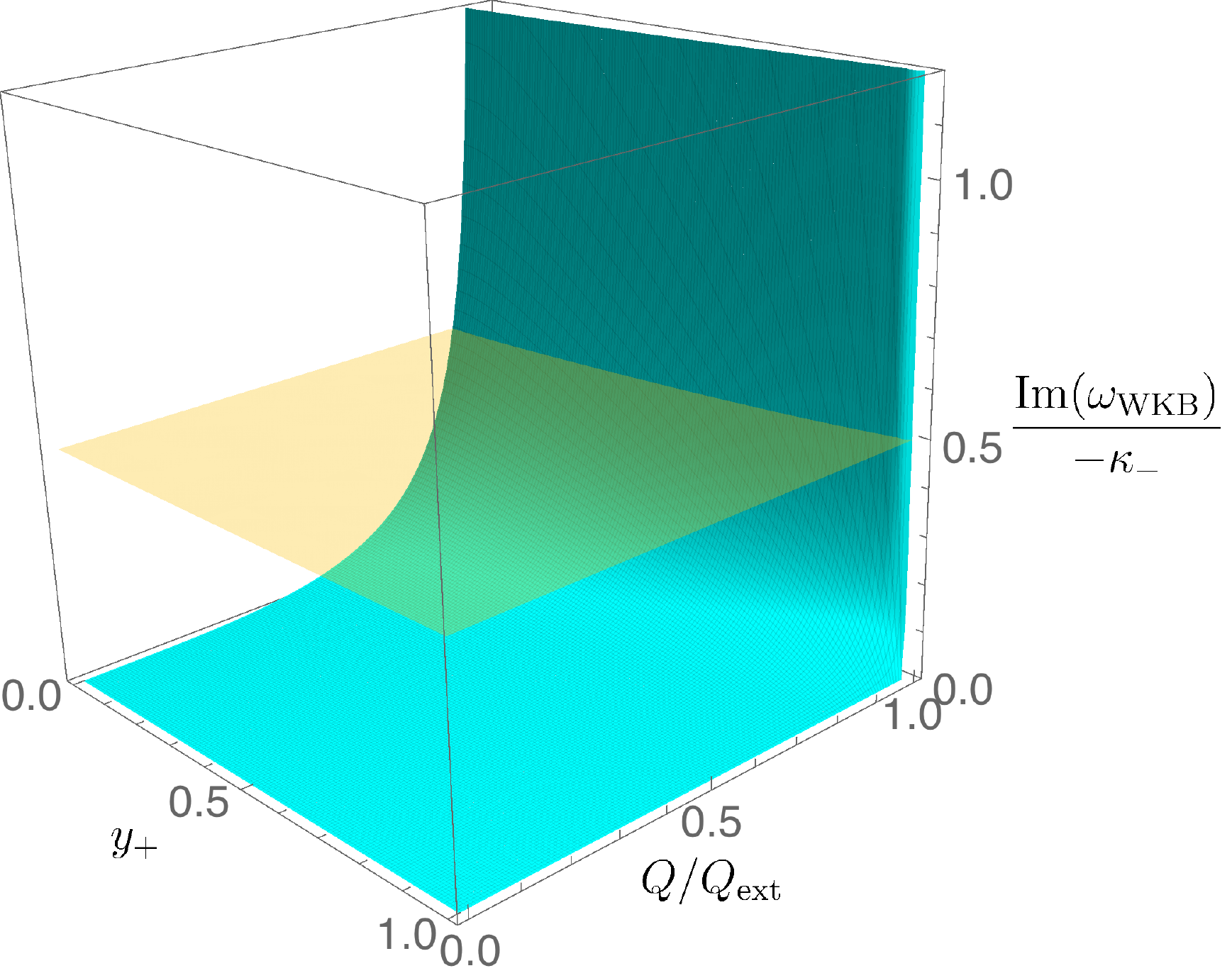}
	\hspace{0cm}
	\includegraphics[width=0.44\textwidth]{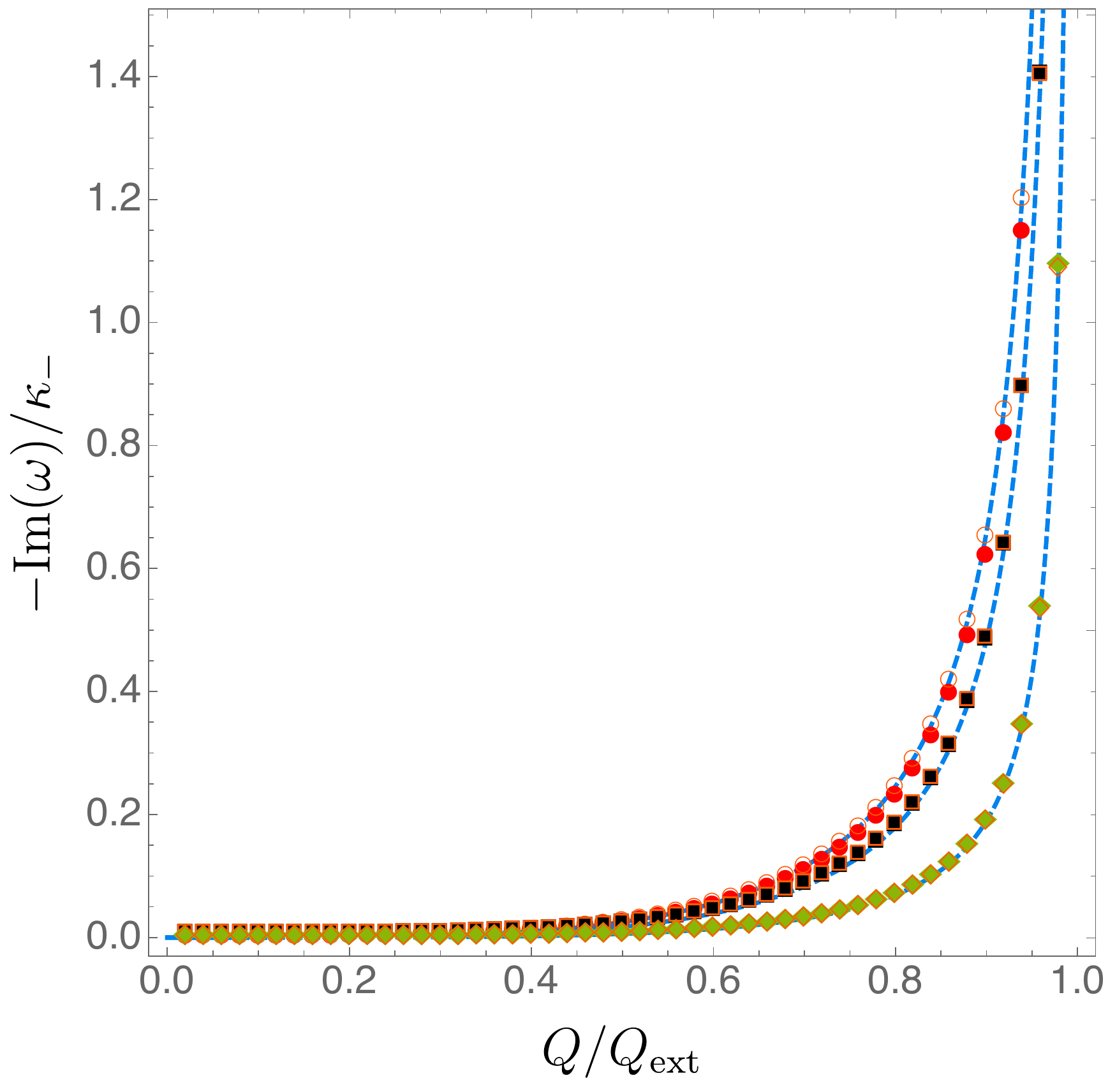}
	\caption{{\bf Left panel:} Photon sphere quasinormal modes in the geometric optics/WKB approximation  as a function of charge ratio $Q/Q_{\rm ext}$ and $y_+=r_+/r_c$ for $n=0$. The yellow plane is $-{\rm Im}(\omega)/\kappa_-=1/2$. {\bf Right panel:} WKB prediction for $-{\rm Im}(\omega)/\kappa_-$ compared with numerical results for $\Phi_-$ photon sphere modes. The curves are for $y_+=0.1$ (top), $y_+=0.4$ (middle) and $y_+=0.8$ (bottom). The dashed blue lines are the $n=0$ geometric optics/WKB prediction $-\mathrm{Im}(\omega_{\mathrm{WKB}})/\kappa_-$. The red disks, black filled squares, green filled diamonds are the numerical results for $\ell=2, n=0$. The empty orange marks (circles, squares, diamonds) are the numerical results for $\ell=10, n=0$.}
	\label{fig:WKBps}
\end{figure}

We can now analyse $-{\rm Im}(\omega_{\mathrm{WKB}})/\kappa_-$. In the left panel of Fig.~\ref{fig:WKBps} we plot this quantity for $n=0$ (which yields the smallest value) as a function of the horizon radii ratio $y_+=r_+/r_c$ and charge ratio $Q/Q_{\rm ext}$. Over most of the RNdS moduli space we have $-{\rm Im}(\omega_{\mathrm{WKB}})/\kappa_-<1/2$. Since we expect our result for to be exact as $\ell \rightarrow \infty$, we must therefore have $\beta < 1/2$ over most of the RNdS moduli space \cite{Cardoso:2017soq}. Thus the Christodoulou version of strong cosmic censorship is respected by most RNdS black holes. However, for any fixed $y_+$, there is always a critical value for $Q/Q_{\rm ext}$  (close to extremality) above which $-{\rm Im}(\omega_{\mathrm{WKB}})/\kappa_->1/2$. So there is the possibility of a violation of strong cosmic censorship by near-extremal RNdS black holes. 

\begin{figure}[th]
	\centering
	\includegraphics[width=0.47\textwidth]{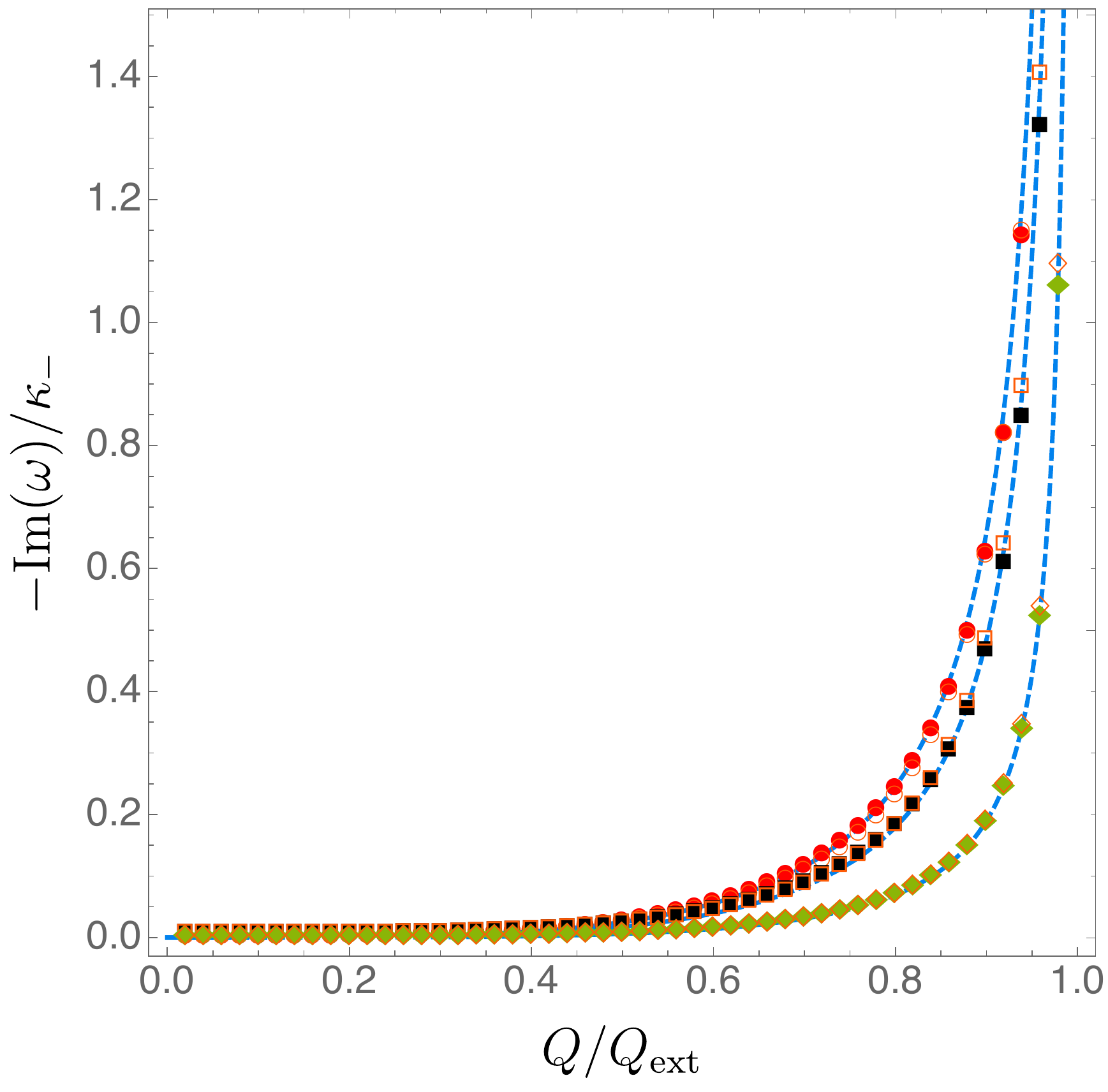}
	\hspace{0.5cm}
	\includegraphics[width=0.47\textwidth]{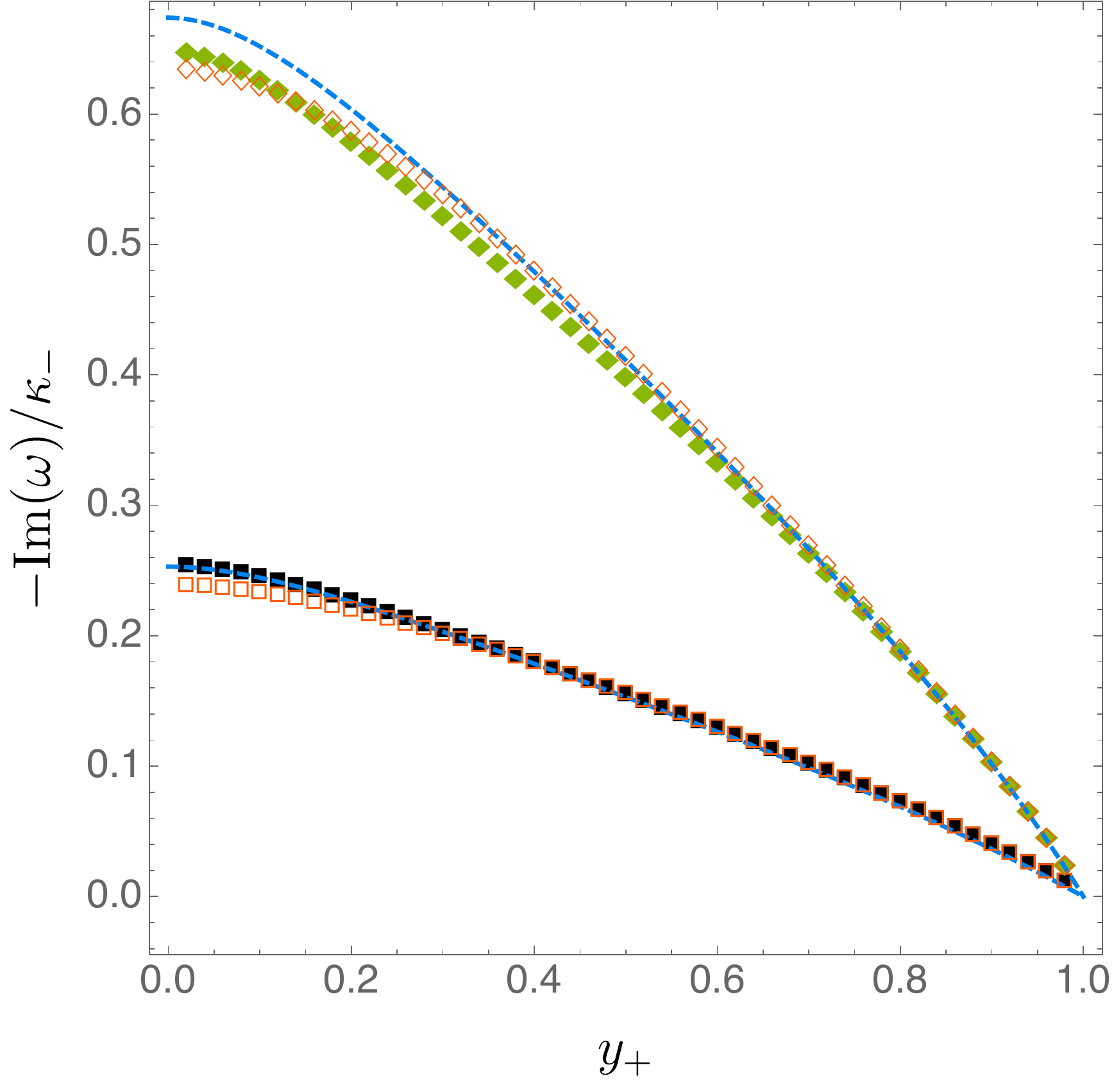}
	\caption{Photon sphere modes for modes $\Phi_+$ with $\ell=1$ (filled marks: disks, squares, diamonds) and $\Phi_-$ with $\ell=2$ (empty orange marks: circles, squares, diamonds). The dashed blue curves are the WKB predictions. 
{\bf Left panel:}  The three curves are for $y_+=0.1$ (top), $y_+=0.4$ (middle) and $y_+=0.8$ (bottom). Note that the empty marks here are the filled marks in Fig.~\ref{fig:WKBps}. {\bf Right panel:} The  black filled/empty squares describe solutions with $Q/Q_{\rm ext}=0.7992$, while the green filled/empty diamonds represent the numerical results for $Q/Q_{\rm ext}=0.8991$.}
	\label{fig:WKBps2}
\end{figure} 

We will now compare the WKB prediction with our numerical results for the quasinormal frequencies of photon sphere modes. In the right panel of Fig.~\ref{fig:WKBps} we compare the $n=0$ WKB result with our numerical results for $-{\rm Im}(\omega)/\kappa_-$ for the  $\Phi_-$ photon sphere quasinormal modes (with $n=0$). From the plot we see that, when $\ell=10$, the WKB prediction is in excellent agreement with our numerical results. In fact even for $\ell=2$ the plot shows that the WKB prediction is in very good agreement with our numerical results. This agreement extends to other values of $y_+$ not shown in the plot. Note that, as expected, the agreement is very good  for the imaginary part of the frequency but not so good for the real part (not shown in the plot). As a check of our numerical computations we have also confirmed that we reproduce some (the ones we searched for in our tests) of the quasinormal frequencies listed in \cite{Mellor:1989ac} (note that this reference only computed what we call photon sphere modes).

Recall that to compute $\beta$ defined in \eqref{betadef} we need to determine the spectral gap $\alpha$. To determine $\alpha$ we need to find the slowest decaying quasinormal mode, {\it i.e.} the one with the smallest value of $-{\rm Im}(\omega)$. We will now discuss which of the photon sphere modes has the smallest value of $-{\rm Im}(\omega)$. There are two types of photon sphere modes: one corresponding to $\Phi_-$ and another to $\Phi_+$. Our numerical results indicate that, for each type, the lower $\ell$ and $n$ modes dominate. Therefore the slowest decaying photon sphere mode must be one of the following (with $n=0$): (1) $\Phi_-$, $\ell=2$, or  (2) $\Phi_+$, $\ell=1$.   

Which of these two modes decays most slowly? For most of the black hole parameter space we find that the  $\Phi_+$ modes with $\ell=1$ decay most slowly. To illustrate this, in the left panel of Fig.~\ref{fig:WKBps2} we  plot $-{\rm Im}(\omega)/\kappa_-$ vs $Q/Q_{\rm ext}$ at fixed $y_+$ for $\Phi_+$ modes with $\ell=1$ (and $n=0$) and  $\Phi_-$ modes with $\ell=2$ (and $n=0$).
We see that $\Phi_+$ modes with $\ell=1$ typically have lower $-{\rm Im}(\omega)/\kappa_-$ (for fixed background parameters) than $\Phi_-$ modes with $\ell=2$. However, there are small islands in the parameter space where the opposite occurs: see curve $y_+=0.1$ (red disks/circles) for $Q/Q_{\rm ext} \lesssim 0.9$. A similar conclusion is reached from the right panel of Fig.~\ref{fig:WKBps2}. Here we plot the same modes but this time for RNdS with fixed $Q$ and varying $y_+$. 
We see that typically the $\Phi_+$, $\ell=1$ modes dominate over the $\Phi_-$, $\ell=2$ modes. However, for small $y_+$ there is a crossover and the $\ell=2$ modes become dominant. 

These crossovers will not be a problem for our purposes. For each RNdS black hole we will compute numerically the two types ($\Phi_\pm$) of photon sphere quasinormal mode and then pick the one with lowest $-{\rm Im}(\omega)$. This can then be compared with the results from the other families (dS and near-extremal) of quasinormal modes in order to calculate the spectral gap.     

\subsection{\label{sec:dSqnm}de Sitter family of modes}

In the de Sitter limit,  $M=0$, $Q=0$, the master equations for $\Phi_+$ and $\Phi_-$ are the same. To find the spectrum, we just need to take  \eqref{vector:masterEOM} or \eqref{scalar:masterEOM} and set $M=0$, $Q=0$. Using the radial coordinate \eqref{def:y} this yields the master equation  
\be \label{vector:masterEOMdS}
\left(1-y^2\right) \Phi_\pm''(y)-2 y \Phi_\pm'(y)+\left(\frac{\tilde{\omega}^2}{1-y^2}-\frac{\ell (\ell+1)}{y^2}\right)\Phi_\pm(y) =0\,,
\ee
where we have introduced the dimensionless frequency $\tilde{\omega}=\omega\,r_c$ (with $r_c=L$ for the dS solution). Note that $\ell=1,2,3,\ldots$ for electromagnetic modes $\Phi_+$ and $\ell= 2,3,4, \ldots$ for gravitational modes $\Phi_-$. 

The general solution of \eqref{vector:masterEOMdS} is  
\begin{eqnarray}
\Phi_\pm&=& A \,y^{\ell +1} \left(1-y^2\right)^{-\frac{i \tilde{\omega}}{2}} \, _2F_1\left(\frac{1}{2} (\ell -i \tilde{\omega} +1),\frac{1}{2} (\ell -i \tilde{\omega} +2),\frac{3}{2}+\ell;y^2\right) \nonumber\\
&& +B \, y^{-\ell }  \left(1-y^2\right)^{-\frac{i \tilde{\omega}}{2}} \, _2F_1\left(\frac{1}{2} (-\ell -i \tilde{\omega} ),\frac{1}{2} (-\ell -i \tilde{\omega} +1),\frac{1}{2}-\ell ;y^2\right) 
\end{eqnarray}
for arbitrary amplitudes $A$ and $B$, with $_2F_1(a,b,c;z)$ being the Gaussian Hypergeometric function. At the origin this solution behaves as $\Phi_\pm \big|_{y=0}\approx A\,y^{\ell+1}+B\,y^{-\ell}$ and regularity at $y=0$ thus requires that we set $B=0$. On the other hand, a Taylor expansion about the cosmological horizon $y=1$ yields
\ba
\Phi_\pm \big|_{y=1} \simeq \frac{i\,A \,\pi \,  \Gamma \left(\ell +\frac{3}{2}\right)}{\sinh (\pi  \tilde{\omega} )}&& \left(\frac{2^{\frac{i \tilde{\omega} }{2}} (1-y)^{\frac{i \tilde{\omega} }{2}}/\Gamma (1+i \tilde{\omega})}{ \Gamma \left(\frac{1}{2} [\ell+1 -i \tilde{\omega}]\right) \Gamma \left(\frac{1}{2} [\ell+2 -i \tilde{\omega}]\right)} \right. \nonumber \\ &-& \left. \frac{2^{-\frac{i \tilde{\omega} }{2}} (1-y)^{-\frac{i \tilde{\omega} }{2}}/\Gamma (1-i \tilde{\omega} ) }{\Gamma \left(\frac{1}{2} [\ell+1 +i \tilde{\omega} ]\right) \Gamma \left(\frac{1}{2} [\ell+2 +i \tilde{\omega} ]\right)}\right).
\ea
Requiring outgoing boundary conditions demands that we discard the $ (1-y)^{i\frac{\tilde{\omega}}{2}}$ solution. This can be done using the property $\Gamma(-n)=\infty$, $n\in \mathbb{N}_0$, {\it i.e.} requiring that $\Gamma \left(\frac{1}{2} [\ell+1 -i \tilde{\omega}]\right) =\Gamma(-n)$ or  $\Gamma \left(\frac{1}{2} [\ell+2 -i \tilde{\omega}]\right)=\Gamma(-n)$ with $n=0,1,2,\ldots$. The former condition embraces the latter and  quantizes the $\Phi_\pm$ quasinormal mode frequencies of de Sitter as 
\be\label{puredSw}
\hbox{de Sitter:} \qquad \omega \,r_c \big|_{\rm dS}=-i (1+\ell+2n)\,, \quad \hbox{for}\quad n=0,1,2,\ldots
\ee  
with $\ell=1,2,3,\cdots$ for $\Phi_+$ modes and at $\ell=2,3,\cdots$ for $\Phi_-$ modes.

\begin{figure}[ht]
	\centering
	\includegraphics[width=0.45\textwidth]{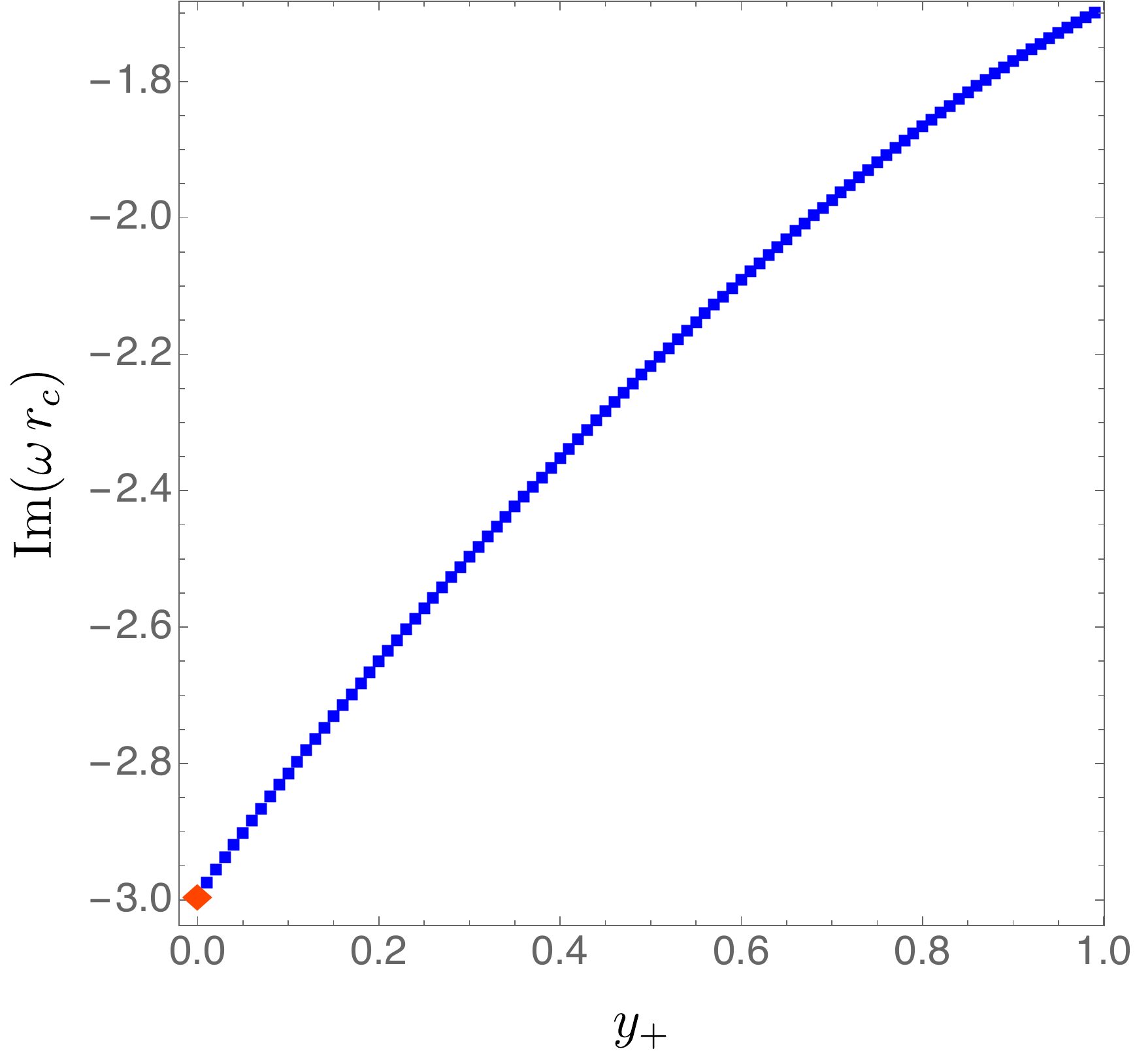}
	\hspace{0.5cm}
	\includegraphics[width=0.45\textwidth]{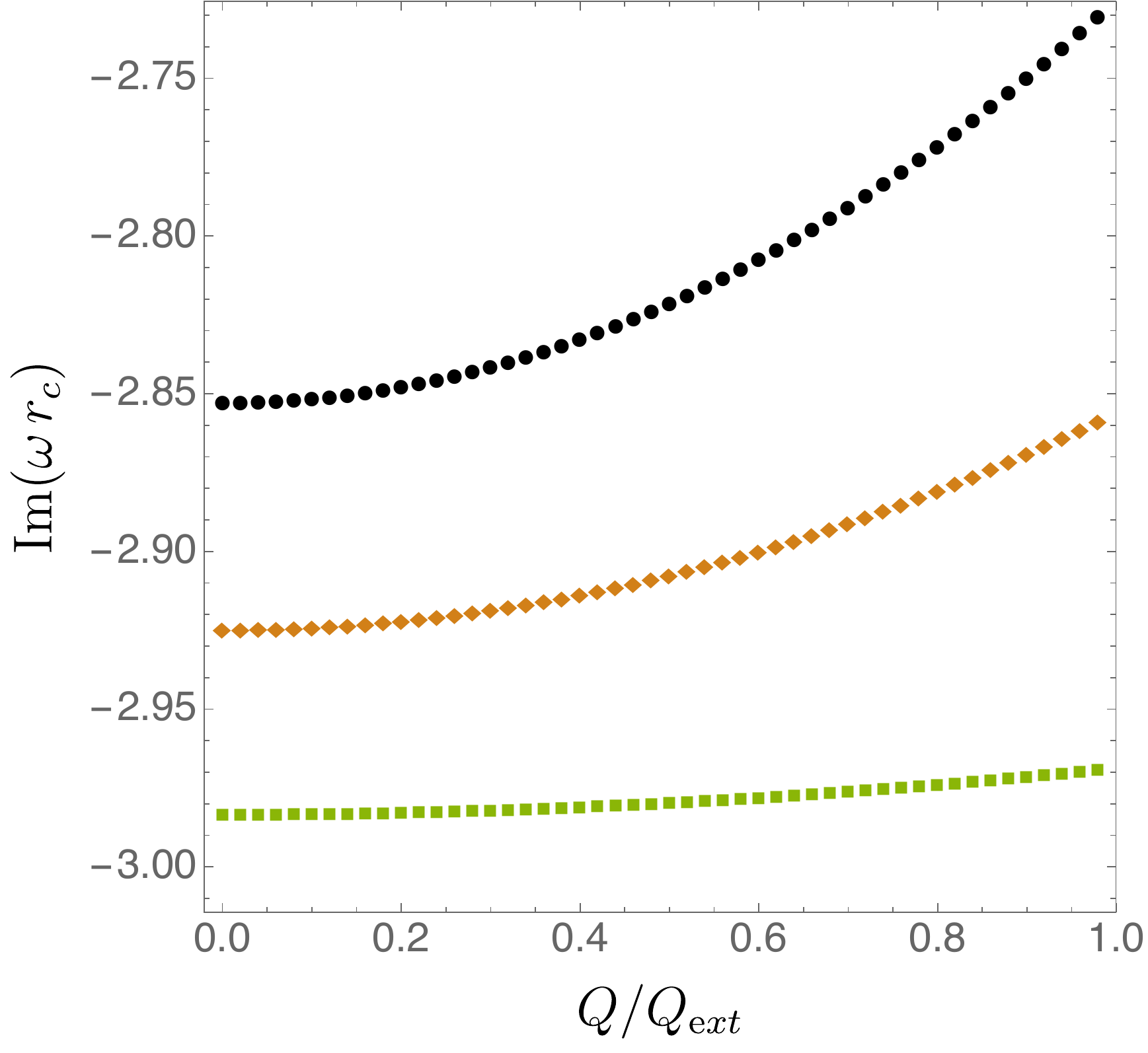}
	\caption{de Sitter gravitoelectromagnetic mode $\Phi_-$ with $\ell=2$ and $n=0$. {\bf Left panel:} de Sitter frequency ${\rm Im}(\omega\,r_c)$ as a function of $y_+$ at fixed $Q/Q_{\rm ext}=0.5$.  The red diamond at $y_+=0$ is the analytical de Sitter gravitational quasinormal mode frequency $\omega\,r_c=-3\,i$. {\bf Right panel:}  Imaginary part of the frequency  as a function of $Q/Q_{\rm ext}$ for fixed $y_+=0.01$ (green squares), $y_+=0.05$ (brown diamonds) and $y_+=0.1$ (black disks).}
	\label{fig:dS}
\end{figure} 

\begin{figure}[ht]
	\centering
	\includegraphics[width=0.47\textwidth]{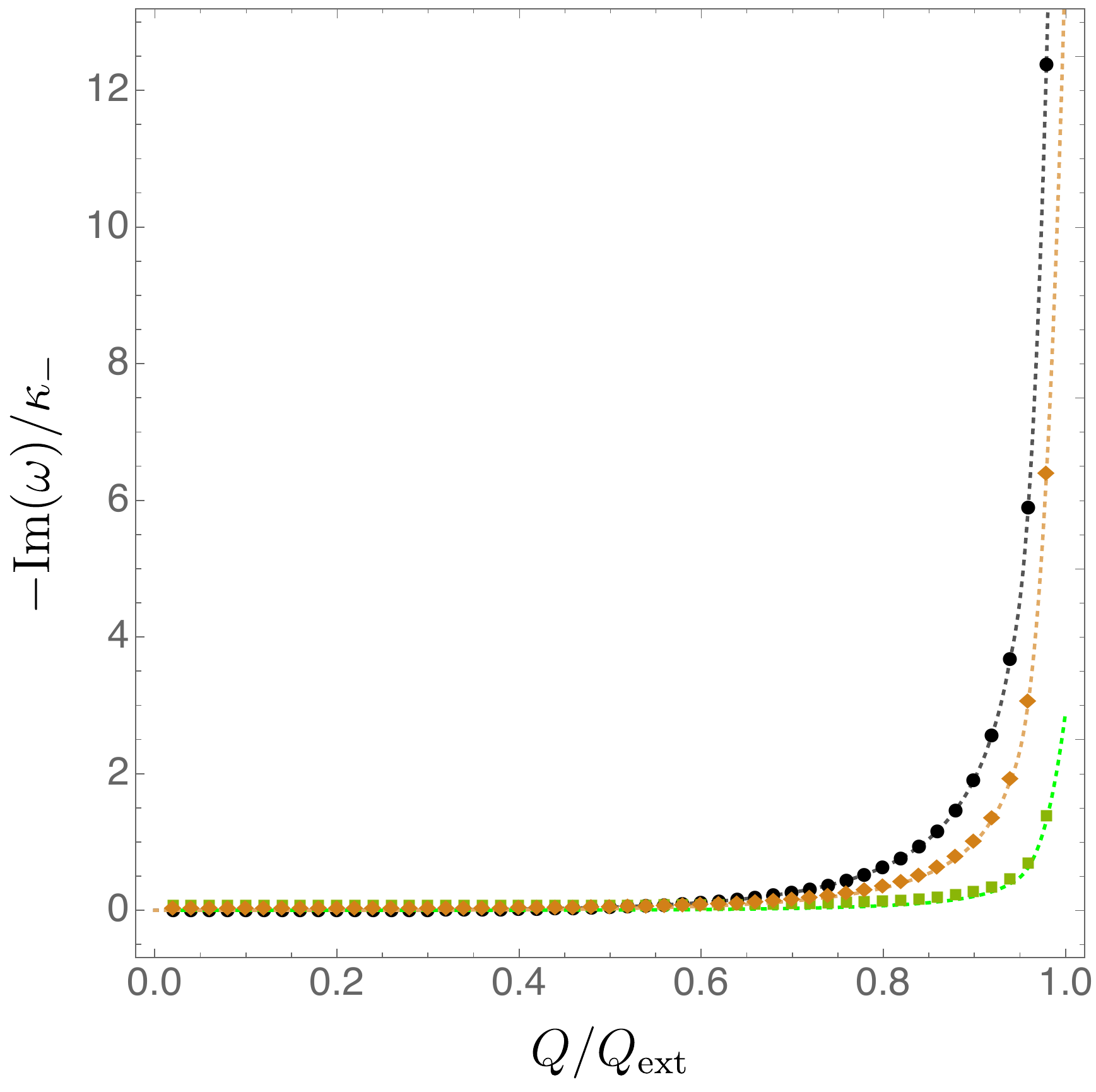}
		\hspace{0.5cm}
	\includegraphics[width=0.48\textwidth]{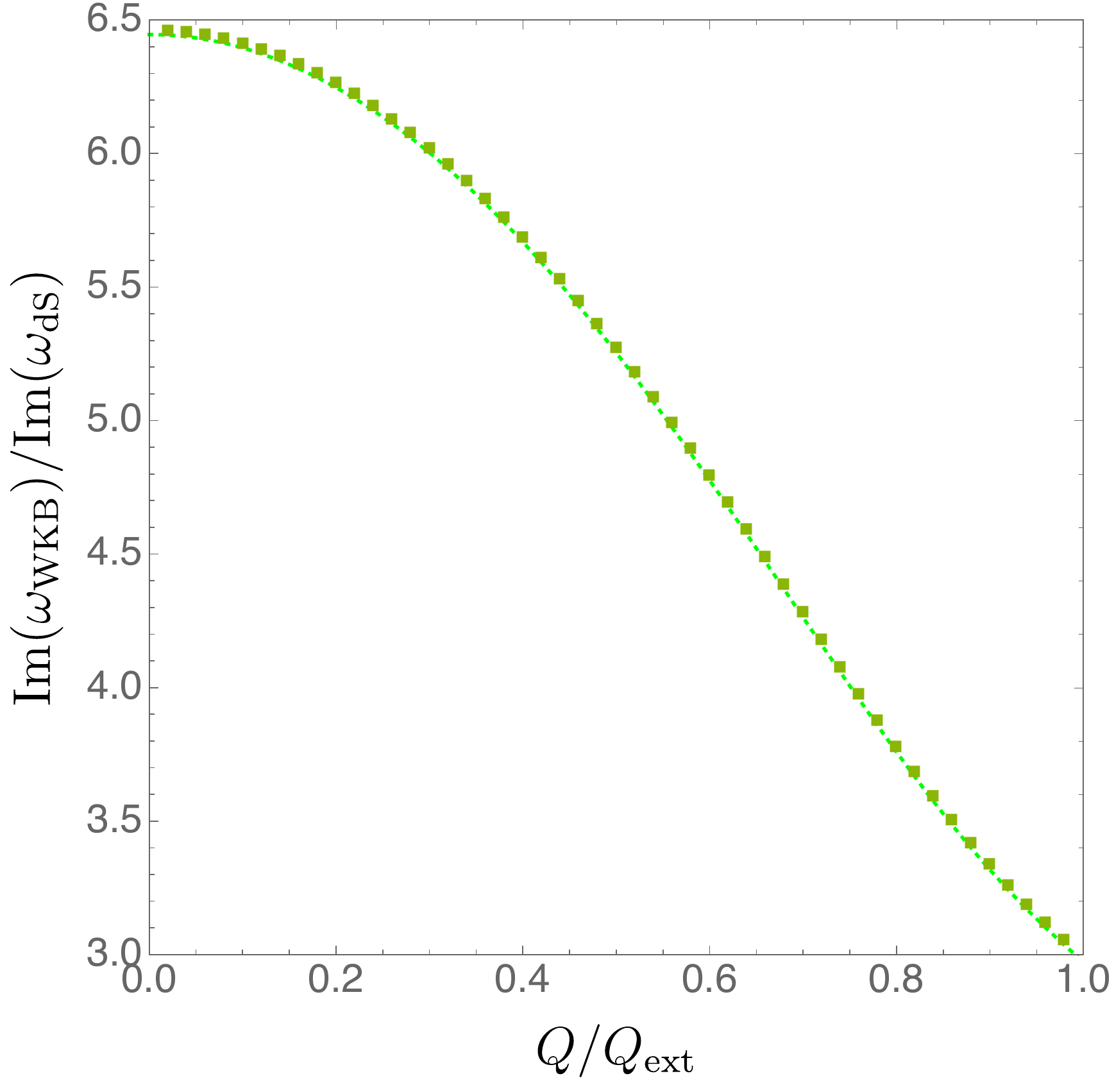}
	\caption{{\bf Left panel:} de Sitter  gravitoelectromagnetic mode $\Phi_-$ with $\ell=2$ and $n=0$: $-{\rm Im}(\omega)/\kappa_-$ as a function of $Q/Q_{\rm ext}$ for fixed $y_+=0.01$ (green squares), $y_+=0.05$ (brown diamonds) and $y_+=0.1$ (black disks). {\bf Right panel:} The ratio between the frequency ${\rm Im}(\omega_{\rm dS})$ of de Sitter mode of the left panel with $y_+=0.01$ and the imaginary part of the geometric  optics WKB frequency prediction  \eqref{wWKB} for the photon sphere modes of the same black holes. We see that, for a small black hole,  $-{\rm Im}(\omega_{\rm dS})$ is smaller than $-{\rm Im}(\omega_{\rm WKB})$ for the full range of $Q/Q_{\rm ext}.$}
	\label{fig:dS2}
\end{figure} 

So far we have restricted our attention to the dS limit ($M=0=Q$) of the RNdS solution. Naturally, RNdS has quasinormal modes $\Phi_\pm$ that in the dS limit reduce to \eqref{puredSw}. These are what we call the {\it dS family} of RNdS quasinormal modes. Numerically we find that these modes have purely imaginary frequencies and their wavefunctions are localized near the cosmological horizon.

Fig. \ref{fig:dS} shows some numerical results for the dS family of modes. For concreteness we do this illustration for  modes $\Phi_-$ with $\{\ell,n\}=\{2,0\}$. In the left panel we fix $Q/Q_{\rm ext}$ and we plot the imaginary part of the frequency ${\rm Im}(\omega\,r_c)$ as a function of the dimensionless ratio $y_+=r_+/r_c$. By definition, dS quasinormal frequencies must approach \eqref{puredSw} as $y_+\to 0$ and this is indeed the case (see red diamond). Note that the frequency changes substantially with $y_+$. However, if we instead fix $y_+$ and vary $Q$ then we find that the frequency does not change that much as $Q/Q_{\rm ext}$ increases from 0 up to 1. This is illustrated in the right panel of Fig.~\ref{fig:dS}.  
This is similar to what was found for massless scalar field quasinormal modes in Ref. \cite{Cardoso:2017soq}. In particular, note that the result \eqref{puredSw} works well for any small ($y_+ \ll 1$) black hole, independently of $Q$.

Ultimately we will be interested in the ratio $-{\rm Im}(\omega)/\kappa_-$. In the left panel of Fig.~\ref{fig:dS2}, we plot this quantity  for the modes displayed in the right panel of Fig.~\ref{fig:dS}. This plot illustrates that for the dS family,  $-{\rm Im}(\omega)/\kappa_-$ can attain large values well above $1/2$ or $1$. The reason we choose to display data with small $y_+$ is because this is the region where the slowest decaying quasinormal modes belong to the dS family (as will be clear later, in Fig.~\ref{fig:spgap}). 

For a small black hole, we can compare our analytical formula \eqref{puredSw} for the slowest decaying ($\ell=1$, $n=0$) de Sitter modes with our WKB prediction \eqref{wWKB} for the photon sphere modes. The latter is strictly valid for $\ell \gg 1$ but we found it worked well even for small $\ell$. We find that in this small black hole limit, the de Sitter modes always decay more slowly than the WKB prediction for the photon sphere modes. This is illustrated in the right panel of Fig. \ref{fig:dS2} for the black hole family with $y_+=0.01$ (the same green square solutions shown in the left panel of the same figure).  Thus for small black holes the $\ell=1$, $n=0$ de Sitter mode is the slowest decaying mode belonging to either the de Sitter or photon sphere families. 

\subsection{\label{sec:NHqnm}Near-extremal family of modes and its near-horizon limit}

The third family of quasinormal modes for RNdS black holes is called the near-extremal family since these modes are continuously connected to modes that can be identified in the near-horizon limit of the (near-)extremal RNdS  solution, {\it i.e.} as  $r_- \to r_+$. The analytical analysis of the near-extremal modes of this subsection (and the near-Nariai modes of the next one) is very much inspired by ideas from Appendix A of \cite{Teukolsky:1974yv} and \cite{Yang:2013uba,Zimmerman:2015trm}. This family of near-extremal modes is also present in the case of massless scalar field perturbations of a RNdS black hole \cite{Cardoso:2017soq}.

In this subsection we will first perform an approximate analytical calculation of the near-extremal quasinormal modes using the near-horizon limit. We will then compare this to numerical results for these modes. 

It is convenient to define the dimensionless quantities
\be\label{NHregime}
x=1-\frac{r}{r_+}\,,\qquad\text{and}\qquad \sigma\equiv1-\frac{r_-}{r_+}\,,
\ee
where $\sigma\geq0$ vanishes at extremality. The idea is to use the manifest SL$(2,\mathbb{R})$ symmetry of the $AdS_2 \times S^2$ near horizon geometry of an extremal RNdS black hole to simplify our calculation. The modes we seek, in the near extremal limit, are supported near the black hole horizon. So the limit we want to take has to accomplish two things: approach extremality, and zoom in near the black hole horizon. This can be achieved by sending $\sigma\to 0$ while keeping $z=x/\sigma$ fixed.  We can anticipate that $\omega$ will vanish linearly as $\sigma$, so we define $\omega\,r_c = \widetilde{\delta\omega}\,\sigma$ and solve for $\widetilde{\delta\omega}$ in what follows.

We set
\be
\Phi_{\pm}=\hat{f}_{\pm}(z)\,, \qquad z = \frac{x}{\sigma}\,,
\ee
and expand \eqref{vector:masterEOM} $-$ or \eqref{scalar:masterEOM} since the vector-type and scalar-type modes are isospectral~$-$ to leading order in $\sigma$. The resulting equation takes a simple form
\be
(1-z)z\frac{\mathrm{d}^2}{\mathrm{d}z^2}\hat{f}_{\pm}(z)+(1-2\,z)\frac{\mathrm{d}}{\mathrm{d}z}\hat{f}_{\pm}(z)+\left[\frac{\hat{\varphi}^2}{z(1-z)}+\hat{\eta}_{\pm}\right]\hat{f}_{\pm}(z)=0\,,
\label{eq:nearGHcs}
\ee
where we defined
\begin{align}
\label{eq:monster}
&\hat{\varphi}=y_+ \Xi \, \widetilde{\delta \omega}\,,\nonumber
\\
& \hat{\eta}_{\pm}=1+\Xi \ell(\ell+1)\pm\sqrt{\left[1+\Xi \ell(\ell+1)\right]^2-\Xi^2  (\ell+2)(\ell+1) \ell (\ell-1) } \\
&\Xi = \frac{1+2y_+ +3 y_+^2}{\left(1-y_+\right) \left(1+3 y_+\right)}. \nonumber 
\end{align}
Note that $\hat{\varphi}$ depends on $\widetilde{\delta \omega}$, but $\hat{\eta}_{\pm}$ does not. The expression for $\hat{\eta}_{\pm}$ is easily shown to be real, and it is then manifestly positive. This will play an important role in what follows.

Equation \eqref{eq:nearGHcs} can be readily solved in terms of Gaussian Hypergometric functions ${}_2F_1$ via the following combination
\begin{multline}
\hat{f}_{\pm}(z)=\hat{C}^{(1)}_{\pm}\,z^{-i\,\hat{\varphi}}(1-z)^{i\hat{\varphi}}{}_2F_1\left(a^{(1)}_{\pm},a^{(2)}_{\pm}\,;\,1-2\,i\,\hat{\varphi}\,;\,z\right)\\
+\hat{C}^{(2)}_{\pm}\,z^{i\,\hat{\varphi}}(1-z)^{i\hat{\varphi}}{}_2F_1\left(a^{(1)}_{\pm}+2\,i\,\hat{\varphi},a^{(2)}_{\pm}+2\,i\,\hat{\varphi}\,;\,1+2\,i\,\hat{\varphi}\,;\,z\right),
\end{multline}
where $\hat{C}^{(1)}_{\pm}$ and $\hat{C}^{(2)}_{\pm}$ are integration constants to be fixed via boundary conditions and
\begin{subequations}
\begin{align}
&a^{(1)}_{\pm}=\frac{1}{2}-\sqrt{\frac{1}{4}+\hat{\eta}_{\pm}}\,,
\\
&a^{(2)}_{\pm}=\frac{1}{2}+\sqrt{\frac{1}{4}+\hat{\eta}_{\pm}}\,.
\label{eq:a2}
\end{align}
\end{subequations}
We want to impose ingoing boundary conditions at the event horizon, \emph{i.e.} regularity in ingoing Eddington-Finkelstein coordinates. This is equivalent to setting $\hat{C}^{(2)}_{\pm}=0$.

Next we need to impose a boundary condition at large $-z$. In principle this should be done by matching to a solution that is outgoing at the cosmological horizon. But we will follow the simpler approach of simply demanding that the solution vanishes at large $-z$. This can be motivated by the observation that near-extremal modes are highly localized near the event horizon and are therefore very small at large $-z$. Ultimately the justification for this boundary condition is that it gives quasinormal frequencies that match very well our numerical results. 

At large negative values of $z$, we get
\begin{multline}
\hat{f}_{\pm}(z)\approx -\frac{e^{-\pi\hat{\varphi}}}{\sqrt{-z}}\hat{C}^{(1)}_{\pm}\;\Gamma(1-2i\hat{\varphi})
\\
\times\Bigg\{\frac{(-1)^{\sqrt{1+4\hat{\eta}_{\pm}}}\Gamma\left(\sqrt{1+4\hat{\eta}_{\pm}}\right)}{\Gamma\left(a^{(2)}_{\pm}\right)\Gamma\left(b^{(2)}_{\pm}\right)}(-z)^{\frac{1}{2}\sqrt{1+4\hat{\eta}_{\pm}}}\left[1-\frac{a^{(1)}_{\pm}}{2}\frac{1}{(-z)}+\mathcal{O}(z^{-2})\right]+
\\
\frac{(-1)^{-\sqrt{1+4\hat{\eta}_{\pm}}}\Gamma\left(-\sqrt{1+4\hat{\eta}_{\pm}}\right)}{\Gamma\left(a^{(1)}_{\pm}\right)\Gamma\left(b^{(1)}_{\pm}\right)}(-z)^{-\frac{1}{2}\sqrt{1+4\hat{\eta}_{\pm}}}\left[1-\frac{a^{(2)}_{\pm}}{2}\frac{1}{(-z)}+\mathcal{O}(z^{-2})\right]
\Bigg\}\,,
\label{eq:largeG}
\end{multline}
where
\begin{subequations}
\begin{align}
&b^{(1)}_{\pm}=a^{(1)}_{\pm}-2i\hat{\varphi}\,,
\\
&b^{(2)}_{\pm}=a^{(2)}_{\pm}-2i\hat{\varphi}\,.
\end{align}
\end{subequations}
The expansion (\ref{eq:largeG}) diverges at large positive values of $(-z)$ because of the term proportional to $(-z)^{\frac{1}{2}\sqrt{1+4\tilde{\eta}_{\pm}}}$. This can be avoided if we set of the Gamma functions in the denominator to have a pole, which occurs for $\Gamma(-n)$, with $n\in\mathbb{N}_0=\{0,1,2,\ldots\}$. In particular, we quantize the frequency by demanding
\be
b^{(2)}_{\pm}=-n\,,
\ee
with $n\in\mathbb{N}_0$. This equation can be readily solved for $\widetilde{\delta \omega}$ and hence for $\omega$:
\be  \label{NEspectrum0}
\omega\,r_c = -i\frac{(1-y_+)(1+3\,y_+)}{2\,y_+\,(1+2\,y_+^2+3\,y_+^2)}(a^{(2)}_{\pm}+n)\sigma\,,
\ee
which simplifies considerably when written in terms of $\kappa_-$:
\be
\label{NEspectrum}
\frac{\omega}{\kappa_-}  = -i \left( n+ \frac{1}{2}+\sqrt{\frac{1}{4}+\hat{\eta}_{\pm}} \right) \,,
\ee
where $\hat{\eta}_{\pm}$ is defined in \eqref{eq:monster}. Note that these quasinormal frequencies are purely imaginary and that they all have $-{\rm Im}(\omega)/\kappa_->1/2$. Which of these modes decays most slowly? The imaginary part of the frequency increases with overtone number $n$ so consider the fundamental ($n=0$) modes. For given $\ell$, we have $\hat{\eta}_-<\hat{\eta}_+$, so the $\Phi_-$ modes decay more slowly than the $\Phi_+$ modes. It can also be checked that, for any $y_+$, $\hat{\eta}_\pm$ is an increasing function of $\ell$. It follows that the slowest decaying modes covered by the above analysis are either the $\Phi_-$ modes with $\ell=2$ or the $\Phi_+$ modes with $\ell=1$ (as there are no $\Phi_-$ modes with $\ell=1$). It is easy to show from \eqref{NEspectrum} that it is always the $\Phi_-$ modes with $\ell=2$ which decay the most slowly.

\begin{figure}[th]
	\centering
	\includegraphics[width=0.45\textwidth]{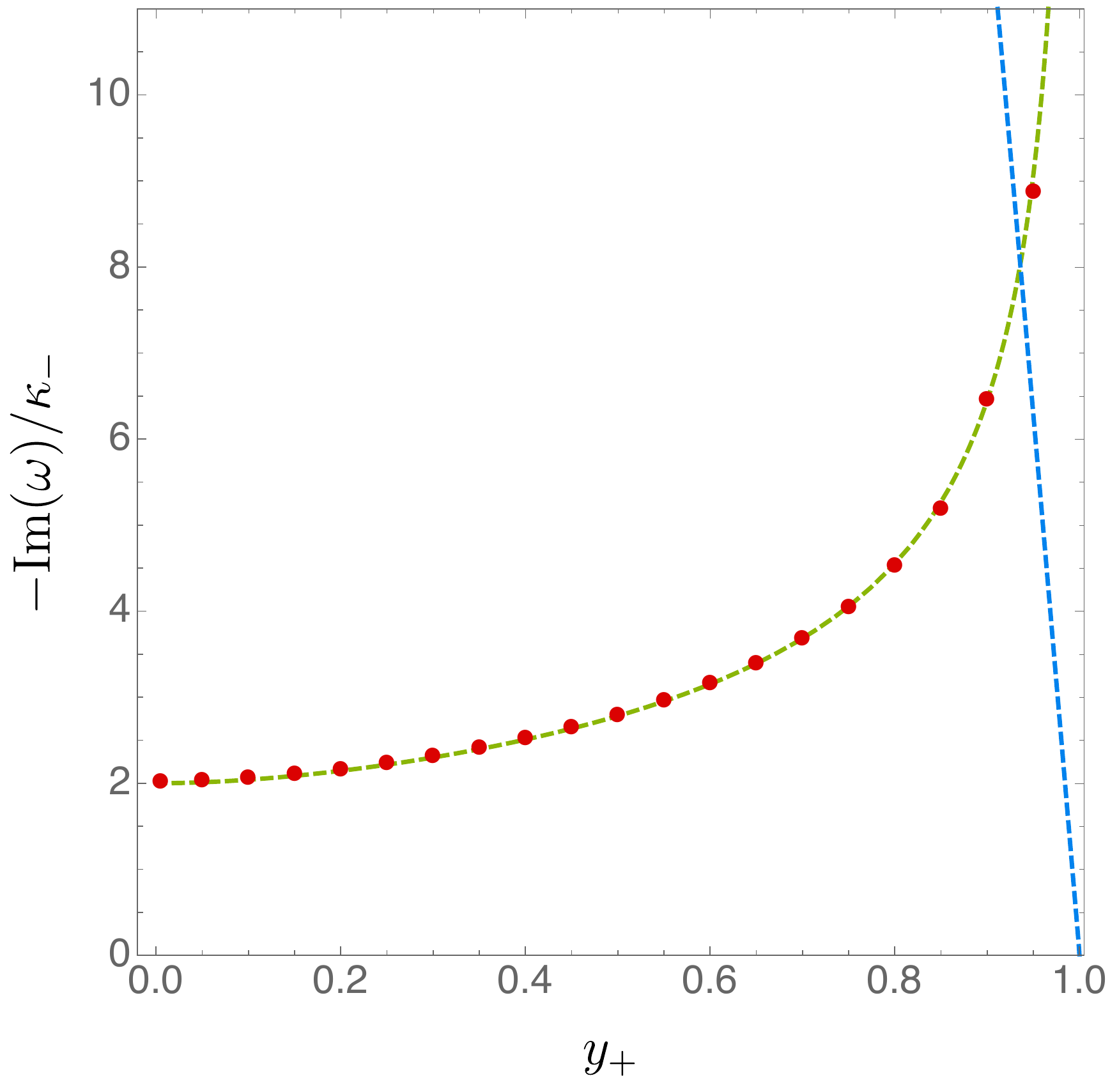}
	\hspace{0.5cm}
	\includegraphics[width=0.46\textwidth]{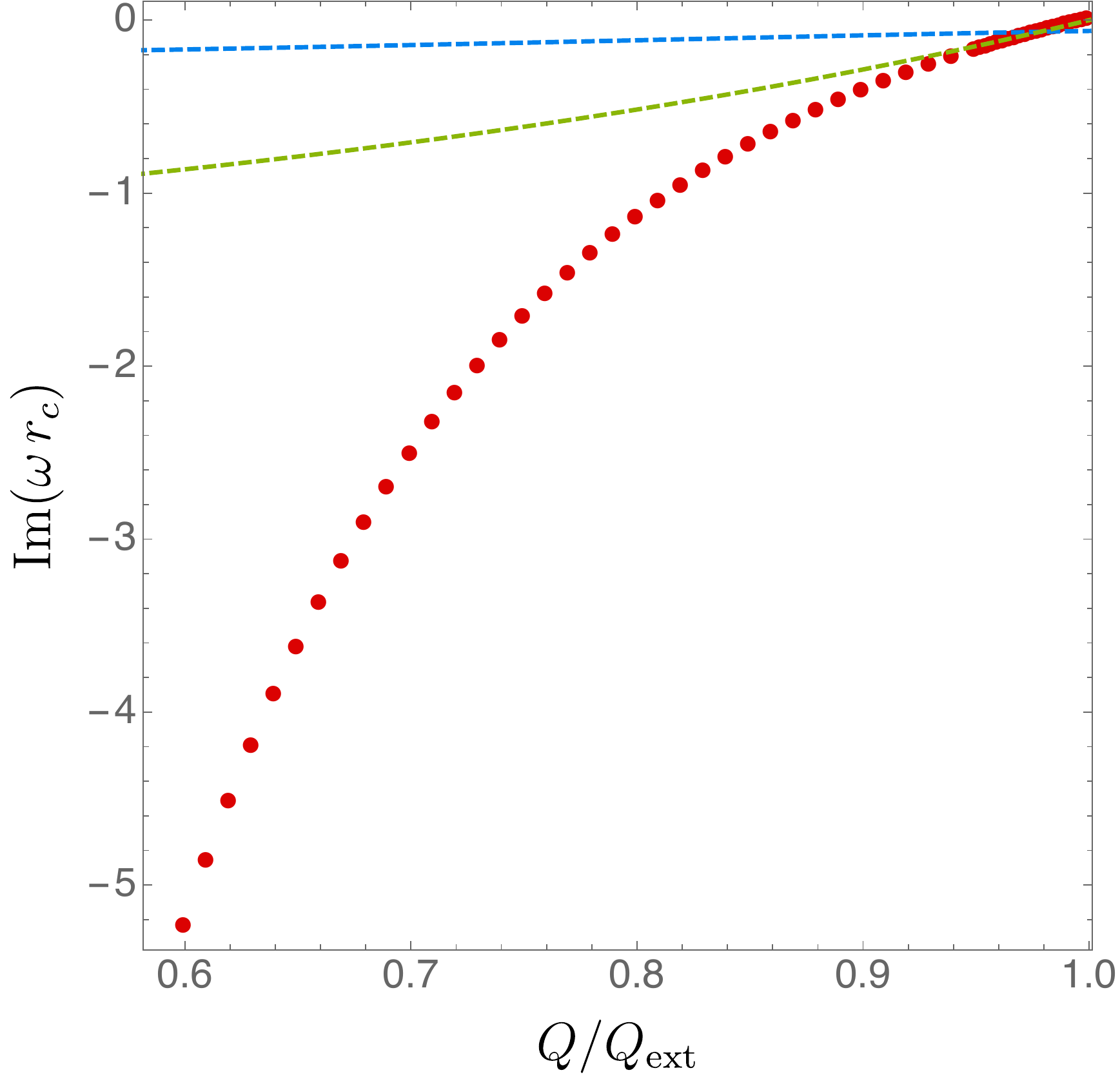}
	\caption{Near-extremal modes for the  gravitoelectromagnetic mode $\Phi_-$ with $\ell=2$ and $n=0$. In both plots, the dashed green line is the analytical prediction \eqref{NEspectrum}, or \eqref{NEspectrum0},  and the red dots are our numerical results. {\bf Left panel:} $-{\rm Im}(\omega)/\kappa_-$  as a function of $y_+$ for near-extremal modes at fixed $Q/Q_{\rm ext}=0.999$. The dashed blue curve is the WKB prediction \eqref{wWKB} for the photon sphere modes (also with $Q/Q_{\rm ext}=0.999$). This WKB blue curve continues to increase monotonically as $y_+$ decreases.  {\bf Right panel:}   ${\rm Im}(\omega\,r_c)$ as a function of $Q/Q_{\rm ext}$ for near-extremal modes at fixed $y_+=1/2$. The dashed blue curve is again the WKB prediction \eqref{wWKB} for the photon sphere modes. We see that for a wide range of charge $Q$ the photon sphere modes decay more slowly than the near-extremal modes but, above a critical charge ratio of $Q/Q_{\rm ext} \sim 0.98$, the opposite happens.}
	\label{fig:NE}
\end{figure}

The above calculation is, at best, valid only in the near-extremal limit, $\sigma \ll 1$, and for small frequencies, $|\omega\,r_c|\ll 1$. In the derivation of \eqref{NEspectrum} we have only used the properties of the RNdS near-horizon geometry but no use of the full geometry or its far region was made. So we might question the validity of this approximation. To address this question, in Fig.~\ref{fig:NE} we compare \eqref{NEspectrum} with the exact numerical data for the quasinormal mode family (with purely imaginary frequency) that we henceforth call the near-extremal modes. For illustrative purposes, we do this for the $\Phi_-$ mode with $\ell=2$ and radial overtone $n=0$. In the left panel of Fig.~\ref{fig:NE}, we fix $Q/Q_{\rm ext}=0.999$ and we plot $-{\rm Im}(\omega)/\kappa_-$ as a function of $y_+$. Since we are very close to extremality we expect that \eqref{NEspectrum} should be a good approximation. This is indeed what we find. The 
red dots representing the numerical data agree very well with the green curve corresponding to \eqref{NEspectrum}.  On the other hand, as expected, the analytical approximation \eqref{NEspectrum} becomes increasingly poor 
as we move away from extremality, {\it i.e.} as $Q/Q_{\rm ext}$ moves further away from unity.  This is illustrated in the right panel of Fig.~\ref{fig:NE}, where we fix $y_+=0.5$ and see that the prediction \eqref{NEspectrum0} (green dashed curve) is an excellent approximation when $Q \approx Q_{\rm ext}$ but quickly becomes a bad approximation as $Q$ decreases.  

The validity of the approximation that leads to \eqref{NEspectrum} was also tested in the following way. The fact that we just use the near-horizon geometry  to get \eqref{NEspectrum} suggests that these quasinormal modes have to be localized near the event horizon and very quickly decay away from it. Our numerical results confirm that this is the case: the numerical near-extremal mode wavefunctions are indeed localized near the event horizon, $r=r_+$, becoming more localized as extremality is approached.

In summary, we find that the analytical prediction \eqref{NEspectrum0} works very well for near-extremal modes of near-extremal black holes. It is interesting to compare this analytical prediction, for the dominant $\Phi_-$, $\ell=2$ modes, to the extremal limit of our WKB prediction \eqref{wWKB} for the photon sphere modes. This comparison is shown in the left panel of Fig. \ref{fig:NE} for $Q/Q_{\rm ext} = 0.999$. If we go even closer to extremality then the blue curve moves to the right, and $-{\rm Im}(\omega_{\rm WKB})/\kappa_-$ diverges in the extremal limit. Thus we see that, sufficiently close to extremality, the near-extremal modes always decay more slowly than the WKB prediction for the photon sphere modes. Thus, to the extent that the WKB prediction is valid at small $\ell$ (and, as we have seen, it seems to work well), our analytical results predict that, in a neighbourhood of extremality, the $\Phi_-$, $\ell=2$ near-extremal modes should be the slowest decaying modes belonging to either the near-extremal or photon sphere families. This is further illustrated in the right panel of Fig.  \ref{fig:NE} where we are at fixed $y_+=0.5$ and vary $Q/Q_{\rm ext}$: as we approach extremality, there is a critical value of the charge ratio above which the near-extremal modes indeed become more slowly decaying than the WKB photon sphere modes. 

We can also compare the near-extremal family of modes to the de Sitter family. For the slowest decaying de Sitter modes, we see from Fig. \ref{fig:dS} (right panel) that ${\rm Im}(\omega r_c)$ does not vary much as we approach extremality. It follows that $-{\rm Im}(\omega)/\kappa_-$ diverges for the de Sitter modes as we approach extremality. This ratio remains finite for the near-extremal modes, hence the near-extremal modes decay more slowly than the de Sitter modes in a neighbourhood of extremality.

In summary, a combination of analytical and numerical calculations indicates that, in a neighbourhood of extremality, the slowest decaying quasinormal mode across all families is the $\Phi_-$ near-extremal mode with $\ell=2$ and $n=0$. Furthermore, we have an analytical prediction from \eqref{NEspectrum0} for the frequency of this mode. Hence \eqref{NEspectrum0} gives us an analytical prediction for the behaviour of $\beta$ as we approach extremality. This is the green curve in the left panel of Fig. \ref{fig:NE}. We will discuss the implications of this below. 

\subsection{\label{sec:Nariaiqnm}Nariai modes}
RNdS black holes have three horizons, $r_-,r_+$ and $r_c$. In the previous subsection we considered the extremal limit where  $r_-\to r_+$. There is however another interesting limit $-$ the Nariai limit $-$ which occurs when $r_+\to r_c$. The surface gravity remains non-zero in this limit. It is natural to wonder wether there is a fourth family of RNdS quasinormal modes that reduce to Nariai quasinormal modes in this limit.  

For massless scalar field perturbations, the results of Ref. \cite{Cardoso:2017soq} suggest that these ``Nariai modes" are a subset of photon sphere modes, rather than constituting a distinct fourth family of modes. In the Appendix, we will show that this is indeed the case for gravitoelectromagnetic modes. Therefore we do not need to consider the Narai modes as a distinct family.

\section{\label{sec:GravResults}Results}

As explained above, for each type of perturbation ($\Phi_+$ or $\Phi_-$) we expect quasinormal modes to fall into three families (dS, photon sphere and near-extremal). Furthermore, from the discussion above, we expect that the slowest decaying quasinormal modes for each family and each type of perturbation to be given by the modes with the lowest allowed value of $\ell$ for that type of perturbation (this will be illustrated later in Table \ref{tab:table} for a particular black hole). Therefore our numerical calculations of quasinormal modes have focused on the two gravitoelectromagnetic sectors $\{\Phi_-,\ell=2\}$ and $\{\Phi_+,\ell=1\}$ since other sectors are expected to give more rapidly decaying modes. 

As an example of how we classify the quasinormal modes emerging from our numerical calculations, we will consider the family of ``lukewarm" RNdS black holes \cite{Mellor:1989gi,Romans:1991nq}. This is the 1-parameter subfamily of RNdS black holes that are in thermal equilibrium since the temperature of the event and cosmological horizons are the same {\it i.e.} $\kappa_+=\kappa_c$ \footnote{Lukewarm black hole are in thermal equilibrium but not in full thermodynamic equilibrium because the chemical potential of the two horizons is not the same.}. It turns out that this is equivalent to $M=|Q|$ \cite{Mellor:1989gi}. For a lukewarm hole
\be
\frac{Q}{Q_{\rm ext}}=\frac{1}{1+y_+}\sqrt{\frac{3 y_+^2+2 y_++1}{1+2 y_+}}
\ee
with $Q/Q_{\rm ext}=1/\sqrt{2}\sim 0.707$ for $y_+=1$ and $Q/Q_{\rm ext}=1$ for $y_+=0$.
We have discretized the lukewarm RNdS family with a numerical grid of 100 points for $0 \leq y_+ \leq 1$, and we searched for the full spectra of frequencies solving each one of the relevant two perturbation equations as a quadratic eigenvalue problem for $\omega^2$. To evaluate the numerical convergence of our results we then took the frequency spectrum of each lukewarm solution and inserted it as a seed in a Newton-Raphson code, and we progressively increased the number of grid points along the radial direction $0\leq y\leq 1$ $-$ see \eqref{def:y} $-$ until we got the desired accuracy for the quasinormal frequency. 

\begin{figure}[th]
	\centering
	\includegraphics[width=0.46\textwidth]{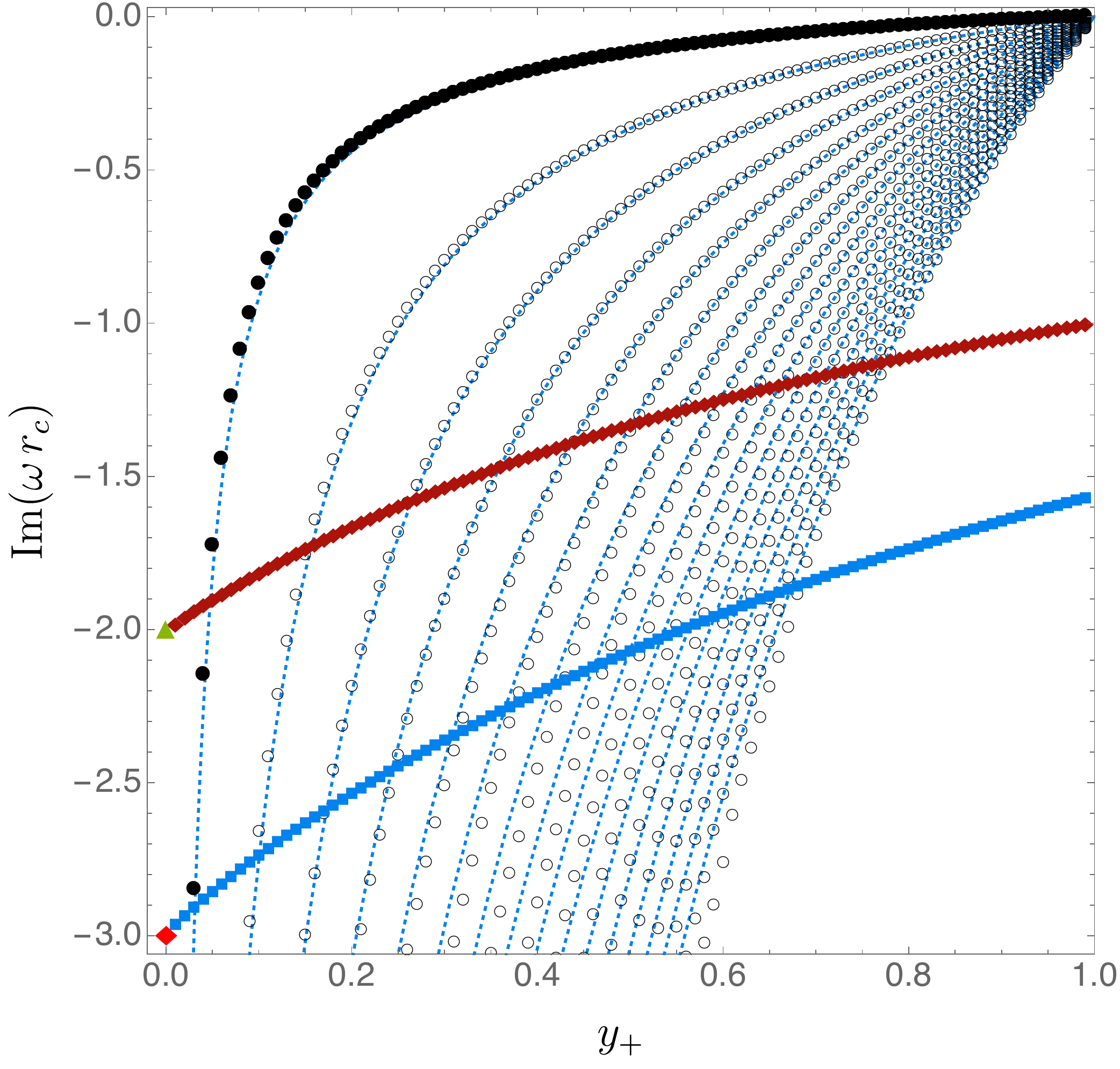}
	\hspace{0.5cm}
	\includegraphics[width=0.45\textwidth]{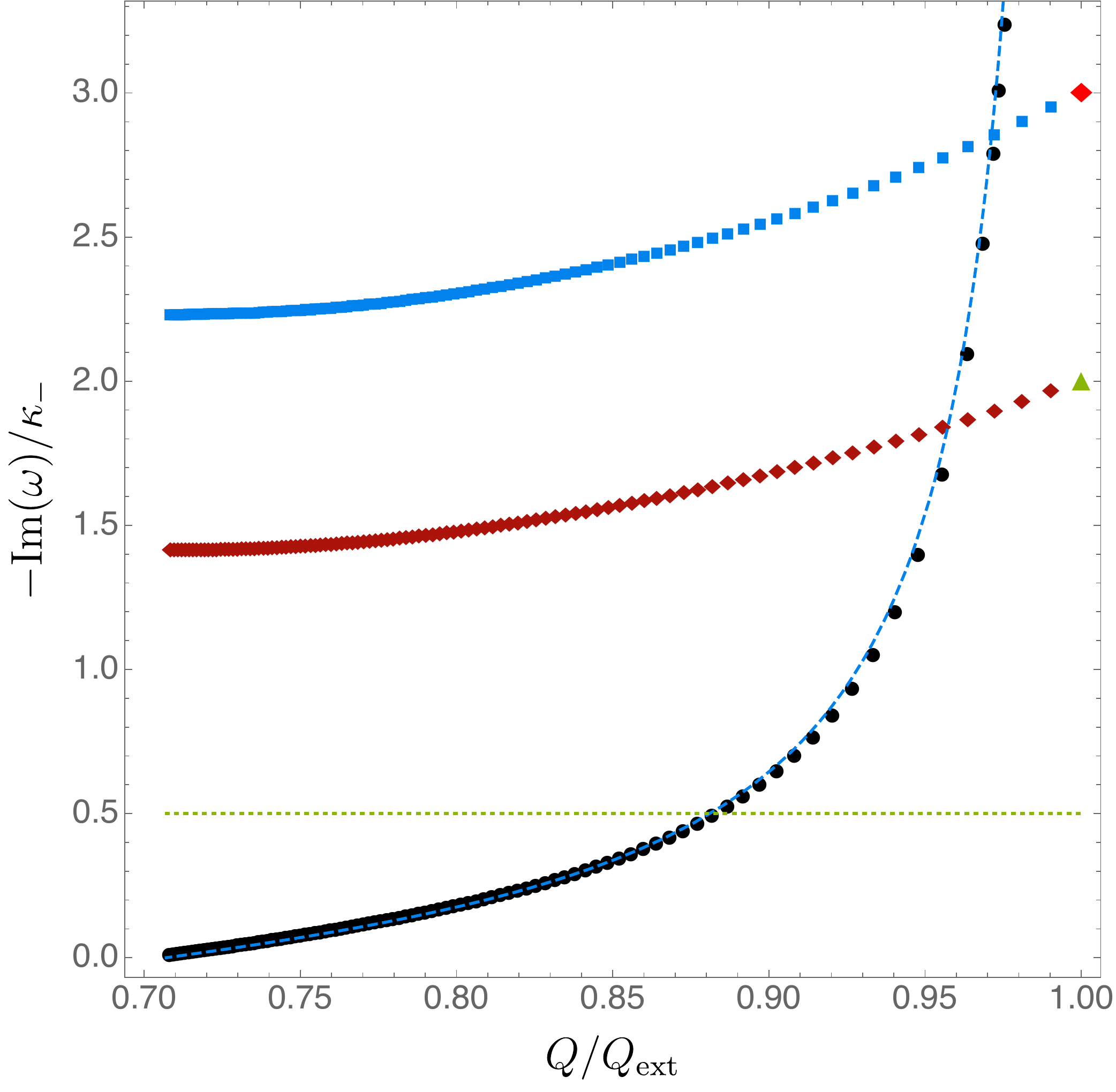}
\caption{Results for the $\Phi_-$ quasinormal modes with $\ell=2$ for lukewarm RNdS black holes.
{\bf Left panel:}  The filled marks identify the fundamental ($n=0$)  modes of the three families, namely  photon sphere (black disks), near-extremal (red diamonds), and de Sitter (blue squares). The black circles represent the next 15 photon sphere overtones ($n=1,\cdots, 15$) and the 16 blue dotted lines represent the  WKB approximation  ${\rm Im} \left(\omega_{\rm WKB}\right)(n)$, $n=0,\cdots, 15$, for the photon sphere modes (valid for $\ell\gg 1$).   The red diamond (in the de Sitter curve) represents the $n=0$ pure de Sitter frequency ${\rm Im}(\omega \,r_c)|_{\rm dS}=-3$.  The green triangle (in the near-extremal curve) represents the $n=0$ analytical approximation  ${\rm Im}(\omega \,r_c)|_{\rm NE}=-2$ in the limit where $Q=Q_{\rm ext}$, which for lukewarm RNdS occurs when $y_+\to 0$. {\bf Right panel:} The three families of fundamental ($n=0$) quasinormal modes. Here we plot $-{\rm Im}(\omega)/\kappa_-$ against $Q/Q_{\rm ext}$. The colour code is the same as for the left panel.}
\label{fig:lukewarmSpectra}
\end{figure} 

As an example, in the left panel of Fig.~\ref{fig:lukewarmSpectra} we give our results for the imaginary part of the frequency  for the  $\Phi_-$ modes with $\ell=2$. The black disks are the fundamental ($n=0$) photon sphere modes. This identification emerges from the fact that they match the geometric optics/WKB approximation \eqref{wWKB} for ${\rm Im}(\omega_{\mathrm{WKB}})$ (blue dotted line). These modes also have ${\rm Re}(\omega\,r_c)\neq 0$ which distinguishes them from the purely imaginary dS and near-extremal modes. In the same figure, below this $n=0$ photon sphere curve, we identify a total of 15 more curves with black circles. From the left/top to the right/bottom these are the photon sphere overtones $n=1,2,\cdots,15$. This identification follows from: 1) the fact that they match the geometric optics/WKB approximation \eqref{wWKB} (see the associated 15 blue dashed curves\footnote{Note that, as expected, the WKB approximation becomes less accurate for higher overtones. It is however remarkable that the $\ell\gg 1$ approximation \eqref{wWKB} is so accurate for $\ell=2$.}), and 2) the number of radial zeros in the real and imaginary parts of the associated eigenvectors increases by one unit as $n$ increases by one unit. For clarity of our presentation we decided not to plot the photon sphere modes with $n\geq 16$. From  the figure the reader can however understand that these accumulate on the right side of the plot.\footnote{Without much effort, {\it i.e.} without increasing the resolution beyond the value required to have the accuracy desired for the leading overtones, we were able to capture the first $\sim$40 photon sphere overtones.} 

Also on the left panel of Fig.~\ref{fig:lukewarmSpectra} we also see a line of red diamonds. This is the fundamental ($n=0$) near-extremal mode of the lukewarm RNdS family.\footnote{The higher, $n\geq 1$, near-extremal overtones have lower ${\rm Im}(\rm \omega\,r_c)$ and are not shown.} This identification emerges from the fact that: 1) these frequencies are purely imaginary, 2) they converge to ${\rm Im}(\omega\,r_c) |_{\mathrm{NE}}=-2$ in the lukewarm extremal limit $y_+\to 0$ (see the green triangle), as dictated by the analytical analysis \eqref{NEspectrum}, and 3) the eigenvectors of these modes (real functions) are very localized near the event horizon.   

Also on the  left panel of Fig.~\ref{fig:lukewarmSpectra} there is a curve of blue squares. This is the $n=0$ de Sitter family of modes because: 1) these modes are purely imaginary, 2) they  converge to ${\rm Im}(\omega\,r_c)|_{\mathrm{dS}}=-3$ as $y_+\to 0$ (see the red diamond), in agreement with the analytical analysis \eqref{puredSw}, and 3) the eigenvectors of these modes (real functions) are very much localized near the cosmological horizon.\footnote{The higher, $n\geq 1$, de Sitter overtones have lower ${\rm Im}(\omega\,r_c)$ and are not shown.} 

To conclude our analysis of the left panel of Fig.~\ref{fig:lukewarmSpectra}, the numerical solution of the quadratic eigenvalue problem gives the full spectrum of eigenfrequencies and associated eigenvectors. We have identified each family of modes that appears in the spectrum using the information discussed in section \ref{sec:analytics}. All the numerical data fits in one of the three classes of modes (de Sitter, photon sphere or near-extremal). Still in the lukewarm family of RNdS, we did a similar analysis for the other relevant  sector of perturbations, $\{ \Phi_+,\ell=1\}$, with similar results. 

Recall, that we are studying the quasinormal spectra of RNdS to find the spectral gap $\alpha$ in order to calculate $\beta$ defined by \eqref{betadef}. To calculate $\alpha$ we need to determine the slowest decaying quasinormal mode across the two types of perturbation ({\it i.e.} $\Phi_+$ and $\Phi_-$) in all three families of quasinormal modes. We can illustrate this with the lukewarm family of RNdS black holes. Focus first on the  sector of perturbations $\{ \Phi_-,\ell=2 \}$ already studied in the left panel of Fig.~\ref{fig:lukewarmSpectra}. Clearly, for our purposes, it is enough to compare the leading ($n=0$) overtone $- {\rm Im}(\omega)/\kappa_-$ for the three families of modes. This is done in the right panel of   Fig.~\ref{fig:lukewarmSpectra}. We see that for lukewarm black holes and in the  $\{ \Phi_-,\ell=2 \}$ sector, photon sphere modes have the lowest $- {\rm Im}(\omega)/\kappa_-$ for $Q/Q_{\rm ext} \lesssim 0.955$. However, for $0.955 \lesssim Q/Q_{\rm ext} \leq 1$ the slowest decaying modes are the near-extremal ones. The de Sitter modes are irrelevant for the spectral gap discussion of lukewarm black holes. This analysis still does not identify $\beta$ for the lukewarm family. For that, we have to repeat the analogue of the right panel of  Fig.~\ref{fig:lukewarmSpectra} for the  other sector $\{ \Phi_+,\ell=1\}$ of perturbations and $\beta$ is then the minimum of $- {\rm Im}(\omega)/\kappa_-$ over the two sectors of quasinormal modes. 

Moving away from the lukewarm family, we will now describe our results for the full moduli space of RNdS black holes. We have spanned the full parameter space $0 \le y_+\le 1$ and $0 \le Q/Q_{\rm ext}\le 1$ using a numerical grid with 100 points along $y_+$ and another 100 points along $Q/Q_{\rm ext}$. That is to say, we have computed the $\{ \Phi_+,\ell=1\}$ and $\{ \Phi_-,\ell=2\}$ quasinormal modes for a total of $10^4$ RNdS black holes. Where necessary we further zoomed in a particular region of parameter space, {\it e.g.} near $Q/Q_{\rm ext}\sim 1$ and/or $y_+ \sim 0$ or $y_+ \sim 1$. Again, all the numerical modes were identified as belonging to one the three families of modes (de Sitter, photon sphere or near-extremal). It is in this sense that we are confident that, for each of the $10^4$ RNdS black holes that we studied, the frequency spectra of quasinormal modes belongs to one of the three families discussed in section \ref{sec:analytics} and no fourth family exists.

\begin{figure}[th]
	\centering
	\includegraphics[width=0.48\textwidth]{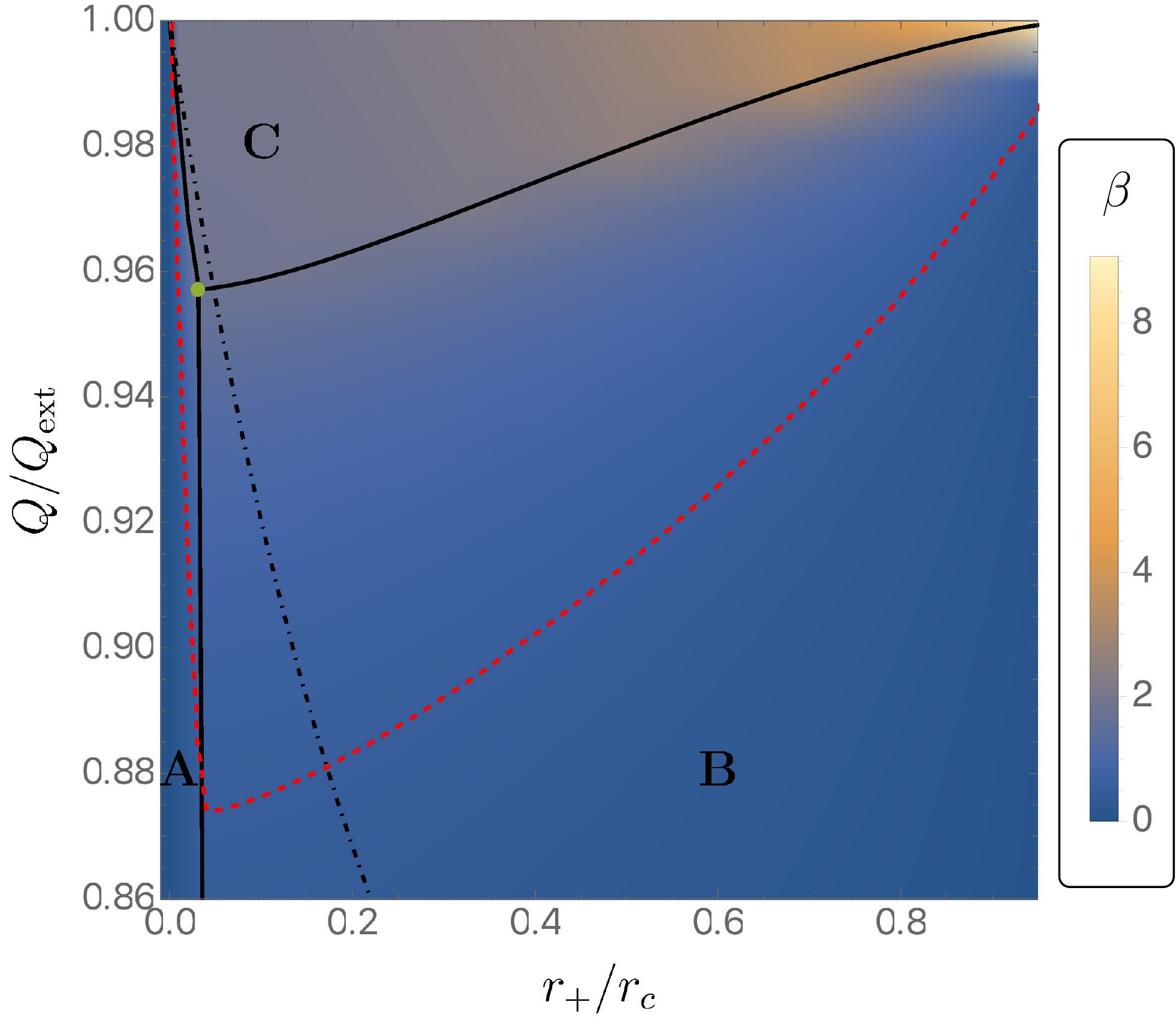}
	\hspace{0.2cm}
	\includegraphics[width=0.49\textwidth]{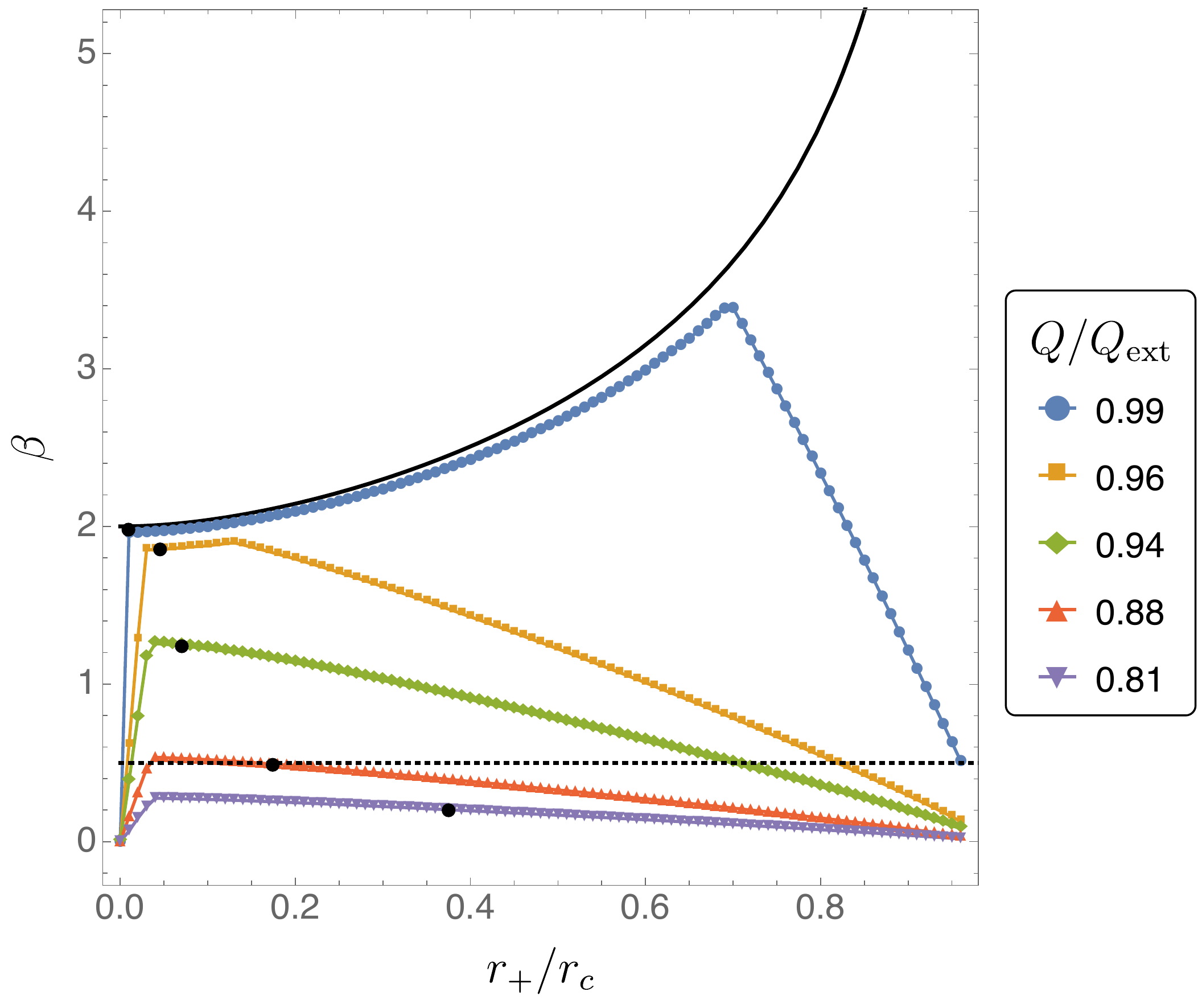}
	\caption{Value of $\beta$ for RNdS black holes. {\bf Left panel.} In region $A$ the dS family dominates (\emph{i.e.} the slowest decaying quasinormal mode is a dS mode), in region $B$ the photon sphere family dominates, and in region $C$ the near-extremal family dominates. The red dashed curve corresponds to the critical value of $\beta=1/2$. Above this line one has $\beta>1/2$ so the Christodoulou formulation of strong cosmic censorship is violated for smooth initial data. The black dashed dotted line corresponds to lukewarm black holes. {\bf Right panel.} $\beta$ against $r_+/r_c$ for different values of $Q/Q_{\rm ext}$. The discontinuities in the derivatives of these curves occur across the boundaries of the different regions $A,B,C$. Note that $\beta>2$ sufficiently close to extremality, and large near-extremal black holes can have arbitrarily large $\beta$. The black curve is the analytical prediction for near-extremal modes and the black disks  correspond to lukewarm black holes.} 
	\label{fig:spgap}
\end{figure} 

Our main results for  the spectral gap are presented in Fig.~\ref{fig:spgap}. In the left panel we show a density plot where we plot $\beta=\alpha/\kappa_-$ as a function of the horizon ratio $y_+=r_+/r_c$ and charge ratio $Q/Q_{\rm ext}$. We identify three regions $A,B,C$ separated by three black curves. In region $A$ the spectral gap is dominated by the de Sitter modes. That is, in this region, the slowest decaying quasinormal mode is a de Sitter mode. This region $A$ extends all the way down to $Q\to 0$, {\it i.e.} de Sitter modes dominate the region of parameter space described by very small values of $y_+$. On the other hand, in region $B$ it is the photon sphere modes that dominate. Finally, in region $C$, {\it i.e.} in a band of parameter space around extremality $Q/Q_{\rm ext}\sim 1$, it is the near-extremal modes that dominate. 

The left panel of Fig.~\ref{fig:spgap}, also shows a red dashed curve. This curve identifies solutions with $\beta=1/2$ and, above it, we have a region of parameter space near extremality where the solutions have $\beta>1/2$ (see also the density plot legend).  It follows from the discussion of section \ref{sec:weak_conclude} that, in this region, the Christodoulou version of strong cosmic censorship is violated (for smooth initial data) by gravitoelectromagnetic perturbations.

These results are similar to the results for massless scalar field perturbations presented in Ref. \cite{Cardoso:2017soq}. However, there is an important qualitative difference between our results and the results for massless scalar field perturbations. In the massless scalar case one always has $\beta<1$  \cite{Cardoso:2017soq}. But in our case we can have $\beta>1$. This is apparent in the right panel of Fig.~\ref{fig:spgap}, which plots $\beta$ against $y_+$ for different values of $Q/Q_{\rm ext}$. The black curve corresponds to the analytical prediction \eqref{NEspectrum} for near-extremal modes  $\Phi_-$ with $n=0$ and $\ell=2$. From the discussion at the end of section  \ref{sec:NHqnm} we expect this analytical prediction to be reliable as we approach extremality. Our plot shows that this analytical result does indeed give an accurate prediction for the value of $\beta$ close to extremality. From the plot we see that, not only that do near-extremal black holes have $\beta>1/2$, but in fact they have $\beta>2$, which implies (section \ref{sec:weak_conclude}) that the {\it $C^2$ version of strong cosmic censorship is violated} for smooth initial data. In fact, for any $r$, by taking $y_+$ large enough we can find a near-extremal black hole for which $\beta>r$ (the appropriate value of $y_+$ can be determined from \eqref{NEspectrum}).  Hence, {\it for any $r$, the $C^r$ version of strong cosmic censorship is violated} for smooth initial data.

\begin{table}[htp]
{\small
\begin{center}
\begin{tabular}{|m{1cm} |m{3.0cm} | m{3.0cm} | m{3.0cm} |}
\hline
\multicolumn{1}{|c|}{$\ell$}  & \multicolumn{1}{c|}{$1$}           & \multicolumn{2}{c|}{$2$}   \\ \hline\hline
\multicolumn{1}{|c|}{sector} & \multicolumn{1}{c|}{$\Phi_+$}      & \multicolumn{1}{c|}{ $\Phi_+$  } & \multicolumn{1}{c|}{$\Phi_-$}  \\ \hline\hline
\multicolumn{1}{|c|}{NE}      &\multicolumn{1}{c|}{$-0.294455\,i$ }&\multicolumn{1}{c|}{ $-0.397250\,i $}& \multicolumn{1}{c|}{$-0.200374\,i       $   }   \\ \hline
\multicolumn{1}{|c|}{PS}      &\multicolumn{1}{c|}{$ 1.90727-0.34830\,i$} &\multicolumn{1}{c|}{$ 3.10274-0.35888\,i$ }&\multicolumn{1}{c|}{ $1.85991-0.36246\,i  $    }   \\ \hline
\multicolumn{1}{|c|}{dS}      &\multicolumn{1}{c|}{ $-1.62849\,i $}&\multicolumn{1}{c|}{$ -2.44384\,i $}&\multicolumn{1}{c|}{$ -2.49383\,i $  }  \\ \hline
\end{tabular}
\end{center}
\begin{center}
\begin{tabular}{|m{1cm} |m{3.0cm}  | m{3.0cm} | m{3.0cm} | m{3.0cm} |}
\hline
\multicolumn{1}{|c|}{$\ell$}  & \multicolumn{2}{c|}{$3$}           & \multicolumn{2}{c|}{$4$}   \\ \hline\hline
\multicolumn{1}{|c|}{sector} & \multicolumn{1}{c|}{$\Phi_+$}  & \multicolumn{1}{c|}{$\Phi_-$}      & \multicolumn{1}{c|}{ $\Phi_+$  } & \multicolumn{1}{c|}{$\Phi_-$}  \\ \hline\hline
\multicolumn{1}{|c|}{NE}     
 &\multicolumn{1}{c|}{$-0.499491\,i$ } &\multicolumn{1}{c|}{$-0.303216\,i$ }  &\multicolumn{1}{c|}{ $-0.601530\,i $}& \multicolumn{1}{c|}{$-0.405476\,i       $   }   \\ \hline
\multicolumn{1}{|c|}{PS}     
 &\multicolumn{1}{c|}{$ 4.24555-0.36316\,i$} &\multicolumn{1}{c|}{$ 3.05802-0.36729\,i$}  &\multicolumn{1}{c|}{$ 5.36805-0.36530\,i$ }&\multicolumn{1}{c|}{ $4.20146-0.36923\,i  $    }   \\ \hline
\multicolumn{1}{|c|}{dS}  
   &\multicolumn{1}{c|}{ $-3.25871\,i $}   &\multicolumn{1}{c|}{ $-3.28702\,i $}&\multicolumn{1}{c|}{$ -4.07350\,i $}&\multicolumn{1}{c|}{$ -4.09286\,i $  }  \\ \hline
\end{tabular}
\end{center}
}
\caption{Gravitoelectromagnetic quasinormal mode frequencies $\omega\,r_c$ of a RNdS black hole with $y_+ = 0.2$ and $Q/Q_{\rm ext}=0.98$. The rows of the table(s) refer to the near-extremal (NE), photon sphere (PS) and de Sitter (dS) families of modes. Only the fundamental ($n=0$) mode is shown in each case. Our analytical near-horizon calculation \eqref{NEspectrum0} gives $\omega\, r_c |_{\rm NE} = -0.280281\, i$ ($\ell=1$, $\Phi_+$),  $\omega\, r_c  |_{\rm NE} = -0.378483\, i$ ($\ell=2$, $\Phi_+$) and $\omega\, r_c  |_{\rm NE} = -0.191033\, i$ ($\ell=2$, $\Phi_-$), $\omega\, r_c  |_{\rm NE} = -0.476091\, i$ ($\ell=3$, $\Phi_+$), $\omega\, r_c  |_{\rm NE} = -0.289223\, i$ ($\ell=3$, $\Phi_-$), $\omega\, r_c  |_{\rm NE} = -0.573483\, i$ ($\ell=4$, $\Phi_+$) and $\omega\, r_c  |_{\rm NE} = -0.386826\, i$ ($\ell=4$, $\Phi_-$) for the NE modes. Our WKB calculation \eqref{wWKB} (valid for large $\ell$) yields ${\rm Im}(\omega\, r_c) |_{\rm WKB}= -0.370369$ for the PS modes. For reference, the de Sitter frequency \eqref{puredSw} $-$ valid strictly for $y_+=0$ and $Q=0$ $-$ yields $\omega\, r_c |_{\rm dS} =-2\, i$ ($\ell=1$), $\omega\, r_c |_{\rm dS}=-3\, i$ ($\ell=2$), $\omega\, r_c |_{\rm dS}=-4\, i$ ($\ell=3$) and $\omega\, r_c |_{\rm dS}=-5\, i$ ($\ell=4$).} 
\label{tab:table}
\end{table}%

Finally, in Table \ref{tab:table} we present detailed numerical results for a particular near-extremal black hole which violates the $C^2$ version of strong cosmic censorship. For this particular example we have computed not just the $\{ \Phi_+,\ell=1\}$ and the $\{ \Phi_-,\ell=2\}$ modes but also the $\{ \Phi_+,\ell=2\}$ modes and both types of mode with $\ell=3,4$. 
From the table we see that the slowest decaying mode for this particular black hole is the $\{ \Phi_-,\ell=2\}$ mode, in agreement with the discussion at the end of section \ref{sec:NHqnm}. This black hole has $\kappa_- r_c = 0.098005$ and so $\beta = 0.200374/0.098005 = 2.04$.

\section{Discussion} 
\label{sec:discussion} 


\subsection{Taking the rough with the smooth}

We have reviewed the reason why quasinormal modes determine the behaviour at the Cauchy horizon of linear perturbations arising from smooth initial data. By calculating the gravitoelectromagnetic quasinormal modes of RNdS black holes we have shown that, the Christodoulou and $C^2$ formulations of strong cosmic censorship are always violated close to extremality, and, for any $r$, the $C^r$ formulation is violated close to extremality for a sufficiently large black hole. Thus gravitoelectromagnetic perturbations exhibit a much worse violation of strong cosmic censorship than the massless scalar field perturbations considered in Ref. \cite{Cardoso:2017soq}.

We emphasize that this violation of strong cosmic censorship in Einstein-Maxwell theory {\it does not} occur in pure Einstein gravity. Ref. \cite{Dias:2018ynt} showed that any non-extremal Kerr-dS black hole has slowly decaying photon sphere gravitational quasinormal modes which ensure that the Christodoulou version of strong cosmic censorship is respected for smooth initial data. 

As we have discussed above, Dafermos and Shlapentokh-Rothman (DSR) have shown that one can rescue strong cosmic censorship for RNdS black holes at the expense of considering rough initial data \cite{Dafermos:2018tha}. We have explained how a lack of smoothness of the initial data is also required to make sense of the older argument of Ref. \cite{Brady:1998au} in favour of strong cosmic censorship.

What are we to make of this? Should we allow rough initial data? In physics we often assume that it is sufficient to work with smooth initial data. However, in some theories, smooth initial data can lead to a rough solution. For example, a shock can form in a compressible perfect fluid. Once we accept the existence of shocks, it is natural to weaken the regularity of our initial data to allow for shocks present initially. So for a fluid it is natural to allow rough initial data. However, in Einstein-Maxwell (-scalar field) theory, if we start with smooth initial data then the solution will remain smooth throughout the domain of dependence of this data. Shocks do not form dynamically. So we are not forced to consider rough initial data. 

On the other hand, rough initial data can be approximated by a sequence of smooth initial data labelled by an integer $n$, and all with the same energy as the rough data. The sequence of smooth solutions arising from such data will be close to the rough solution in a region of spacetime that becomes larger as $n \rightarrow \infty$, and approaches the Cauchy horizon in this limit (this follows from Cauchy stability of the equations of motion). DSR's rough version of strong cosmic censorship indicates that one can find a sequence such that the energy at the Cauchy horizon diverges as $n \rightarrow \infty$. Hence, even for smooth perturbations, the energy at the Cauchy horizon is not bounded by the initial energy. Even if the energy of a smooth perturbation does not diverge at the Cauchy horizon, it can still become arbitrarily large there. 

Maybe for some reason one would want the initial data not just to have finite energy but also that the first $k$ derivatives are square integrable, {\it i.e.} the initial data has finite $H^k$ norm. For example, such a condition might arise from the requirement that the leading higher derivative corrections to the equations of motion are negligible initially. DSR's rough version of strong cosmic censorship implies that there exist smooth initial data whose $H^k$ norm on a spacelike surface intersecting the Cauchy horizon is not bounded by the $H^k$ norm of the initial data. This suggests that generic smooth initial data for which the leading higher derivative corrections are negligible will give a solution for which the leading higher derivative corrections become large near the Cauchy horizon. This does seem to capture the physics of the strong cosmic censorship hypothesis, namely that there is always a breakdown of effective field theory at a Cauchy horizon. 

\subsection{Comments on quantum effects}

The analysis of this paper has been entirely classical. In this section we will discuss the role of Hawking radiation \cite{Hawking:1974sw} in enforcing strong cosmic censorship. Recall that the behaviour at the Cauchy horizon is determined by the late-time behaviour of the black hole solution. So we need to discuss the effects of Hawking radiation on this late time behaviour. In de Sitter spacetime, we have to account for Hawking radiation both from the black hole horizon and from the cosmological horizon \cite{Gibbons:1977mu}. 

Consider first pure Einstein-Maxwell theory. In this case there are no charged particles and so Hawking radiation cannot change the charge of the black hole. If the black hole has a higher temperature than the cosmological horizon then it will radiate photons and gravitons and its temperature will decrease. If it has a lower temperature than the cosmological horizon then it will absorb photons and gravitons emitted by the cosmological horizon and the black hole temperature will increase. Thus Hawking radiation will drive the black hole towards a lukewarm solution for which the black hole and the cosmological horizon have equal temperatures, {\it i.e.} $\kappa_+ = \kappa_c$ \cite{Mellor:1989gi}. 

We can approximate the late time solution as a (slightly perturbed) lukewarm solution and the behaviour near the Cauchy horizon will be determined by the behaviour near the Cauchy horizon of a lukewarm black hole. Fig.~\ref{fig:spgap} (right panel) shows that small lukewarm black holes have $1/2 < \beta < 2$ and so (in pure Einstein-Maxwell theory) they violate the Christodoulou formulation of strong cosmic censorship (for smooth initial data) but not the $C^2$ formulation. Thus it appears that Hawking radiation does not rescue the Christodoulou version of strong cosmic censorship in pure Einstein-Maxwell theory.

However, there is another way in which quantum effects can influence the geometry, namely via vacuum polarization. At late time, one would expect the quantum state of fields outside the black hole to approach the Hartle-Hawking state in the lukewarm black holes background. In this state, the results of calculations in a 2d toy model \cite{Birrell:1978th} (with conformally coupled quantum fields) indicate that $\langle T_{\mu\nu} \rangle$ diverges at the Cauchy horizon. This divergence is proportional to $(-V_-)^{-2}$, which is not locally integrable at the Cauchy horizon hence one cannot make sense of the semi-classical Einstein equation $G_{\mu\nu} = 8 \pi \langle T_{\mu\nu} \rangle$ there, even in the sense of weak solutions. This suggests that quantum effects may rescue strong cosmic censorship. It would be interesting to confirm this with a calculation of $\langle T_{\mu\nu} \rangle$ in the Hartle-Hawking state near the Cauchy horizon of a lukewarm black hole. 

Of course, in the real world there exist charged particles {\it e.g.} electrons, that an electrically charged RNdS black hole can emit as Hawking radiation, and thereby decrease its charge. If the radiation of charged particles is rapid compared to the radiation of uncharged particles then the black hole will first lose most of its charge, and then evaporate away completely. If the radiation of charged particles is slow compared to the radiation of uncharged particles then the latter would tend to push the black hole onto the lukewarm family of solutions as above. The emission of charged particles would then cause the charge gradually to decrease whilst remaining within the lukewarm family. But ultimately the black hole would evaporate away completely. Note that this conclusion does not depend on the mass of the charged particles. This is because, unlike in flat spacetime, particles of any mass can escape the black hole by tunnelling through the potential barrier separating the event horizon from the cosmological horizon. In other words, the mass of the particle is redshifted away at the cosmological horizon. 

It seems that Hawking radiation of charged particles will ensure that strong cosmic censorship is respected. However, one could also imagine a magnetically charged RNdS hole, perhaps formed by pair creation in de Sitter spacetime \cite{Mellor:1989gi}. By performing an electromagnetic duality rotation, our results on gravitoelectromagnetic perturbations of electrically charged RNdS holes map to identical results for magnetically charged holes. If there are no magnetically charged particles then such black holes will evolve via Hawking radiation to lukewarm holes, which will behave as discussed above. 

\subsection*{Acknowledgments}

We thank M. Dafermos, C. Kehle and C. Warnick for helpful conversations. We are grateful to F. Eperon for producing Fig. \ref{fig:penrose}. OJCD is supported by the STFC Ernest Rutherford Grants No. ST/K005391/1 and No. ST/M004147/1, and by the STFC ``Particle Physics Grants Panel (PPGP) 2016" Grant No. ST/P000711/1. HSR and JES were supported in part by STFC Grants No. PHY-1504541 and ST/P000681/1.


\appendix

\section{Nariai modes}

In this appendix we will consider quasinormal modes which are continuously connected to quasinormal modes of the Nariai solution, {\it i.e.} the  $r_+ \rightarrow r_c$ limit of RNdS. To explore the Nariai limit, we introduce the dimensionless quantities 
\be\label{Nariairegime}
X=\frac{r_c-r}{r_c}\,,\qquad \delta=\frac{r_c-r_+}{r_c}\,,\qquad \mu=\frac{Q}{r_+}\,,\qquad \widetilde{\omega}=\omega\,r_c\,.
\ee
We are interested in low frequency perturbations $ \widetilde{\omega}\to 0$ in the near-horizon limit about the cosmological horizon, $X\to 0$, of near-Nariai solutions, $\delta\to 0$. The second relation in \eqref{Nariairegime} can be used to express $y_+$ as a function of $\delta$, $y_+=1-\delta$.

The procedure described in section \ref{sec:NHqnm} also applies to the current Nariai analysis as long as we do the identifications $x \to X$ and $\sigma \to \delta$ in these formulas. We want modes that are regular at $X=0$ (which corresponds to have outgoing boundary conditions at the cosmological horizon in the full geometry) and the condition that the solutions should decay at large $X$ quantizes the frequencies. The latter condition is poorly motivated but it gives results that agree well with our numerics.

Recall, from Section \ref{sec:isospectral}, that the  vector-type and scalar-type sectors of perturbations are isopectral in the Nariai limit. We find that the near-Nariai frequency spectrum is given by
\begin{eqnarray} \label{Nariaispectrum}
\hspace{-0.2cm}\omega\,r_c \big|_{{\rm Nariai}}&\simeq&
\left[ \frac{1}{4} \left(2 \mu ^2-1\right)
\sqrt{\frac{4 \ell  (\ell +1)-5+10 \mu ^2+4 s \sqrt{4 \mu ^2 \left(\mu ^2+\ell ^2+\ell -1\right)+1}}{1-2 \mu ^2}} \right. \nonumber \\
&-& \left. \frac{1}{4} i \left(1-2 \mu ^2 \right) (2 n+1) 
\right] \frac{r_c-r_+}{r_c}  +\mathcal{O}\bigg( \Big(\frac{r_c-r_+}{r_c}\Big)^2 \bigg), 
\end{eqnarray}
where $ n\in \mathbb{N}_0$ is the overtone of the mode with angular quantum number $\ell$ and $s=\pm 1$ for the modes $\Phi_\pm $, respectively. Note that $\omega\,r_c \big|_{\rm Nariai}\to 0$ as $r_+\to r_c$,  {\it i.e.} as $y_+\to 1$.

The frequencies \eqref{Nariaispectrum} of near-Nariai modes have a real and imaginary part. This analytical approximation is strictly valid in the near-Nariai limit, $\delta \ll 1$ ({\it i.e.}  $y_+ \to 1$), $Q\ll Q_{\rm ext}$ and for small frequencies, $|\omega\,r_c|\ll 1$. 
So what are these modes? Do they represent a fourth family of modes in RNdS? 

To answer this question we attempted different strategies. In one of them we fix the black hole parameters and the quantum number $\ell$ and we solve the perturbation master equation as an eigenvalue problem to find the frequencies that are allowed in the background. After identifying the frequencies $-$ including the first few overtones $n\geq 0$ $-$ that describe  the 1) de Sitter, the 2) photon sphere and 3) near-extremal modes we do not find evidence of a new fourth family of modes. In a second approach, we use a Newton-Raphson algorithm whereby we give directly \eqref{Nariaispectrum} as a seed (in a region of parameter space, {\it i.e.} $y_+\sim 1$, where it is a good approximation). Again, such a code does not converge to a new fourth family of modes. Instead, this Newton-Raphson code always converges for the family of modes that we have already identified as being the photon sphere of modes. Moreover, this happens not only when we search for the leading radial overtone, $n=0$ in the seed  \eqref{Nariaispectrum}, but also for the first few other overtones that we attempted ($n=1,2,3$). 

\begin{figure}[th]
	\centering
	\includegraphics[width=0.45\textwidth]{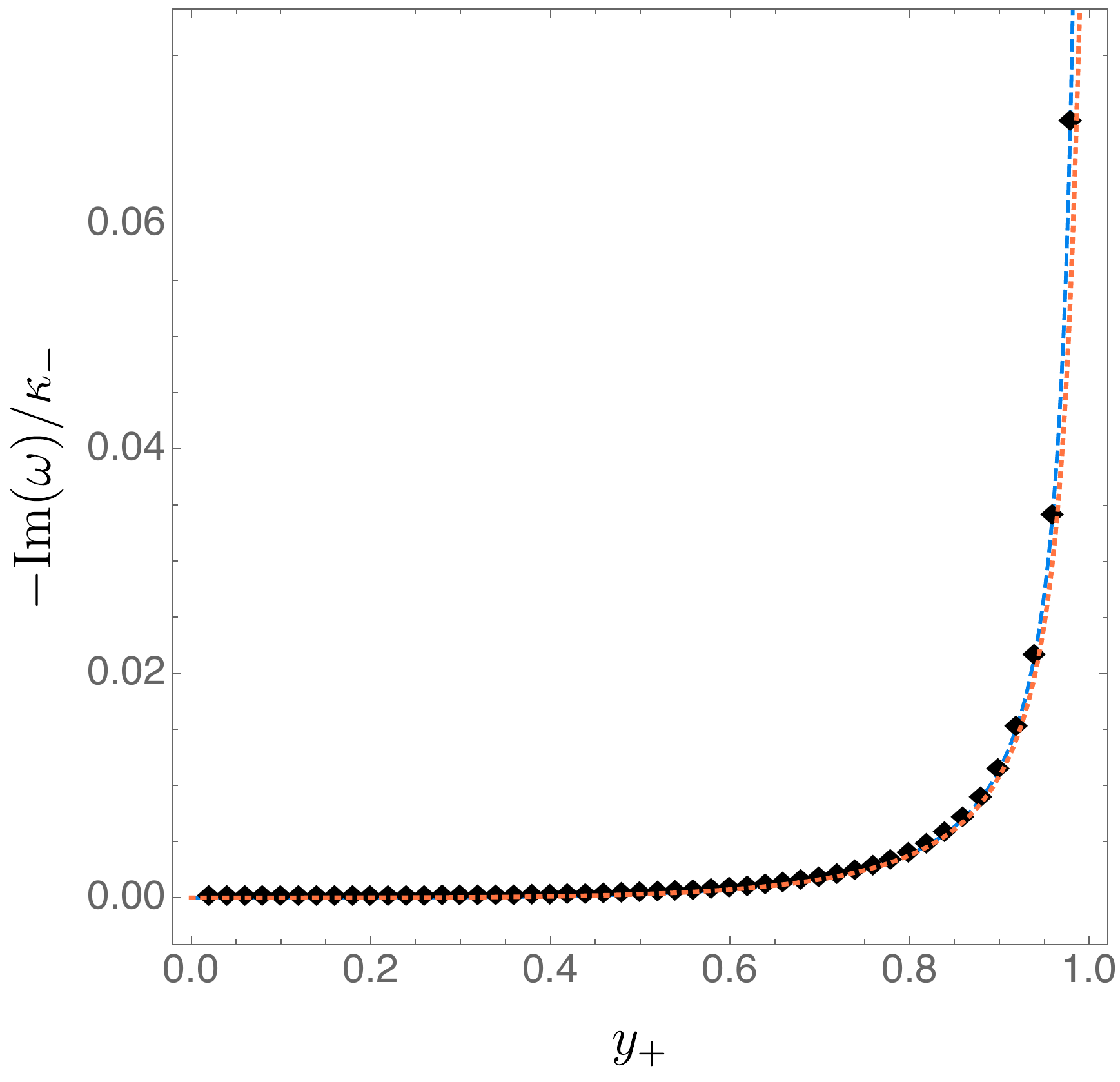}
	\hspace{0.5cm}
	\includegraphics[width=0.46\textwidth]{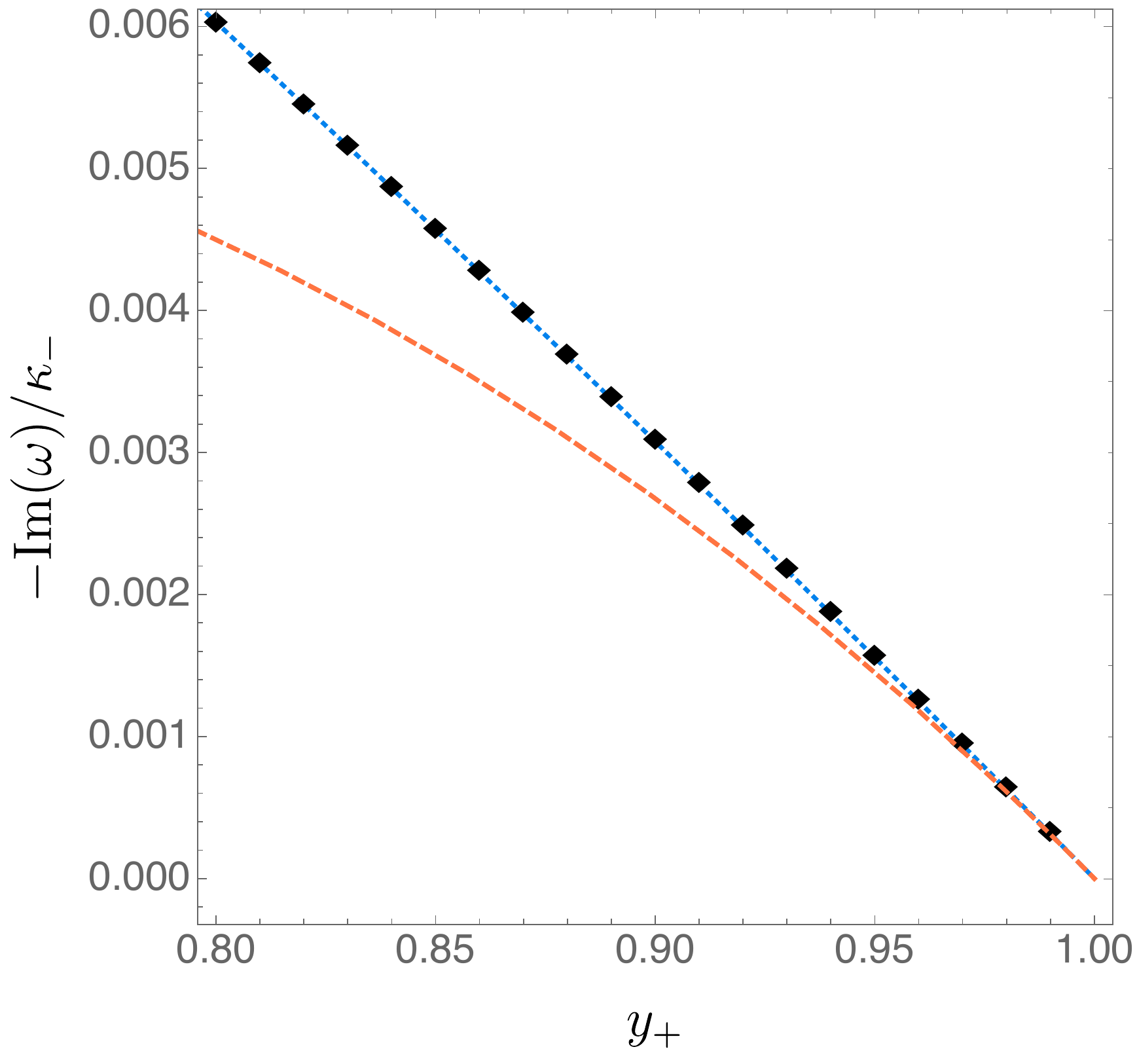}
	\caption{Photon sphere  family of  modes $\Phi_-$ with $\ell=2$, $n=0$ and the Nariai limit.  In both plots, the dashed orange line refers to the Nariai analytical prediction \eqref{Nariaispectrum}  for  $-{\rm Im}(\omega_{\rm Nariai})/\kappa_-$  (with $\ell=2, n=0$), while the dotted blue curve is the analytical photon sphere prediction \eqref{wWKB} for $-\mathrm{Im}(\omega_{\mathrm{WKB}})/\kappa_-$ (with  $n=0$). {\bf Left panel:}  $-{\rm Im}(\omega)/\kappa_-$ as a function of  $Q/Q_{\rm ext}$  at fixed $y_+=0.99$, {\it i.e.} $r_+=0.99 \,r_c$.  {\bf Right panel:}  $-{\rm Im}(\omega)/\kappa_-$ as a function of $y_+$ at fixed $Q/Q_{\rm ext}=0.4995$. Note that, as expected, the Nariai analytical prediction is a good approximation only near $y_+\sim 1$. It seems to describe the $y_+\sim 1$ limit of the photon sphere modes (black diamonds and WKB dotted blue line).} 
	\label{fig:Nariai}
\end{figure} 

We consider that our experiments give good evidence to support the claim that there is no fourth family of quasinormal modes that can be associated to a Nariai origin. Instead, the Nariai frequencies simply give a good approximate description of photon sphere modes in the limit where $y_+ \to 1$. These conclusions are best illustrated in Fig.~\ref{fig:Nariai}. In the left panel, we fix $y_+=0.99$ and the dashed orange curve describes the analytical Nariai expression \eqref{Nariaispectrum} prediction whereas the dashed blue line is the WKB prediction \eqref{wWKB} for the photon sphere modes. The black diamonds represent the outcome of our Newton-Raphson search when we give the Nariai frequency \eqref{Nariaispectrum} as a seed. Both the near-Nariai and WKB photon sphere predictions agree very well with the numerical data although, as expected, the near-Nariai result works less well at large $Q/Q_{\rm ext}$. In the right panel of Fig.~\ref{fig:Nariai}, we fix $Q/Q_{\rm ext}$ and vary $y_+$. Again, the black diamonds represent the outcome of our Newton-Raphson search when we give the Nariai frequency \eqref{Nariaispectrum} as a seed. As expected, the near-Nariai prediction \eqref{Nariaispectrum} is very good for $1-y_+\ll 1$ but quickly gets worst as $y_+$ decreases. The black diamonds turn out to be exactly the photon sphere modes that we had already found in an independent analysis. This is confirmed by the agreement with the WKB prediction \eqref{wWKB}. The results presented in this plot are qualitatively the same for any other value of the charge ratio $Q/Q_{\rm ext}$ (and we did a fine-tunned search which spanned the full interval $0<Q/Q_{\rm ext}<1$). 

\begin{figure}[tb]
	\centering
	\includegraphics[width=0.47\textwidth]{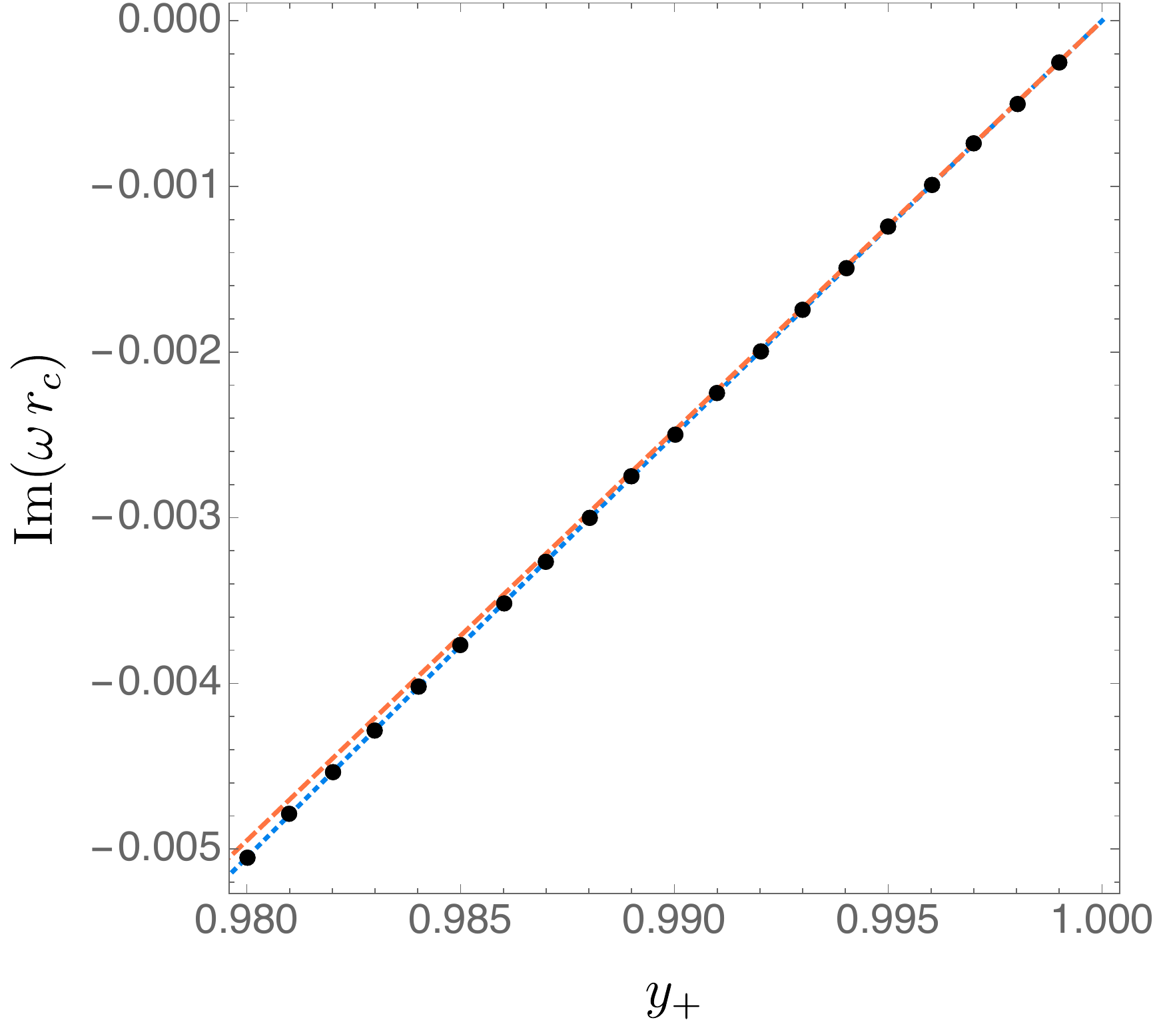}
	\hspace{0.5cm}
	\includegraphics[width=0.46\textwidth]{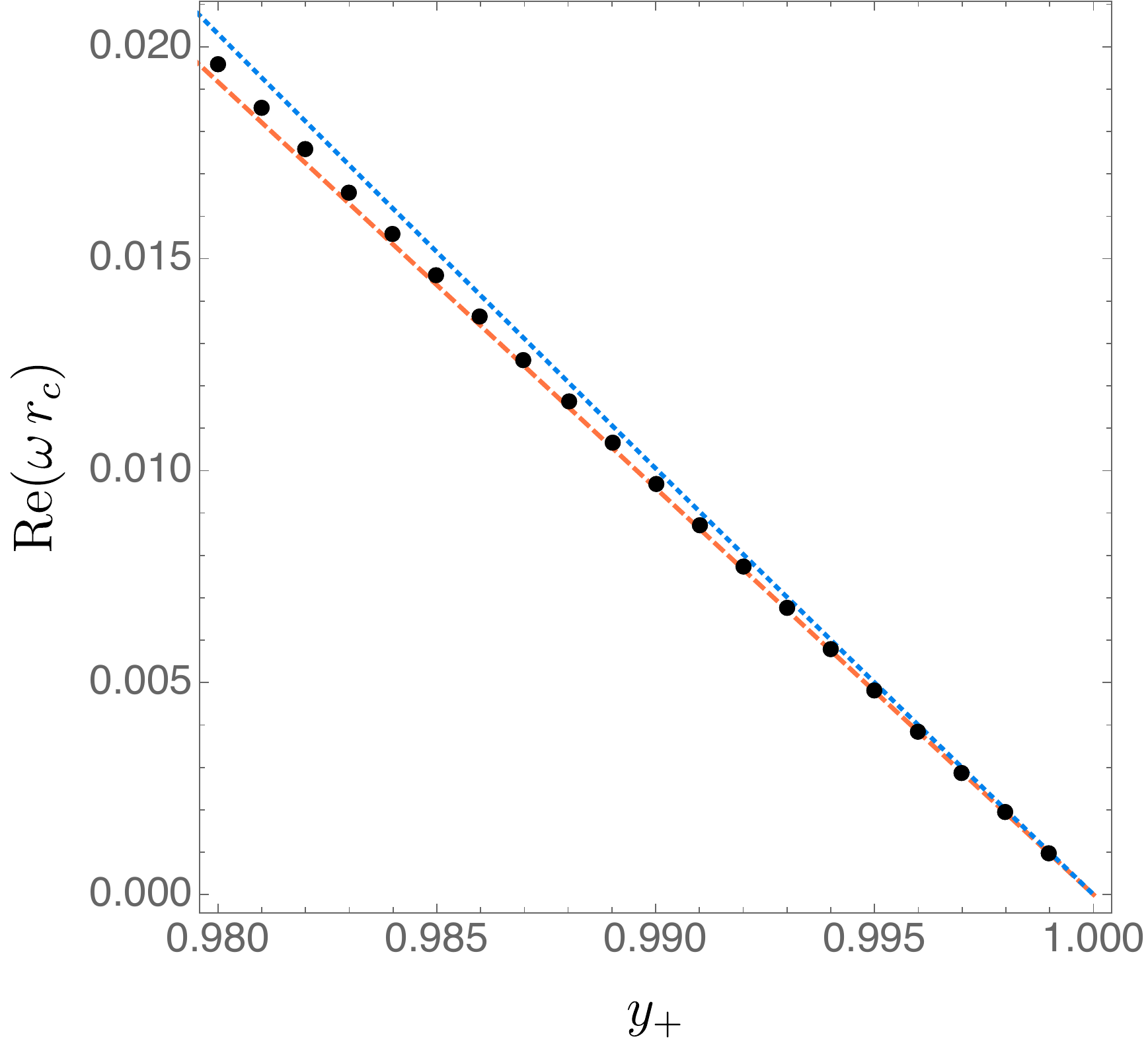}
	\caption{Photon sphere  family of  modes $\Phi_-$ with $\ell=2$, $n=0$ and the Nariai limit for RNdS black holes with $Q/Q_{\rm ext}=0.0999$.  In both plots, the dashed orange line refers to the Nariai analytical prediction \eqref{Nariaispectrum}  for $\ell=2, n=0$, while the dotted blue curve is the analytical geometric optics/WKB photon sphere prediction \eqref{wWKB} for $\ell=2, n=0$. {\bf Left panel:}  ${\rm Im}(\omega \, r_c)$ as a function of  $y_+$  close to the Nariai limit $y_+\sim 1$.  {\bf Right panel:}  ${\rm Re}(\omega \,r_c)$ as a function of  $y_+$  close to the Nariai limit $y_+\sim 1$.} 
	\label{fig:NariaiZoom}
\end{figure}

Analysis similar to the one displayed in Fig.~\ref{fig:NariaiZoom} further reinforce our conclusion. In this figure, we take $Q/Q_{\rm ext}=0.0999$ and we focus our attention in the interval $0.98<y_+<1$, {\it i.e.} very close to the Nariai limit $y_+ \to 1$. We display the modes we obtain with a Newton-Raphson search when we give analytical Nariai expression \eqref{Nariaispectrum}  as a seed. In the left panel we plot the imaginary part of the frequency, while in the right panel we plot the real part of the frequency. The left panel exemplifies again what we already know: as discussed in the previous cases, both the WKB expression (blue dotted curve) and the Nariai expression (orange dashed curve) give good approximations for ${\rm Im}(\omega\, r_c)$ when $y_+\sim 1$ and as we move away from the Nariai limit, the analytical expression  \eqref{Nariaispectrum} starts being a less good approximation. On the other hand, the right panel of Fig.~\ref{fig:NariaiZoom} shows that the analytical Nariai expression \eqref{Nariaispectrum}  yields an approximation for the real part of the frequency that is actually even better than the WKB approximation  \eqref{wWKB}, as long as $1-y_+\ll 1$. However, we would expect that including higher order terms in $1/\ell$ would improve the accuracy of the WKB prediction, which is already remarkably accurate given that we are working with $\ell=2$.

To conclude, we have shown that the Nariai result \eqref{Nariaispectrum} simply describes the photon sphere family of quasinormal modes in the $y_+\to 1$ limit. In the case of a massless scalar field perturbation of RNdS, the analysis of \cite{Cardoso:2017soq} reached the same conclusion.

\bibliographystyle{JHEP}
\bibliography{rnds}{}

\end{document}